\documentclass[5p,twocolumn,10pt,times]{elsarticle}
\usepackage{amsmath}
\usepackage{hyperref}
\usepackage[utf8]{inputenc}
\usepackage{graphicx}
\usepackage{subcaption}
\graphicspath{{images/}}
\usepackage{amsmath,bm}
\usepackage{comment}
\usepackage[normalem]{ulem}
\usepackage{xcolor}
\usepackage{algorithm}
\usepackage{algorithmic}
\usepackage{yhmath}

\begin{document}
\baselineskip11pt

\begin{frontmatter}

\title{Baseline Skinning for Point Sets of Articulated Bodies}

\author{Tong Fu}
%\cortext[mycorresponingauthor]{Corresponding author}
%\ead{first@firstuni.edu}
\author{Raphaëlle Chaine}
\author{Julie Digne}
\address{Université de Lyon, UCBL, CNRS, France}
%\author{Third Author}https://www.overleaf.com/project/5fa12a4922e8f7192ef8f22b
%\address{Department of Mathematics, Stanford University}

%\author{Submission \#}

%\teaser{
%    \centering
%    \includegraphics[width=0.8\linewidth]{images/teaser1.png}
%    \caption{Dancing dragons in the cloud. We change the pose of a point set of dragon with our baseline skinning method. The first dragon at left is the initial point set. }
%    \label{fig:my_label}
%}

\begin{abstract}
%General skinning techniques aim at deforming the surface of a character \raph{following the pose change of an inner skeleton}. \raph{Because of their rapidity,} they are mainly used for real-time  animation. Although popular skinning algorithms are simple, they \raph{suffer from geometric artefacts that are not realistic}. Some research domains, such as virtual restoration of statues \raph{may reuse parts of existing character surfaces after pose change. They} require a more realistic skinning result \raph{without integrating muscle modelling but rather respect for the artist's style}. We propose a novel skinning method that \raph{encode the point set details above a bundle of baselines that cover the surface.} \raph{To accompany the different rotational movements of a bone or a joint, we propose a geometrical model of the evolution of baselines but also of the directions in which the detail points are reported.} Our approach improves artefacts of classical skinning method and gives a result \raph{that preserves the geometric detail}. It works directly on point sets in order to preserve the accuracy of the initial sampling. \raph{The baseline skinning approach does not involve weights to be defined for each points.} We demonstrate plausible deformation results of our approach on several point set of statues. 

General skinning techniques aim to deform the surface of an articulated model following the pose change of a skeleton. Their rapidity makes them ideal tools for real-time animation purposes. However, popular skinning algorithms are simple, but they tend to generate undesirable geometric artefacts.
In our work, we consider skeletons given in the form of sphere-mesh models controlling both the pose and morphology of the shape that is either described as a mesh or a raw point set.
We propose a novel skinning method that encodes the point set details above a bundle of baselines covering the sphere-mesh. In particular, we propose a geometrical model of the baseline and detail direction evolution during bone twisting and joints bending rotations.
Our approach works directly on point sets and thus preserves the accuracy of the initial sampling. It further avoids computing a weight per point or a costly explicit muscle modelling step. We evaluate our method on several articulated body point sets, showing that it creates fewer artefacts than classical methods.
\end{abstract}


\end{frontmatter}

%\linenumbers
\section{Introduction}
The main idea of skinning is to deform a surface following an underlying skeletal animation. Each surface point is influenced by one or more skeleton bones, and weights describe the influence of relevant bones on each point (usually a mesh vertex). Skinning techniques are widely used in computer animation or shape modeling to change a character's pose and shape.

Many widely used geometric skinning methods, such as Linear Blend Skinning~\cite{Magnenat-thalmann88joint-dependentlocal} or Dual Quaternion skinning~\cite{Kavan2008GeometricSW}, suffer from volume collapse problems and need a careful 
%manual 
design of per-vertex weights. Example-based methods~\cite{Mohr:2003:DMI:641480.641488} and physics-based methods do not suffer from such artefacts. However, they are computationally expensive and require additional input data.

Our goal is to provide a point set skinning technique that preserves the geometric detail and keeps the initial sampling accuracy. We assume that a skeleton, in the form of a sphere-mesh model, controls the pose and the dimensions of the overall shape of the point set.
Our approach encodes the shape details as a vector field above a set of baselines covering the sphere-mesh model's surface. We propose to associate a base point on the articulated sphere-mesh model to each of the original shape points. Each base-point is located on a baseline over the sphere-mesh model, and the position of a base-point after deformation follows its baseline's motion. We then deduce the new position of an input point from its corresponding base-point's new position by adapting its initial detail field value.
Our approach's originality is not to require a mesh with fixed connectivity whose triangles quality may be altered by deformations related to poses and anatomy changes, possibly creating triangle slivers and self-intersections. Meanwhile, our point set skinning process does not need to determine and compute weights.

Applications of our method include virtual artwork restoration. In virtual archaeology, one way of restoring statues is to combine parts from different statues after bringing them to a common pose and morphology and require hence artefact-free skinning results while avoiding the explicit simulation of muscles that may not respect the artistic style.
Unlike an animation of 3D characters, skinning for virtual restoration is more sensitive to preserving the initial surface details. Besides, it requires not only rigid transformations but also affine transformations such as scaling. In this context, our baseline skinning method gives relevant results and preserves the details of the input shape.

To summarize, our main contributions are:
\begin{itemize}
	\item A skinning method working directly on a point set.
	\item An approach to encode a point set above a sphere-mesh using a specific detail direction field.
	\item A full geometric model of the evolution of a detail field following a sphere-mesh model deformation.
	\item A skinning method able to handle multi-layered details deformation. 
	\item A method that can handle articulated bodies such as human shapes as well as imaginary creatures.
\end{itemize}

%\begin{itemize}
%    \item A skinning method working directly on a point set.
%    \item A full geometric model of the evolution of a detail field following a sphere-mesh model deformation.
%    \item A method that can handle articulated bodies such as human shapes as well as imaginary creatures.
%    \item A skinning method able to handle multi-layered details deformation. 
%\end{itemize}

\section{Related work}
The most famous and simple skinning method is Linear Blend Skinning (LBS for short)~\cite{Magnenat-thalmann88joint-dependentlocal}. A vertex on a mesh surface is transformed by a linearly weighted combination of the motions related to the moving bones it is attached to. Despite well-known limitations (the candy wrapper effect and the elbow collapse effect), Linear Blend skinning is still the standard skinning method for real-time animation purposes. Many skinning approaches base on the principle of LBS and improve its limitations, such as Pose Space Deformation~\cite{Lewis:2000:PSD:344779.344862}, Log-matrix blending~\cite{Alexa2002LinearCO,Thalmann04}, Optimized centers of rotation~\cite{leCOR16}, Multi-Weight Enveloping~\cite{Wang:2002:MEL:545261.545283}, spherical Skinning~\cite{Kavan:2005:SBS:1053427.1053429} and Dual Quaternions Skinning~\cite{Kavan2008GeometricSW}. Jacobson and Sorkine \cite{STBS:2011} proposed to add extra endpoint weights to the LBS formulation so that it can handle properly stretch and twist. The endpoint weight of one vertex for one bone varies from 0 to 1, which indicates the original vertex’s position relative to the bone’s endpoints. The twist rotation’s angle of this vertex on the bone is a linear combination of the twisting angle of the two endpoints. The bend rotations are carried out afterwards on the positions resulting from the twist. Kavan and Sorkine~\cite{Kavan2012ElasticityinspiredDF} then proposed a joint-based deformer that differentiates bend and twist motions at each joint. The deformer applies a spherical interpolation of the twist (as DQS) and then a linear interpolation of the bend (as LBS). The weights for bend and twist are different which are precomputed separately as their energy-minimizing weights for linear blend skinning.
Fu et al.~\cite{h.20201290} further improved the result by handling bend rotations anisotropically. But there are still some artefacts at joints such as a bulge effect.

Furthermore, setting the suitable weights is an essential question for these skinning methods: while the profile of the weights is generally sketched by graphic designers~\cite{Mohr:2003:DMI:641480.641488}, there exist automatic weighting techniques that, for example, use heat diffusion~\cite{Baran:2007:ARA:1276377.1276467, Thiery13, viper}, geodesic voxel binding \cite{Dionne2013GeodesicVB} or bounded biharmonic weights~\cite{Jacobson2014BoundedBW}. Applying suitable weights on the transformation matrices of bones gives a smooth transition at bones joints when the joints are bending, but they are not sufficient to handle twisting or bone stretching. Our method does not require such weights computation and can deal with all these deformations in a unified way.
A recent skinning method \cite{Le:2019:DDM:3306346.3322982} corrects artefacts of Linear Blend Skinning by locally estimating the rigid transformation that best restores the relative position of a vertex concerning its neighbors using Laplacian differential coordinates. This method, designed for meshes, involves a definition of details in terms of Laplacian differences. In our approach, we instead define the detail as the residual over the registered anatomical model.
Instead of using a skeleton, some methods~\cite{cage1,cage2} rely on a cage deformation which can deal with more general deformation like twisting or stretching. The deformations are controlled using a flexible cage that consists of a closed three-dimensional mesh. Cage deformations are designed for unrealistic articulated characters such as cartoon characters or objects and can work in 2D or 3D.

Going in a different direction, physics-based methods simulate the growth of skeletal muscles and fat tissues using a biological model. Hamadi et al.~\cite{Ali-Hamadi:2013:AT:2508363.2508415} transfers the volume delimited by a mesh to the interior of another mesh by minimizing some harmonic energy. To do so, a deformation field is computed between the two meshes, based on the nearest point matching which is updated at each iteration \cite{gilles:inria-00516374}. However, the process is not fully automatic.  Recent researches ~\cite{Saito:2015:Computational,Ichim2017PhacePF,kadlecek-16-reconstructing} achieve the desired deformation of the human body or face by direct control of each muscle. These methods are computationally expensive and need to define the physical model from physiological data. They are not suitable for artistic shapes since the morphology, and the representation of muscles in artwork may differ from a biological model.

Taking a different perspective on the problem, Implicit skinning~\cite{Vaillant:2013:ISR:2461912.2461960} uses an additional implicit formulation of the surface that better supports pose changes and re-projects skinned vertices on the implicit model after each pose change. In this paper, we also use a proxy model, but it is explicit. 
Volume preserving skinning methods~\cite{Funck2008VolumepreservingMS,Rohmer2009ExactVP,Angelidis2007KinodynamicSU} correct for volume changes through the generation of extra bulges and wrinkles. They use vector fields induced by skeletal motion to describe the skin deformation. Some skinning methods involve a rough modeling of muscles combined with implicit skinning \cite{Roussellet2018DynamicIM}.

Example-based methods~\cite{Sloan2001ShapeBE,Wang2007RealtimeEW,Mohr2003BuildingEA,Kry2002EigenSkinRT} produce more realistic results but require extra training data. They have a limitation for a given range of deformations. In addition, these methods can only be as good as their training dataset is, and for unrealistic characters (imaginary creatures or characters with unrealistic body proportions), relevant datasets are nontrivial to build.

\section{Method}

As defined above, \emph{Skinning} processes allow deforming the skin following an underlying skeletal animation. In most skinning methods, the skin is a 3D mesh whose vertices are attached to a set of bones using different weights, and the skeleton is a tree whose nodes represent joints of the skeleton and edges represent bones.
In this work, we propose to consider the skin of the model as a set of detail points, encoded as a set of displacements on top of a sphere-mesh articulated model and transferred back on this model after pose or morphology changes. The originality of our approach is that the direction in which the detail is encoded does not necessarily correspond to the normal, especially in concave areas of the sphere-mesh. Furthermore, this detail direction may evolve after a pose change. The detail direction field is hence not a simple heightfield \cite{Policarpo2005RealtimeRM}.
%Ajouter une réference sur les papiers qui décrivent les détails comme un champs de hauteur

\begin{figure*}[ht]
    \centering
        \begin{subfigure}[ht]{0.25\textwidth}
            \centering
            \includegraphics[width=\textwidth]{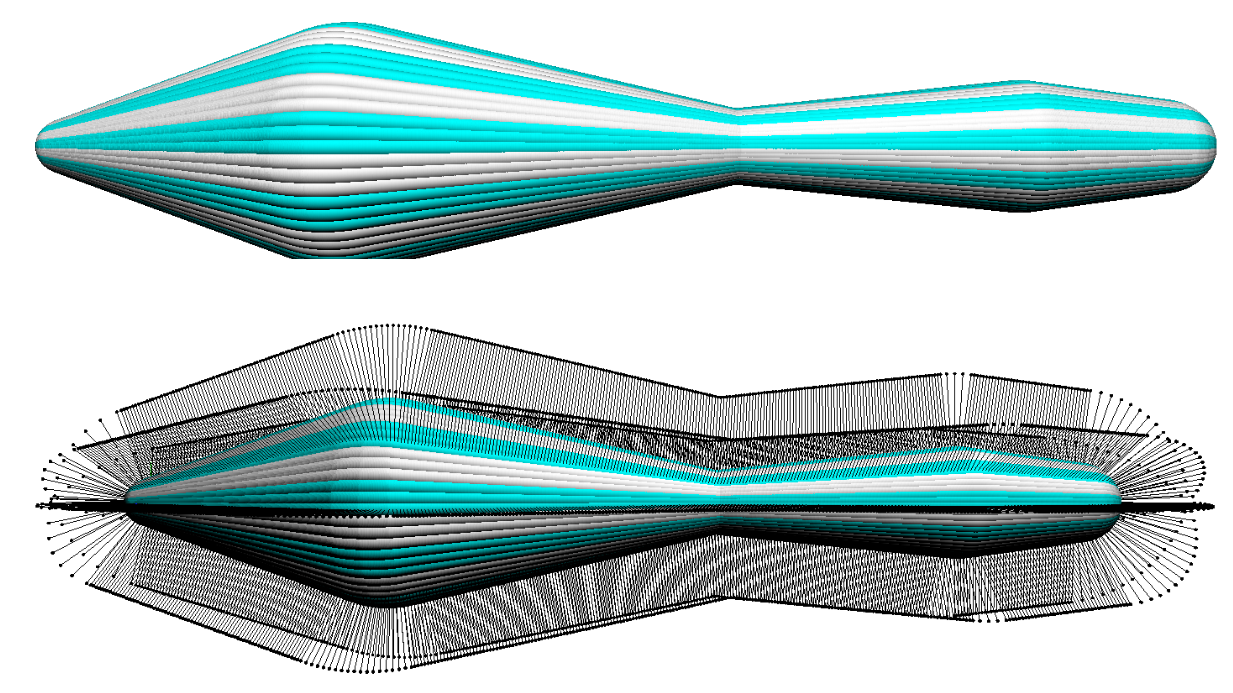}
        \caption{}
        \end{subfigure}
        ~
        \begin{subfigure}[ht]{0.25\textwidth}
            \centering
            \includegraphics[width=\textwidth]{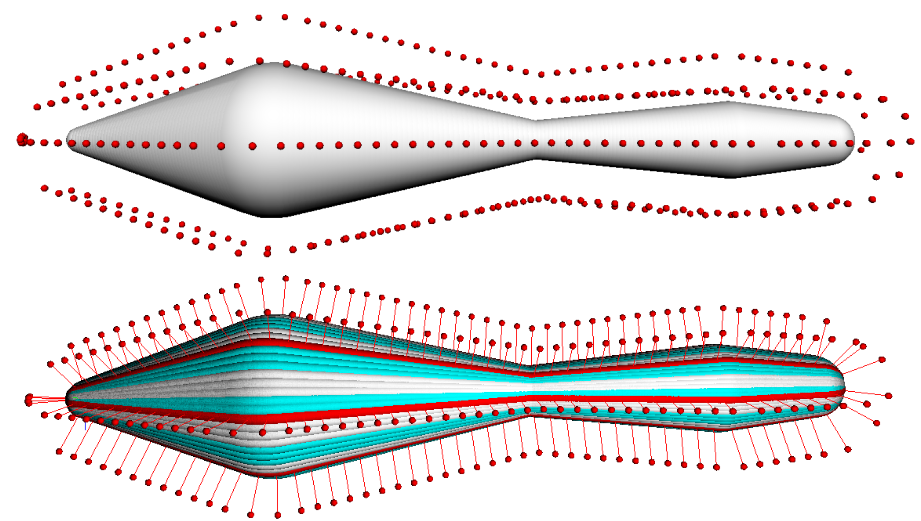}
        \caption{}
        \end{subfigure}
        ~
        \begin{subfigure}[ht]{0.25\textwidth}
            \centering
            \includegraphics[width=\textwidth]{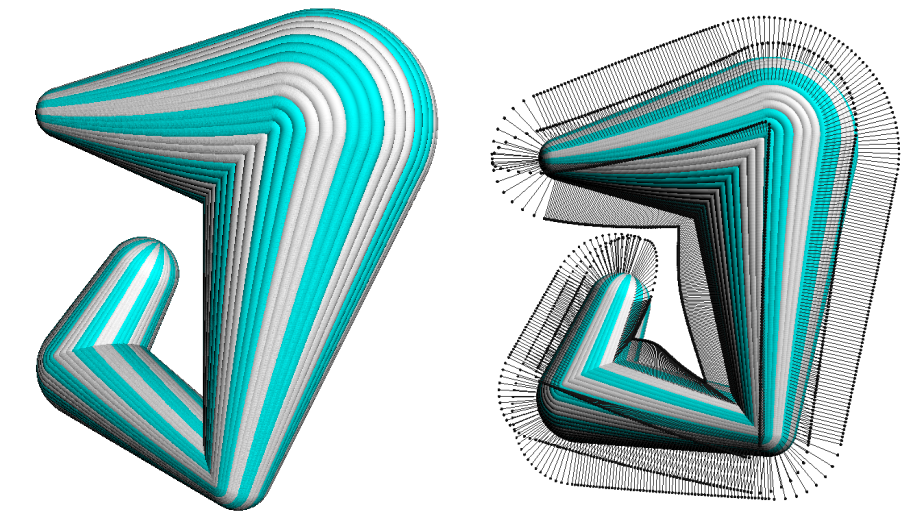}
        \caption{}
        \end{subfigure}
        ~
        \begin{subfigure}[ht]{0.13\textwidth}
            \centering
            \includegraphics[width=\textwidth]{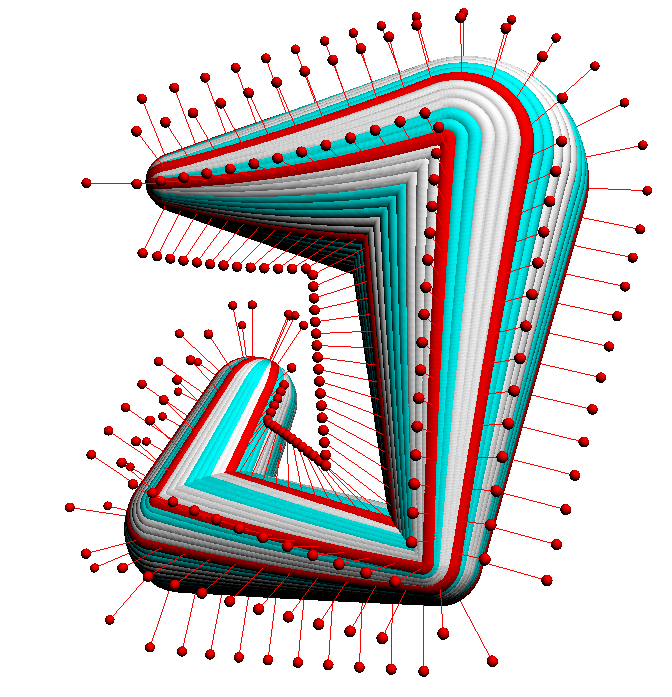}
        \caption{}
        \end{subfigure}
    \caption{The pipeline of baseline skinning: (a) Baseline and detail direction field over the sphere-mesh model; (b) Point set encoding using the detail direction field; (c) Baseline segments are deformed after the bend of joints or the twist of bones; (d). Detail points are updated accordingly by sliding along the deformed baselines and being lifted in new detail directions. }
    \label{fig:pipeline}
\end{figure*}

Starting from a shape point set with a registered sphere-mesh model, we associate each input point to a base point on the sphere-mesh model to have a relative position with respect to the model.
The principle of our baseline skinning is that each base point lies on a baseline, \emph{i.e.} a line defined on the sphere-mesh model, whose profile evolves continuously with the deformations of the model. After applying skinning on base points, we lift them by adding the detail field in the directions corresponding to the new positions to recover the detailed point set after deformation. The pipeline of our approach is illustrated in figure \ref{fig:pipeline}.
%\raphq{Ajouter un pipeline avec une chaine de 3 os rayée, dont on se sert pour modéliser un nuage de points et que l'on déforme, comme sur les slides de présentation.} 
Along with the bones of the sphere-mesh model, the baseline motion can be thought of as mimicking longitudinal muscle's change, and base points locally slide on the moving baseline to mimic the sliding of the skin on and around joints, although these motions are not driven by an explicit physical model. Similarly, the direction in which the detail is reported will change according to the sphere-mesh motion, as if details were carried by hairs that push against each other and lean in the concavity of a joint. Therefore, our skinning process is different from the skinning method presented in the original sphere-mesh paper~\cite{Thiery13, Thiery2016AnimatedMA}, which remains in the classical skinning framework (i.e. LBS).

In the following, we briefly describe the sphere-mesh model introduced in \cite{Thiery13}. Then we explain how to find the baseline and the base point for a given input point. Finally we derive our baseline skinning algorithm.

\subsection{Registered sphere-mesh model}

\begin{figure}[ht]
    \centering
    \begin{subfigure}[ht]{0.22\textwidth}
        \centering
        \includegraphics[trim = 0mm 12mm 0mm 12mm, clip, width=\textwidth]{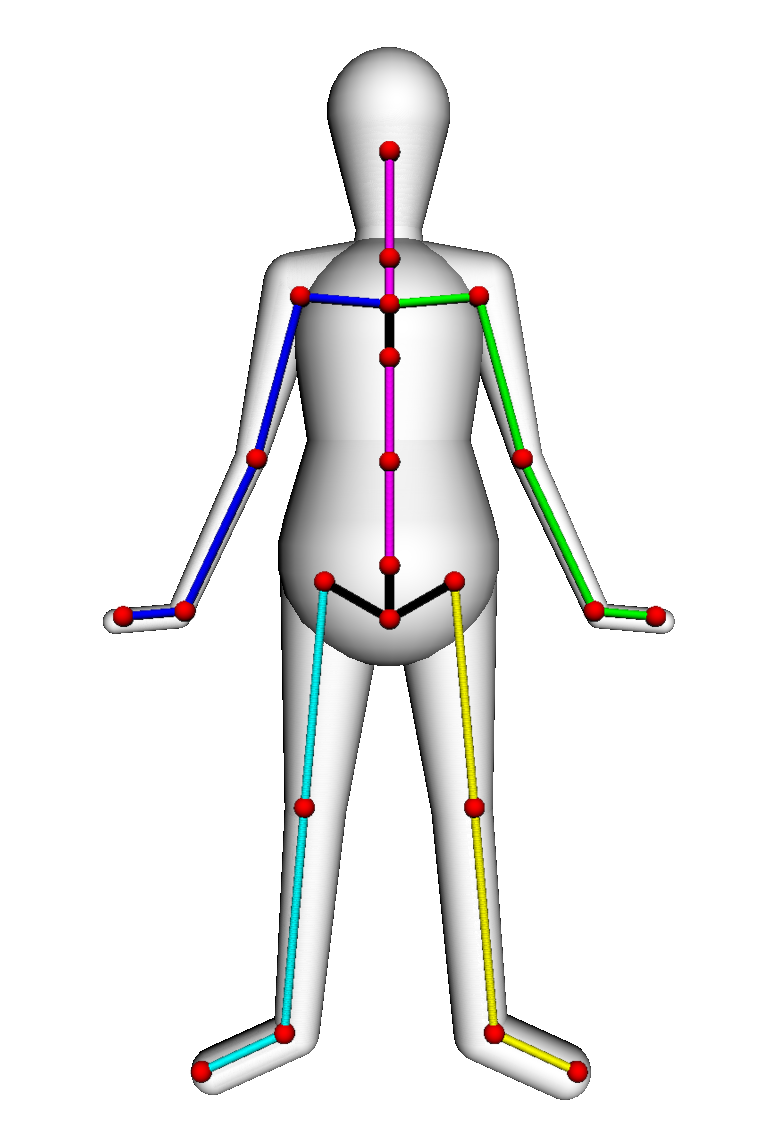}
        %\caption{control skeleton and sphere-mesh model}
        %\label{fig:human_model}
    \end{subfigure}
    ~
     \begin{subfigure}[ht]{0.22\textwidth}
        \centering
        \includegraphics[trim = 0mm 12mm 0mm 12mm, clip, width=\textwidth]{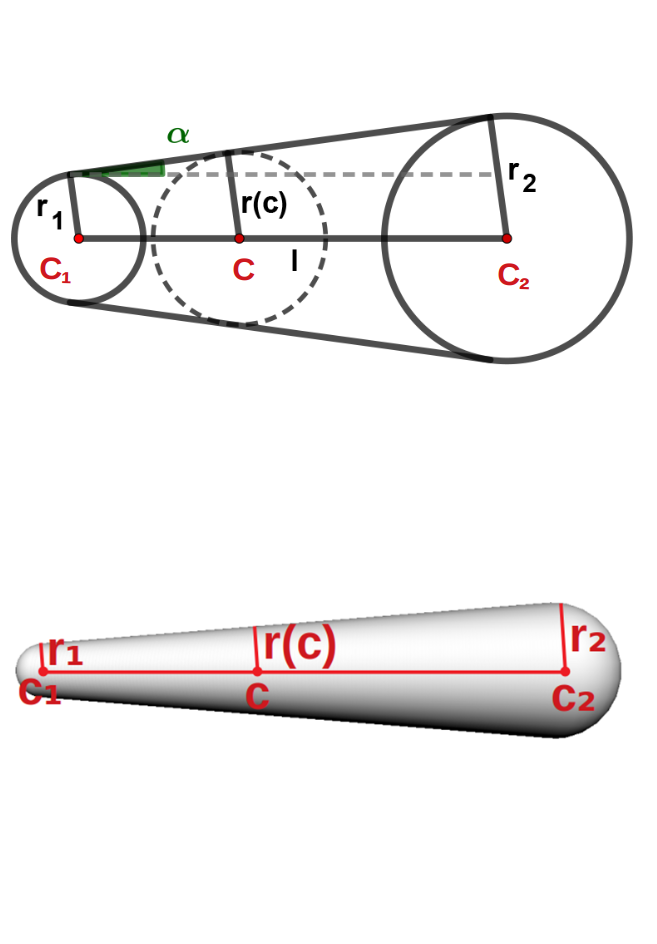}
        %\caption{2D Cross-section and 3D sphere-mesh of a bone}
        %\label{fig:spmesh_bone}
    \end{subfigure}
    \caption{Left figure shows the sphere-mesh model for human-like body. Right figure shows a 2D cross-section and a 3D sphere-mesh of a bone.}
    \label{fig:model}
\end{figure}

Our method uses a \emph{sphere-mesh model}~\cite{Thiery13} as a skeleton guiding the shape deformation. A volumetric sphere-mesh is a union of spheres centered on a simplicial complex, with linearly varying radius between vertices. In our approach, we only consider 1D bones, which means that each bone $B(l,\mathbf{r})$ corresponds to the union of a set of spheres centered on a segment with a linearly varying radius (Figure \ref{fig:model}). The bone model is controlled by the length $l= \|C_1C_2\|$ and the pair of sphere radii $\mathbf{r}=\{r_1,r_2\}$ where $C_1$, $C_2$ are the two end sphere centers and $r_1$, $r_2$ are the associated radius respectively. The segment $C_1C_2$ is the medial axis of the bone. For each sphere center $C \in C_1C_2$,  the radius of the sphere centered at $C$ is $r(C) = (1- \rho ) r_1 + \rho r_2$, with $ \rho = \frac{\|C_1C\|}{\|C_1C_2\|}$. We denote by $\alpha$ the angle of the conic part of the bone, as illustrated on Figure \ref{fig:model}. A chain of bones is a sequence of bones with each bone sharing a common end sphere with the next bone. The skeleton that we use is a tree composed of several chains of bones connected at some junctions spheres.
We further assume that this sphere-mesh model is registered to the input point set, \emph{i.e.} the model is calibrated to match the input shape, and we know which point is associated to which bone. In practice, to get this initial registration, we use the FAKIR algorithm~\cite{Fu2020FAKIRAA}.

\subsection{Initial baselines and detail direction field}

\paragraph{Baseline definition}

Given a chain of bones, a baseline is a continuous curve defined on the surface of the sphere-mesh. Before a pose change, it is composed initially of a constrained sequence of segments and circular arcs. The segments are carried by the conical parts of the bones (generators), and the circular arcs are carried by the sphere caps at the articulation between two bones. 
The constraints on the circles and segments composing the baseline are as follows:

\begin{itemize}
    \item There is one segment per bone.
    \item The segments belonging to two consecutive bones are constrained to belong to the same plane containing the apexes of the two cones. If they do not share a common endpoint, these two segments are connected in a $C^1$ manner by a circular arc also belonging to the plane of the two segments.
\end{itemize}  

\begin{figure}[ht]
    \centering
        \begin{subfigure}[ht]{0.22\textwidth}
            \includegraphics[width=\textwidth]{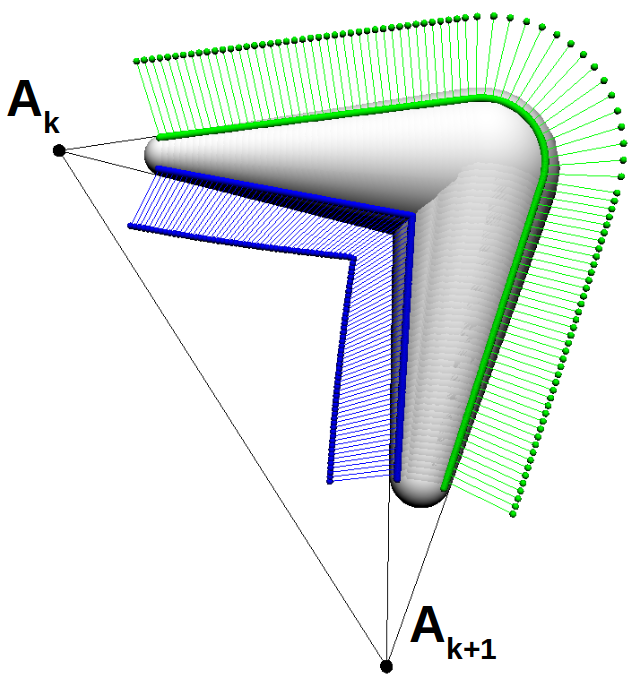}
            \caption{}
            \label{fig:baseline_3d_1}
        \end{subfigure}
        \begin{subfigure}[ht]{0.24\textwidth}
            \includegraphics[width=\textwidth]{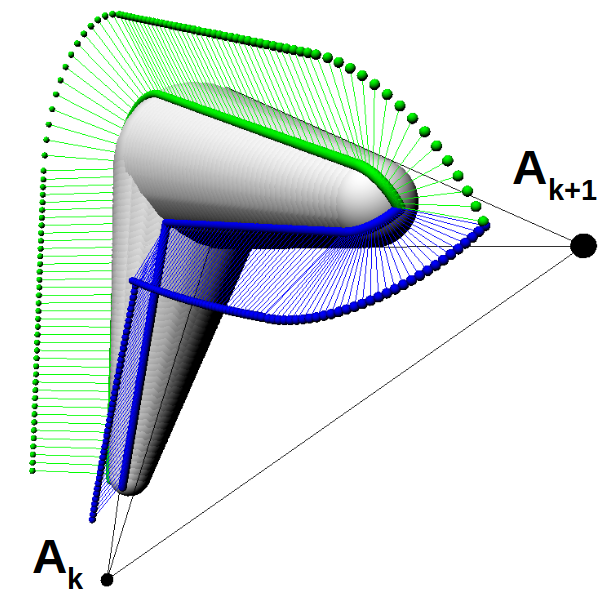}
            \caption{}
            \label{fig:baseline_3d_2}
        \end{subfigure}
        %\caption{\raph{Baselines and direction fields associated to the plane $\tilde{p}A_1A_2$}}
    \caption{(a) Pair of baselines defined in a support plane passing through $A_k$ and $A_{k+1}$ with their associated direction fields. The direction field along the blue baseline is not included in the support plane but each vector is inversely oriented towards the axis of its corresponding cone. (b) baselines closure on a free extremity of a bone.} 
    \label{fig:baseline_3d}
\end{figure}

Figure \ref{fig:baseline_3d} illustrates two portions of baselines on a two bones chain. The one in green contains a circular arc, while the one in blue illustrates the case where two segments directly share a common endpoint.

% A baseline is a curve defined on the sphere-mesh model. Initially, before a pose change, it is composed of a continuous succession of segments on the sphere-mesh cone parts and circular arcs on the sphere-mesh spherical caps. Each circular arc of a baseline curve is coplanar with the segment that precedes it and the segment that follows it (green curve in Figure \ref{fig:baseline_3d}). Two consecutive segments are not necessarily separated by a circular arc, e.g. if they connect on the cone parts of the bones (blue curve in Figure \ref{fig:baseline_3d}). 

%\paragraph{Baseline property}
\paragraph{Building a baseline}
Starting from a point on a cone part of the sphere-mesh, we consider the cone generatrix that passes through that point: the first segment is the part of the generatrix corresponding to the sphere-mesh surface restricted to that cone. Then, the rest of the baseline is built using the following principle:
Given a baseline segment on one bone $B_k$, the corresponding baseline segment on subsequent bone $B_{k+1}$ is obtained by intersecting the sphere-mesh surface restricted to $B_{k+1}$ with the plane containing the segment on $B_k$ and the cone apex $A_{k+1}$. By construction, this plane also contains the cone apex $A_k$. There are two connected components in the intersection of the plane with the two truncated cones and their connecting spherical cap. We choose the segment of $B_{k+1}$ that lies in the same connected component of our initial segment on $B_k$. 
If these two segments do not have a common endpoint, they are connected in a continuously differentiable ($C^1$) way by a circular arc. The latter is a circular portion of the intersection between the plane of the two segments and the spherical cap shared by the two bones. The same construction process is valid for extending a baseline segment of a bone onto the preceding bone.

In degenerate cases where one of the bones is, in fact, a cylinder,  a baseline segment on that bone is a segment parallel to the cylinder axis, and the corresponding segment in an adjacent bone is included in the plane defined by the segment on the cylinder and the apex of the other cone. If both bones are cylinders, the plane containing the baseline segments on those bones is parallel to the axis of both cylinders.

%In the following, \raph{many constructions will be using} the sheaf of planes passing through the apexes $A_k$ and $A_{k+1}$ of two consecutive cones (or parallel to the axis of the cones that are degenerated into cylinders).

If one bone corresponds to a free extremity of the sphere-mesh, a baseline segment on that bone is extended by a circular arc joining the endpoint to the furthest point of the sphere cap that is aligned on the bone axis (see Figure \ref{fig:baseline_3d_2}).

\paragraph{Bundle of baselines}
It is essential to notice that the baselines built for a chain of bones cannot cross each other. Hence, the surface of each chain of bones is covered by a bundle of non-intersecting baselines.

\paragraph{Bundle of baselines at junctions}
A sphere-mesh is usually composed of several chains of bones, which join at a common sphere end. It is possible to extend the construction of the bundle of baselines by subdividing the sphere into several portions in the manner of a Voronoi diagram. Each cell comprises the points that are closer to a tangential cone portion than to another tangential cone portion. On the sphere, the median between two cones is a circular arc in the plane that holds the intersection of the two cones. It is indeed a plane because the cones are tangent to a common sphere. At the junction, baselines connect coplanar segments of two cones if the plane holding these two segments also contains a common point to the two cones on the union of all the cones truncated with each other.
%\raphq{Ajouter la figure illustrative. Une image avec les arcs de cercle sur la sphere et une image avec l'union des cones tronqués pour connaître la nature de la baseline (quels os elle relie).}
\begin{figure}[ht]
    \centering
    \includegraphics[width=0.25\textwidth]{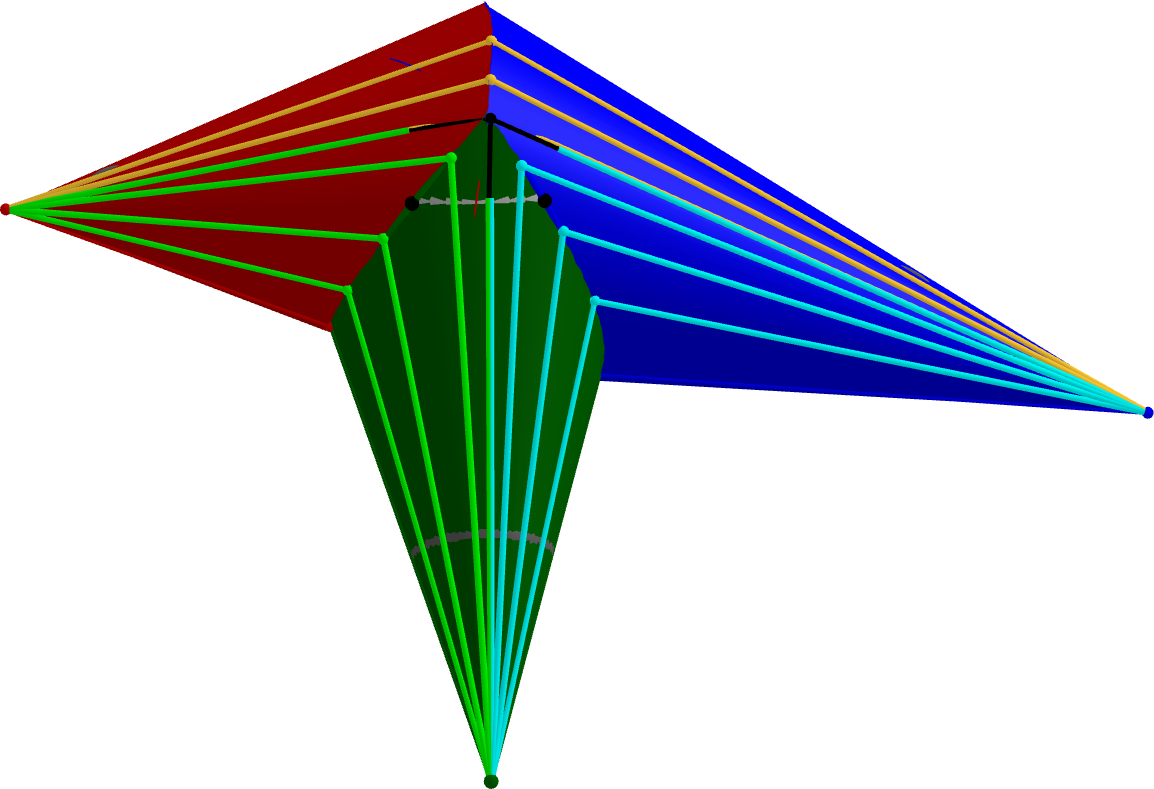}
    \includegraphics[width=0.22\textwidth]{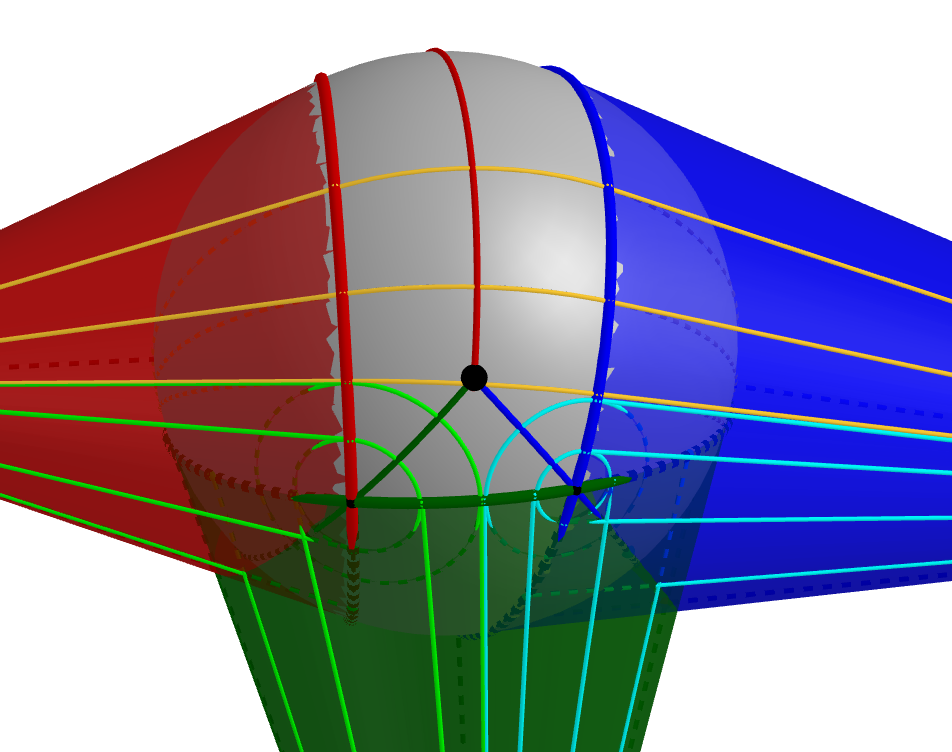}
    \caption{Bundle of baselines at the junction of three chains. Green baselines pass through red cone and green cone, blue baselines pass through green cone and blue cone, yellow baselines pass through red cone and blue cone. In the left figure, we extend segments on the union of truncated cones to identify the kind of baseline they belong to.}
    \label{fig:jonction}
\end{figure}
\paragraph{Detail direction field}
Armed with this baseline definition, we now define how a continuous vector field of detail directions can be defined over the sphere-mesh using the system of baselines. Those directions will be used to encode the detail points above the sphere-mesh.

First, along a circular arc, the direction of detail always coincides with the normal direction. The segment case is slightly more demanding: we first define the detail directions at the endpoints and then linearly interpolate along the segment.
\begin{itemize}
    \item If the endpoint corresponds to the connection of the segment and a circular arc, the detail direction corresponds to the endpoint's normal, which is well defined because of the $C^1$ junction property.
    \item If the endpoint corresponds to the intersection of two cones tangent to a common sphere, the detail direction corresponds to the outward direction from the sphere's center to the intersection point. 
\end{itemize}

% \begin{figure}[ht]
%    \centering
%    \includegraphics[width=0.4\textwidth]{baseline_direction_detail.png}
%    \caption{Detail directions along a baseline.}
%    \label{fig:direction}
%\end{figure}

Figure \ref{fig:baseline_3d} illustrates the profile of the detail direction field above a baseline. Note that along a baseline segment, the direction field vectors are coplanar by construction. The detail direction field along a segment is entirely contained in the plane defined by the segment and the axis of the bone it belongs to. In the next section, we use this property to define a non-orthogonal projection of a point onto the sphere-mesh. Note that the plane including details changes from one segment of the baseline to another or from one segment of the baseline to a circular arc.  

\subsection{Projection of a point on the sphere-mesh using the detail direction field}

Given a point of the input shape, we propose to project it on the sphere-mesh using detail directions induced by baselines instead of using the orthogonal projection. Hence, we can obtain a more natural and continuous way of modelling the details in the concave parts of the sphere-mesh. More precisely, given a point $p$, we are confronted with the inverse problem of determining a \emph{base-point} $b_p$ of the sphere-mesh such that $p$ can be encoded as a displacement in the detail direction above $b_p$.  Fortunately, this inverse problem is made easy because, at a joint and its two incident bones, the segments and circular arcs of a baseline are coplanar, and by the fact that detail directions are also coplanar above each baseline segment. 

 \begin{figure}[ht]
    \centering
    \includegraphics[width=0.4\textwidth]{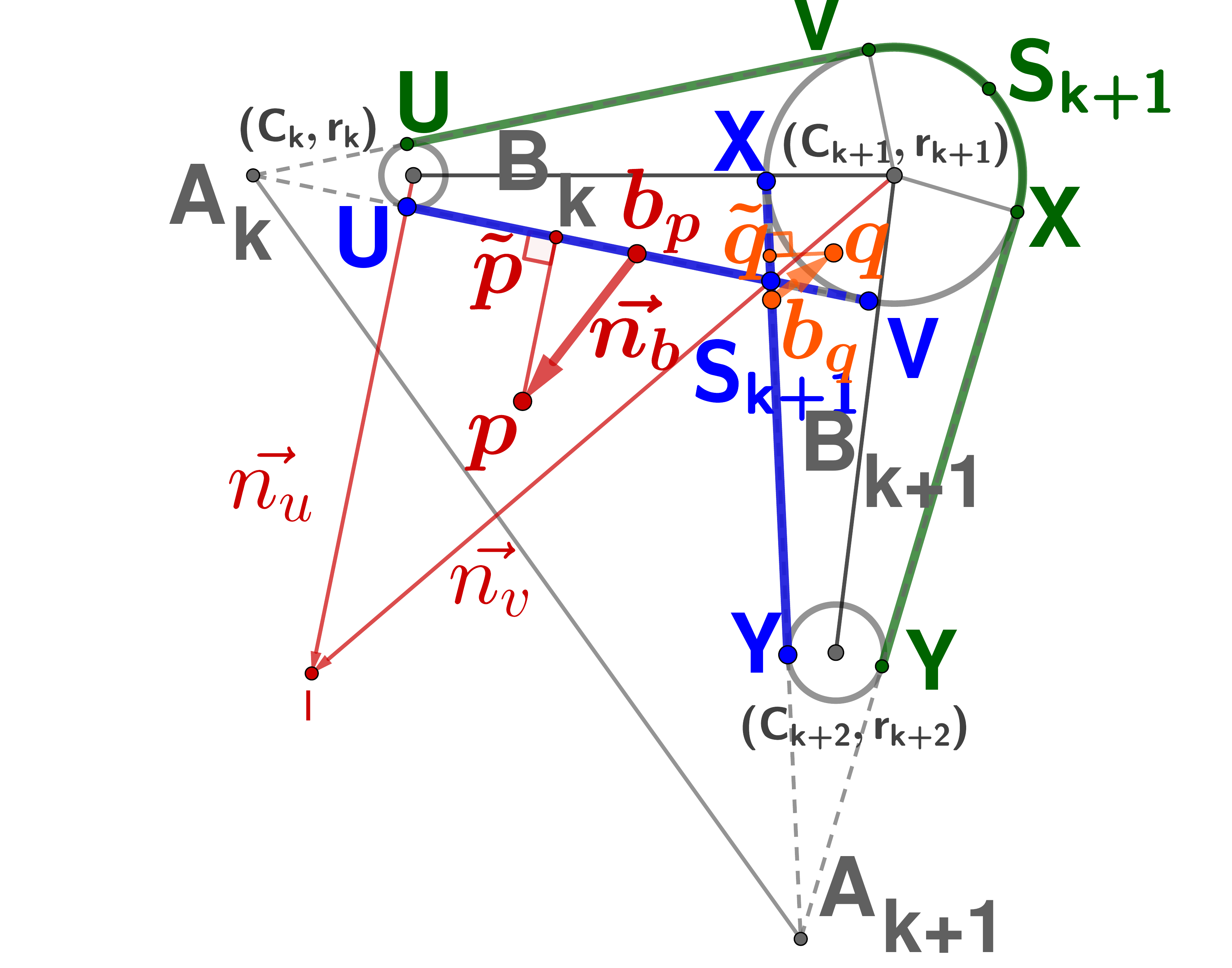}
    \caption{2D Cross-section along the plane $\tilde{p}A_kA_{k+1}$ ($p$ and $\vec{n_b}$ do not belong to that plane but are superimposed for illustration purpose). $\tilde{p}$ is the orthogonal projection of a given point $p$ on its closest bone $B_k$ and base-point $b_p$ is such that $p=b_p+h\vec n_b$, with $\vec n_b$ the direction field at $b_p$, and $h$ the corresponding detail amplitude. }
    \label{fig:ex_baseline_basepoint}
\end{figure}

\paragraph{Baseline determination}
Given $p$, we first have to identify the baseline portion that contains base-point $b_p$. 
Assuming that the local variation of a detail point after a skeletal movement is only influenced by its closest bone, the predecessor bone and the successor bone, the portion of baseline associated to $p$ only needs to be computed over at most three bones. 
First, let us first deal with the easy case where point $p$ projects orthogonally on a spherical cap of the closest bone. In this case, the base-point $b_p$ coincides with the orthogonal projection of $p$ onto the sphere and the detail direction $\vec n_b$ corresponds to the normal to the sphere at this point. Hence the base-point is trivial to find, and the local portion of the baseline is built accordingly on the two adjacent bones.
In the nontrivial case where the orthogonal projection of $p$ lies on a cone part of the closest bone, the base-point must be determined on the baseline associated with $p$. Our method processes the bones by pairs: by first considering the portion of baseline along $p$'s closest bone and its successor bone to deal with one joint and the portion of baseline on the predecessor bone and the closest bone with the other joint. Hence, the following explanations will be illustrated on a sphere-mesh corresponding to a single pair of bones. 
We refer to Figure \ref{fig:ex_baseline_basepoint} for the following notations. The example sphere-mesh model is composed of a sphere joint $(C_{k+1},r_{k+1})$ and its two adjacent bones $B_k$ and $B_{k+1}$. Let $A_k$ and $A_{k+1}$ denote the apexes of the cones underlying $B_{k}$ and $B_{k+1}$. 

$\tilde{p}$ is the orthogonal projection of $p$ on its closest bone ($B_{k}$ in the example), except if $p$ is located within the intersection of the two bones $B_{k}$ and $B_{k+1}$. In the latter case we project $p$ on the more distant of the two bones $B_k$ and $B_{k+1}$ for continuity sake, as illustrated in Figure \ref{fig:ex_baseline_basepoint} with $q$ being projected on $B_{k+1}$ even if $B_{k}$ is its closest bone. Importantly enough, the orthogonal projection $\tilde{p}$ is to be distinguished from the base-point $b_p$. 
We consider the plane containing line $A_kA_{k+1}$ and $\tilde{p}$, and take its intersection with bones $B_k$ and $B_{k+1}$ and their common joint but excluding the spherical caps at the other end of the two bones. If plane $\tilde{p}A_kA_{k+1}$ is not tangent to sphere $(C_{k+1},r_{k+1})$, this yields two possible pieces of baseline curve, instead of a single one (blue and green curves in Figure \ref{fig:ex_baseline_basepoint}). Each piece is composed of two segments (($UV$ and $XY$ in green or $US_{k+1}$ and $S_{k+1}Y$ in blue in Figure \ref{fig:ex_baseline_basepoint}) that are possibly joined by a circular arc ($\wideparen{VX}$ in green in Figure \ref{fig:ex_baseline_basepoint}). Note that, both pieces of baseline may possibly enclose a circular arc. Out of these two possible baselines, we select the one that contains $\tilde{p}$.
Hence, in Figure \ref{fig:ex_baseline_basepoint} the blue curve is the baseline piece associated with point $p$.

Two exceptional cases must be considered. If one of the adjacent cones is a cylinder (both its end-sphere radii are equal), we consider instead the plane parallel to the cylinder axis, containing the apex of the remaining cone and point $\tilde p$. The second case is when both cones degenerate to cylinders, in which case we consider the plane parallel to both cylinders axis and containing $\tilde{p}$.

\paragraph{Anchor points of a baseline}
If there is a circular arc in the baseline portion associated with point $p$ on the joint sphere $(C_{k+1},r_{k+1})$, we denote by $S_{k+1}$ the intersection point of the circular arc and the plane enclosing the intersection of the two cones. This plane is called the \textit{separator plane} of the two bones.
$S_{k+1}$ is different from the midpoint of the arc $\wideparen{VX}$.
If there is no circular arc, $S_{k+1}$ is the intersection of the two \textit{extended segments} $UV$ and $XY$ of the underlying cones, where $V$ and $X$ both belong to the sphere joint $(C_{k+1},r_{k+1})$ in Figure \ref{fig:ex_baseline_basepoint}). Here again $S_{k+1}$ belongs to the \textit{separator plane} of $B_{k}$ and $B_{k+1}$. $S_{k+1}$ is the \textit{anchor point} of the baseline on joint $(C_{k+1},r_{k+1})$.

%Points $U$, $V$, $X$, $Y$ belong to circles connecting cones to joint spheres. Let call them  the control points of the portion of baseline on $B_{k}$ and $B_{k+1}$. 

\paragraph{Base-point determination}
The portion of baseline associated to one point $p$ on its closest bone and its two adjacent bones is composed of three segments that are possibly joined by one or two circular arcs. In the example of Figure \ref{fig:baseline_3os_0}, the piece of baseline contains segments $MS_k$, $S_kS_{k+1}$ and $S_{k+1}Y$, where $S_k$ and $S_{k+1}$ are the intersections of the extended segments $MN$, $UV$ and $XY$.

If $\tilde p$ lies on a circular arc of sphere $(C_k,r_k)$, the detail direction is given by the unit vector pointing from $C_k$ towards $\tilde p$, and $\tilde p$ is the base-point of $p$.
If $\tilde p$ lies on a segment part supported by a bone $B_k$, the detail direction encoding $p$ is enclosed in the plane containing $B_k$'s axis and $\tilde p$. 
To determine the corresponding base-point we use the fact that detail directions are obtained by coplanar linear interpolation between the detail directions at the segment endpoints.
%(possibly followed by a direction normalisation).
If the segment is adjacent to two circular arcs on the baseline curve, the detail direction locally corresponds to the normal to the cone and the base point $b_p$ still coincides with $\tilde p$. 
However, if a segment on $B_k$ is adjacent to another segment on $B_{k+1}$ (resp. $B_{k-1}$) the direction of detail at their common point $S_{k+1}$ (resp. $S_k$) is given by the unit vector pointing from the center $C_{k+1}$ (resp. $C_{k}$) of their common joint sphere toward $S_{k+1}$ (resp. $S_k$).
This case is illustrated in Figure \ref{fig:ex_baseline_basepoint}. 
Introducing an additional point $I$, lying at the intersection of line $(C_k,\vec n_U)$ and $(C_{k+1},\vec n_V)$, $b_p$ is found as the intersection of lines $pI$ and $UV$, and the detail direction is $\vec n_b =\frac{b_pp}{\|b_pp\|}$, which corresponds to a linear interpolation along the segment between $\vec n_U$ and $\vec n_V$.

%Then, the detail direction $\overrightarrow{n_b}$ at $p_b$ is computed by a linear interpolation between $\overrightarrow{n_u}=\overrightarrow{c_1U}$ and $\overrightarrow{n_v}=\overrightarrow{c_2S}$ where $c_1$ and $c_2$ are the centers of the end spheres. 
%$$\overrightarrow{n_b} = (1-t)\overrightarrow{n_u} + t\overrightarrow{n_v}$$ with $t = \frac{Up_b}{US}$.

\subsection{Encoding a point set above a sphere-mesh}
\begin{figure}[ht]
    \centering
    \includegraphics[width=0.3\textwidth]{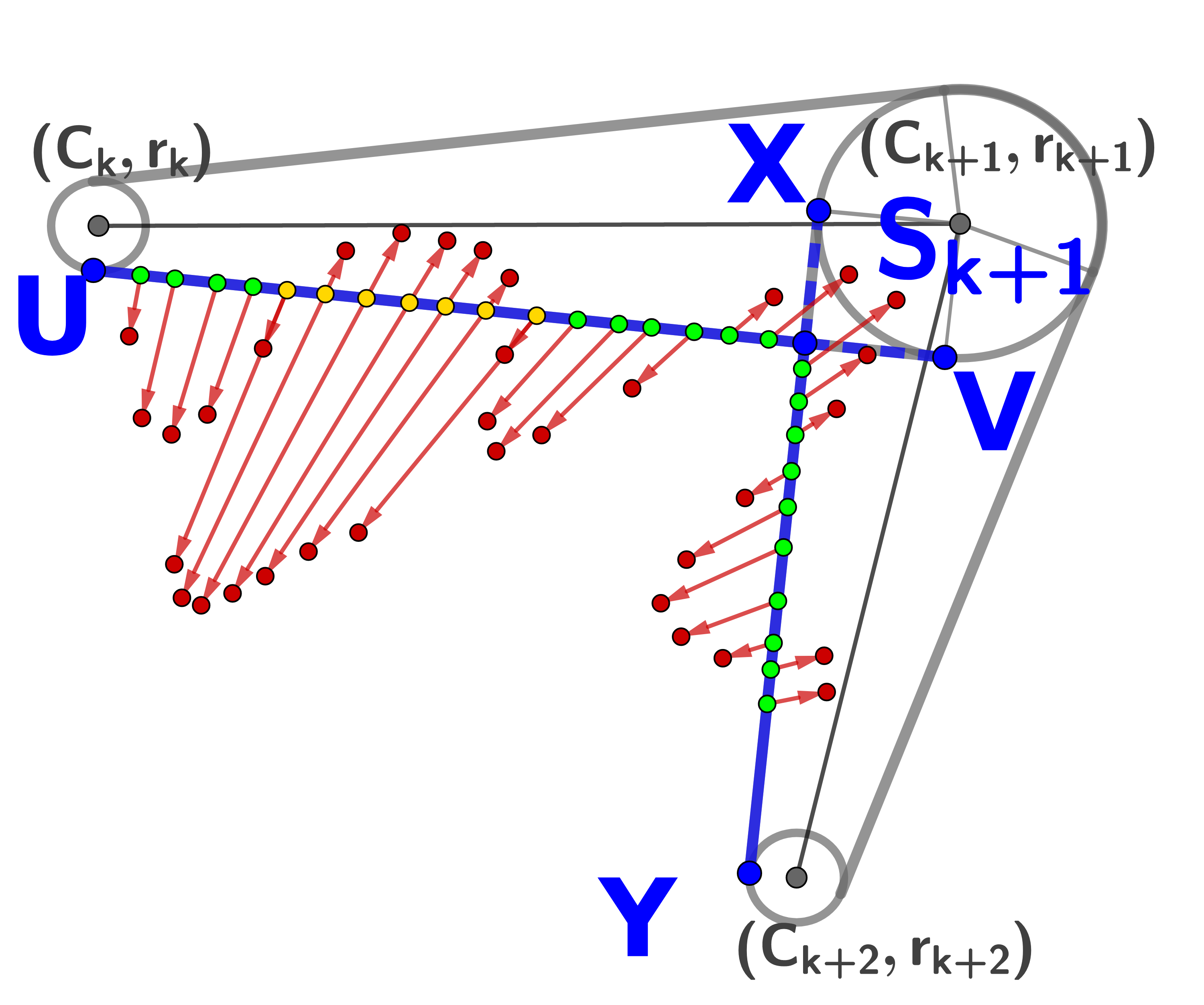}
    \caption{Illustration of a point set (in red) encoded as a detail set over a baseline. The base-points are shown in green and in yellow. The yellow base-points have more than one detail point over them. The detail may be negative when it is oriented inward.}
    \label{fig:encodage}
\end{figure}

Each point is projected onto the sphere-mesh to determine its base-point and its local amplitude of detail in its detail direction. The projection process is independent for each point and can therefore be parallelized.  The input point set is then described by a set of sphere-mesh base-points with a height as illustrated in Figure \ref{fig:encodage}. There can be several occurrences of a same base-point when the starting shape is multi-layered or has plumb lines.

\subsection{Baseline and detail direction field after pose change}
\label{sec:baseline_after_change}

After skeletal movement, segments %and circular arcs 
composing baselines will be modified differently depending on whether a bone is twisted or a joint is folded or unfolded. 
%The same remark applies also to the deformation fields defined above the various elements of a baseline. 
In our deformation approach, the segments are transformed into curves on the conic part of the bones. 
The deformed baselines will be composed of pieces of these curves such that they can still be connected by circular arcs as in the original baselines.
%as well as arcs of circles as in the original baselines.

\paragraph{Twist} 
When a bone $B_k$ is twisted around its axis with $B_{k-1}$ remaining fixed, the points belonging to a baseline segment of $B_k$ will be deformed by rotations around the cone axis. The angle of rotation of a point $\tilde{p}$ on the segment depends on its normalized distance $d \in [0,1]$ to the endpoint closer to $B_{k-1}$, in a similar way of the approach proposed in \cite{STBS:2011,h.20201290}. 
%\raphq{Verifier que c'est similaire au deux. Dans la thèse dire que c'est similaire à ce que l'on faisait déjà au chapitre d'avant.}. 
Normalized distances are measured before twisting $B_k$. After twisting, the segment deforms into a curve depending on the profile of the angle of rotation as an increasing function of $d$. The circular arcs remain unchanged. Figure \ref{fig:twist} illustrates the deformation of a segment after twisting $B_k$ with an angle $\tau_{max}$. The angle of rotation applied along the segment is computed by using a cubic function: $\tau(d) = -2\tau_{max}d^3+3\tau_{max}d^2, d\in[0,1]$. But any other profile provided by an artist or resulting from learning would be possible. The advantage of the cubic profile is that the tangent to the curve at the endpoints of the segment is preserved by twisting.

 \begin{figure}[ht]
    \centering
    \begin{subfigure}[ht]{0.22\textwidth}
        \centering
        \includegraphics[width=\textwidth]{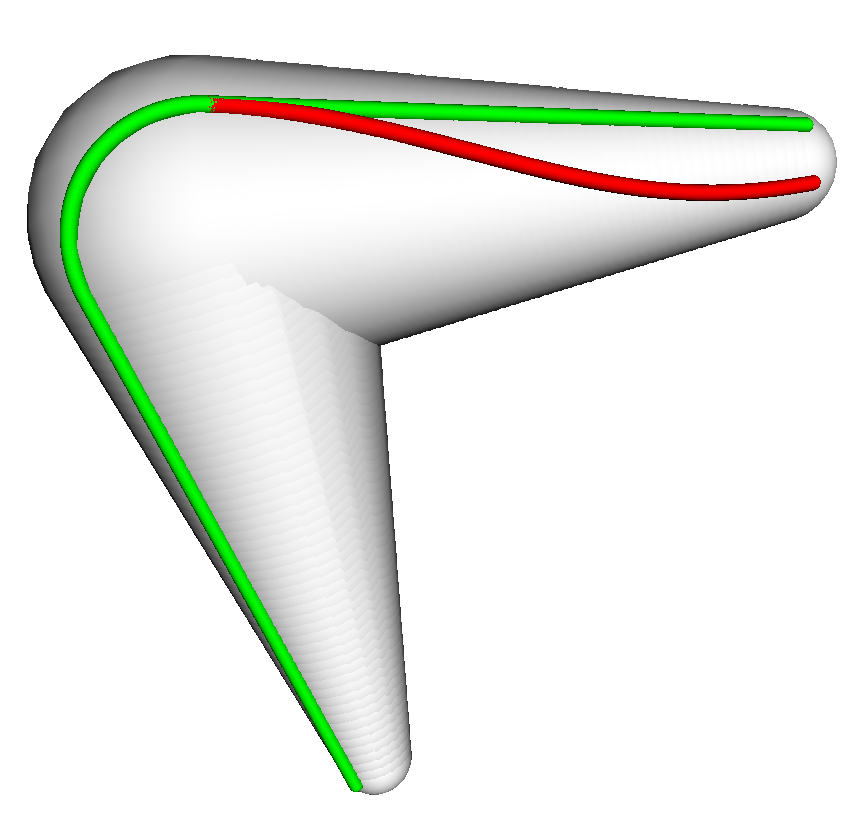}
        \caption{Twisting movement}
        \label{fig:twist}
    \end{subfigure}
     \begin{subfigure}[ht]{0.23\textwidth}
        \centering
        \includegraphics[width=\textwidth]{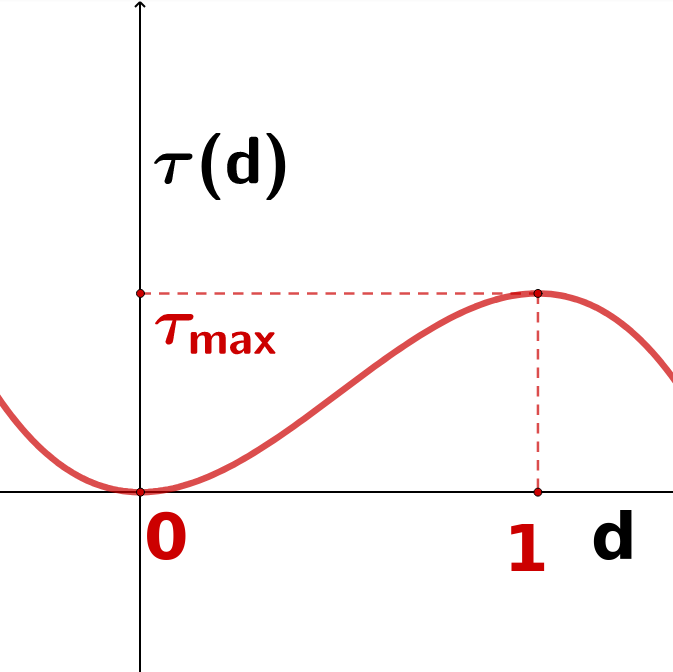}
        \caption{Profile of the twisting angle.}
        \label{fig:profile}
    \end{subfigure}
    \caption{The right bone is rotated around its axis by the angle $\tau_{max}$. The baseline before (green) and after (red) the twist are superimposed. The right image shows the cubic profile of the twisting angle applied to the points: $\tau(d) = -2\tau_{max}d^3+3\tau_{max}d^2, d\in[0,1]$. }
    \label{fig:twist}
\end{figure}

% Raph Bend rotation at a joint : On va là aussi se retrouver avec une rotation des points du segment paramétrée par t autour de l'axe du cone. Cela est du au fait que les segments de baseline liés à des os consécutifs ne peuvent pas rester coplanaires quand les deux sommets $A1$ et $A2$ s'écartent ou se rapprochent : Evolution du plan du faisceau selon l'os de gauche différente de l'évolution du plan du faisceau selon l'os de droite.

\begin{figure}[ht]
    \centering
    \begin{subfigure}[ht]{0.23\textwidth}
        \centering
        \includegraphics[width=\textwidth]{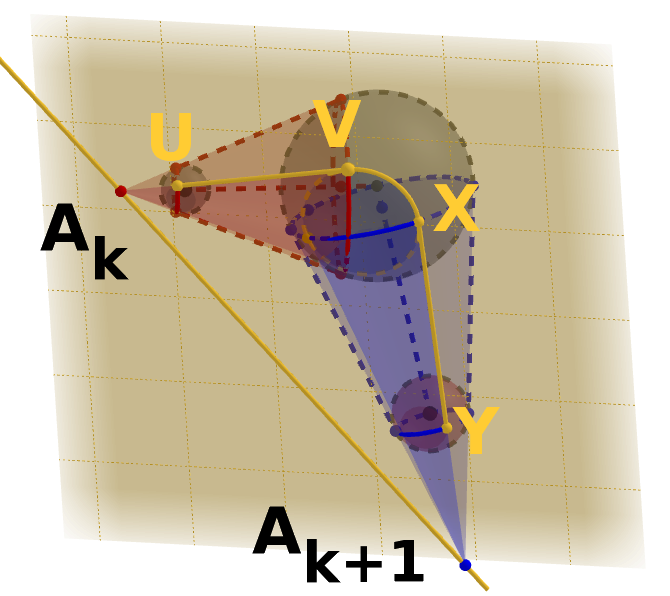}
        \caption{Initial position: the segments are coplanar.} 
        \label{fig:cut_avant}
    \end{subfigure}
    \begin{subfigure}[ht]{0.23\textwidth}
        \centering
        \includegraphics[width=\textwidth]{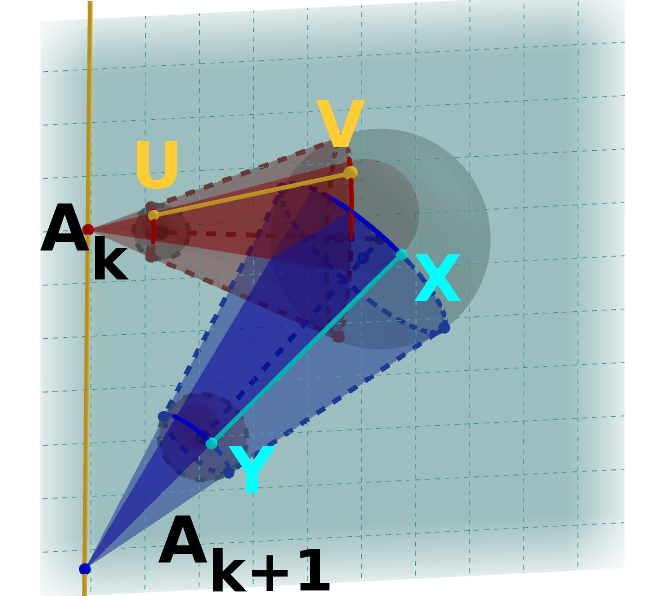}
        \caption{New positions of the segments after a bend rotation of the joint}
        \label{fig:cut_apres}
    \end{subfigure}
    \caption{Positions of baseline segments  before and after a joint is bent.  Circles of tangency on $B_k$ and $B_{k+1}$ are represented in blue and red respectively.}
    \label{fig:cut_change}
\end{figure}

\begin{figure}[ht]
    \centering
    \begin{subfigure}[ht]{0.22\textwidth}
        \centering
        \includegraphics[width=\textwidth]{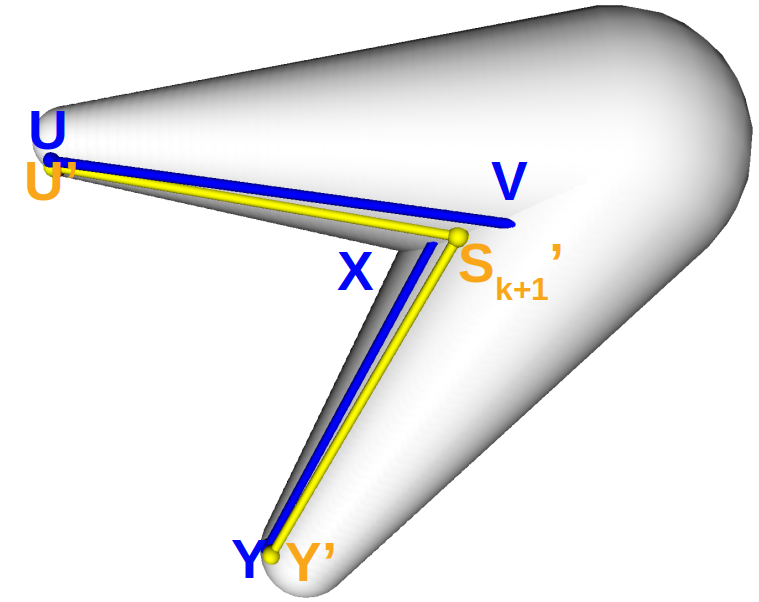}
        \caption{}
        %\caption{Initial segments of baseline after bend rotation.}
        \label{fig:baseline_after}
    \end{subfigure}
    ~
     \begin{subfigure}[ht]{0.24\textwidth}
        \centering
        \includegraphics[width=\textwidth]{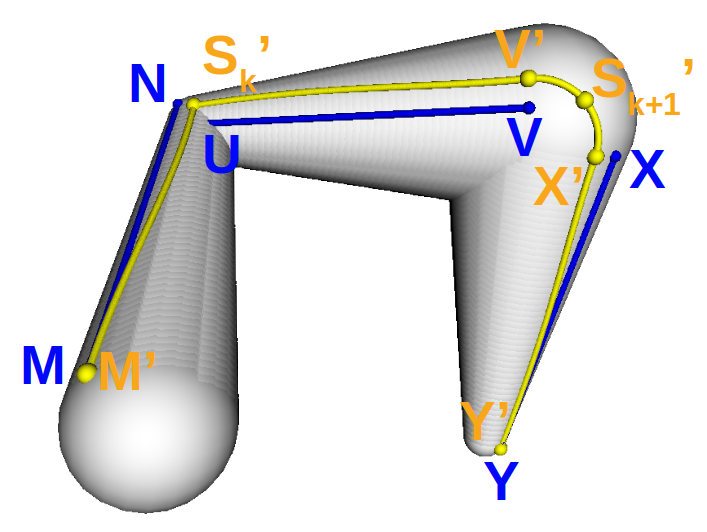}
        %\caption{Baseline after transformation for 3 bones.}
        \caption{}
        \label{fig:baseline_3os_after}
    \end{subfigure}
    \caption{Baseline deformation after bend rotation of joints. The rotated segments (in blue) are superimposed with the deformed baseline (in yellow): (a) bend of a single joint between two bones, $U=U'$ and $Y=Y'$ (b) bend of two consecutive joints. 
    %\toq{J'ai un peu changé figure b pour que ça corresponde à ton texte }
    %\raphq{Je ne sais pas si tu as vu mais j'ai deux textes possibles dans Deformation of a complete baseline on a chain of bones.}
    } 
    \label{fig:baseline_after_trans}
\end{figure}

\paragraph{Bend}
When a joint between two consecutive bones $B_k$ and $B_{k+1}$ is bent, the cone of $B_{k+1}$ is rotated around the axis of the joint,
%passing through the center $c_{k+1}$ of the sphere 
and $B_k$ remains fixed. Given a baseline, the rotated segment $YX$ of $B_{k+1}$ is no longer coplanar with $UV$ on $B_k$. Initially, the two segments were in the same plane of the sheaf passing through $A_kA_{k+1}$. After bend rotation, each segment is in a different plane of the sheaf as illustrated in figure \ref{fig:cut_change}. The planarity loss is also illustrated on figure \ref{fig:baseline_after_trans}. Therefore we propose to deform segments on the surface of cones so that they can be connected again by a circular arc defined by a plane of the sheaf. Let us first explain where the new endpoints of these curves $V'$ and $X'$ are positioned on the sphere.

In the following descriptions, we usually illustrate the case where $UV$ and $XY$ do not intersect initially and are connected by a circular arc in the initial baseline. However, the case where initial segments $UV$ and $XY$ intersect is treated in the same way, by performing the deformation on the extended segments $UV$ and $XY$. %and not on the truncated  ones ($US_{k+1}$ and $S_{k+1}Y$)}. \toq{à voir}

\begin{figure}[ht]
    \centering
    \begin{subfigure}[ht]{0.23\textwidth}
        \centering
        \includegraphics[width=\textwidth]{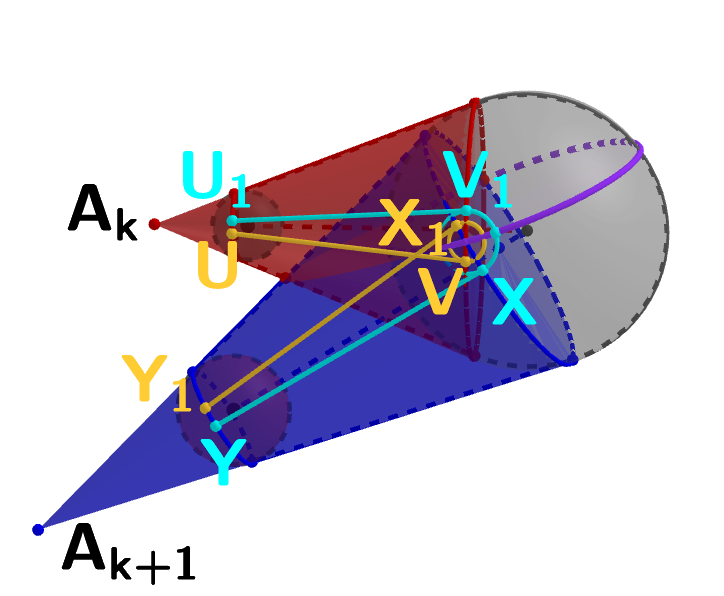}
        \caption{Baselines $U_1V_1XY$ and $Y_1X_1VU$ suggested by two different planes of the sheaf at a joint.}
        \label{fig:intermediate}
    \end{subfigure}
    ~
    \begin{subfigure}[ht]{0.22\textwidth}
        \centering
        \includegraphics[width=\textwidth]{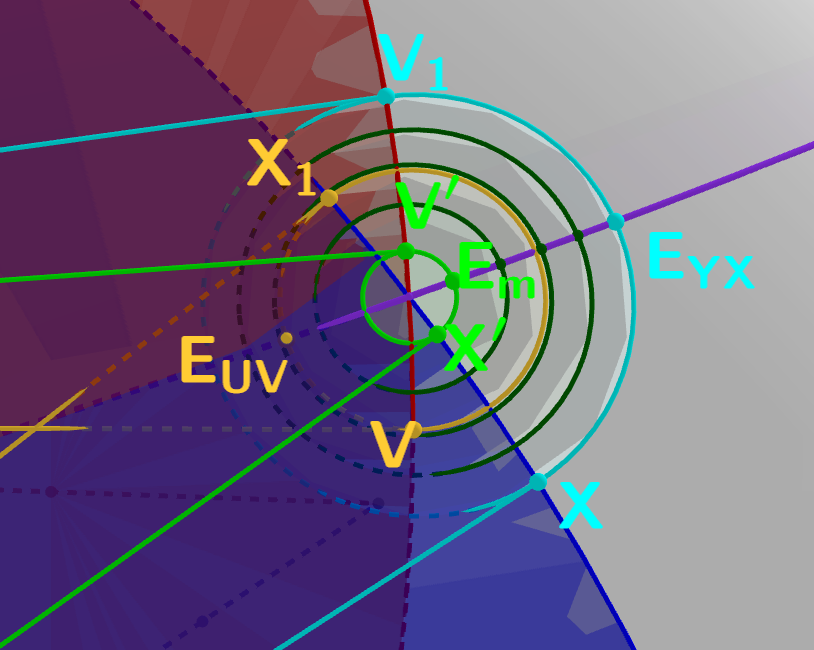}
        \caption{Close-up. Green baseline propositions are in the possible intermediate planes of the shear between the two planes containing baselines $U_1V_1XY$ and $Y_1X_1VU$. 
        %Green circles corresponding to different planes of the sheaf are possible intermediate positions between the blue circle and the yellow circle. The target plane of the sheaf corresponds to the light green circle.
        }
        \label{fig:intermediate_zoom}
    \end{subfigure}
    \caption{Conjoint determination of target positions $V'$ and $X'$ for coupled endpoints at a joint. The pivot circle (in purple) defined by the separator plane of the cones and the sphere is involved.
    %\raphq{Attention j'ai changé les noms de $E_1$ et $E_2$ dans le texte.}
    }
    \label{fig:intermediate_plane}
\end{figure}

By definition, $V$ and $X$ belong to the circles that connect their respective cones to the common sphere. Let us call these circles the \textit{circles of tangency} (see figure \ref{fig:cut_change}). After rotation, we propose to move the endpoints $V$ and $X$ on their respective circles of tangency so that they are brought in a common plane of the sheaf containing $A_kA_{k+1}$ and the displacements of $V$ and $X$ on their respective circles are similar in a sense that takes the position and the radius of the circles of tangency into account. Figure \ref{fig:intermediate_plane} illustrates the conjoint determination of the coupled target positions $V'$ and $X'$ for endpoints of segments being deformed. The rotated segment $YX$ can be completed into a $C^1$ continuous planar baseline proposition using circular arc $\wideparen{XV_1}$ and segment $V_1U_1$ (in blue) in the plane passing through $X$, $A_k$ and $A_{k+1}$. 
Likewise, segment $Y_1X_1$ and circular arc $\wideparen{X_1V}$ complete the segment $UV$ into a $C^1$ continuous planar baseline proposition (in yellow) within a plane passing through $V$, $A_k$ and $A_{k+1}$.  $V$ and $V_1$ (resp. $X$ and $X_1$) are on the circle of tangency which connects the red cone (resp. blue cone) to the common sphere. The target endpoint $V'$ and $X'$ must be located on the circular arcs $\wideparen{VV_1}$ and $\wideparen{XX_1}$ respectively. 

In the corresponding subset of the sheaf of planes passing through  $A_kA_{k+1}$, we are seeking the relevant intermediate plane defining the target endpoints $V'$, $X'$ that balances the distances of $V_1V'$ and $VV'$ but also $X_1X'$ and $XX'$. For this we cannot work directly on the circle of tangency of $B_k$, nor on the circle of tangency of $B_{k+1}$, but on a \textit{pivot circle} at the intersection between the separator plane of the two cones and the joint sphere. This pivot circle is illustrated in purple in figure \ref{fig:intermediate_plane}. Let $E_{YX}$ (resp. $E_{UV}$) be the intersection of the baseline arc $\wideparen{V_1X}$ (resp. $\wideparen{X_1V}$) and the pivot circle. $E_m$ denotes the middle point of the circular arc $\wideparen{E_{YX}E_{UV}}$ on the pivot circle. The plane of the sheaf passing through $E_m$ is the intermediate plane. $V'$ and $X'$ are defined by the intersections of that plane with the tangential arcs $\wideparen{VV_1}$ and $\wideparen{XX_1}$ respectively.

\begin{figure*}[h]
    \centering
    \begin{tabular}{c c c c c c}
        \includegraphics[width=0.15\textwidth]{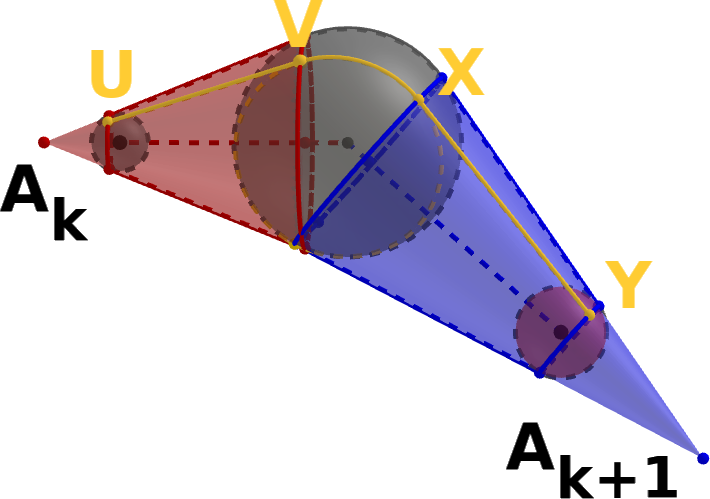}&
        \includegraphics[width=0.15\textwidth]{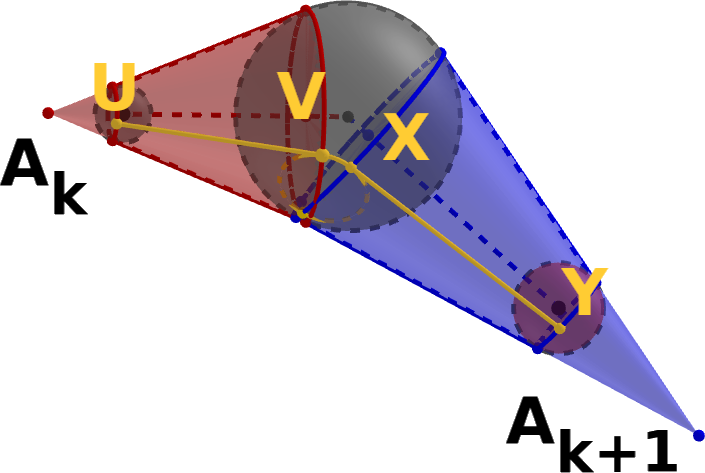}&
        \includegraphics[width=0.15\textwidth]{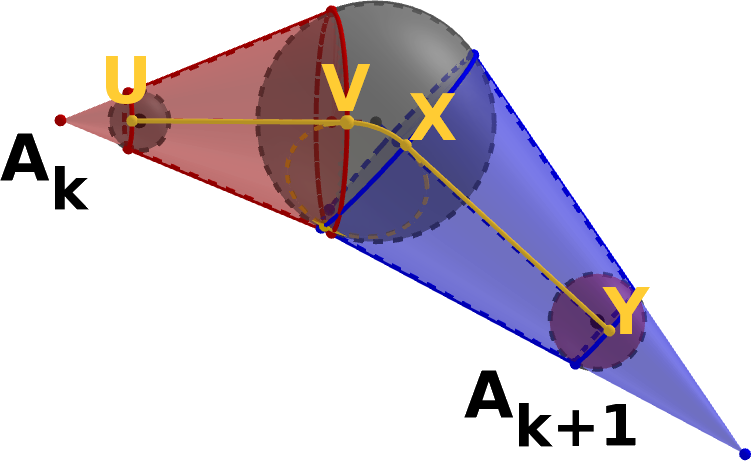}&
        \includegraphics[width=0.15\textwidth]{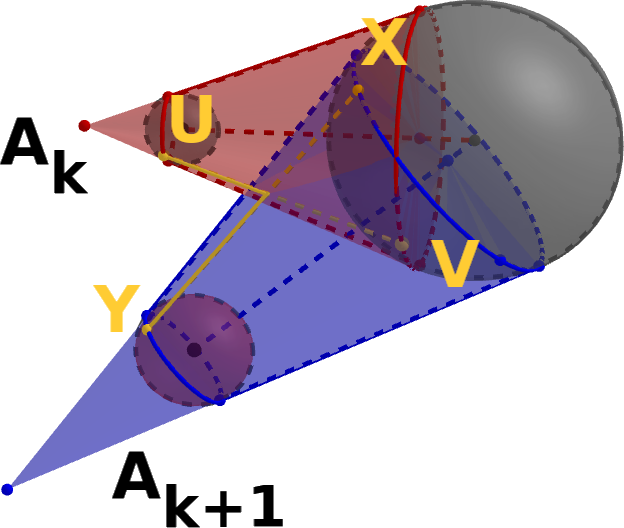}&
        \includegraphics[width=0.15\textwidth]{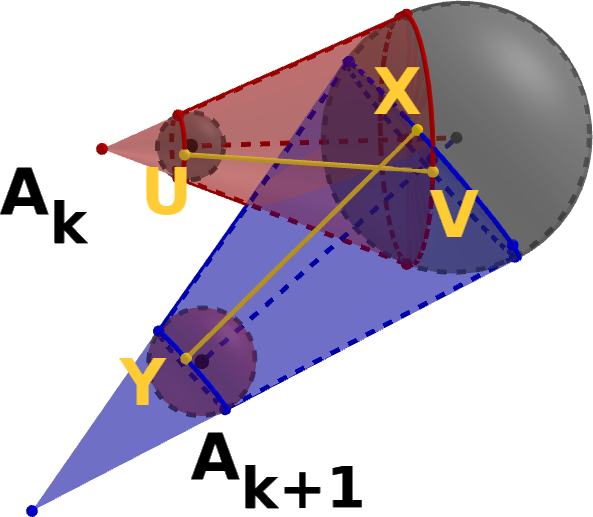}&
        \includegraphics[width=0.15\textwidth]{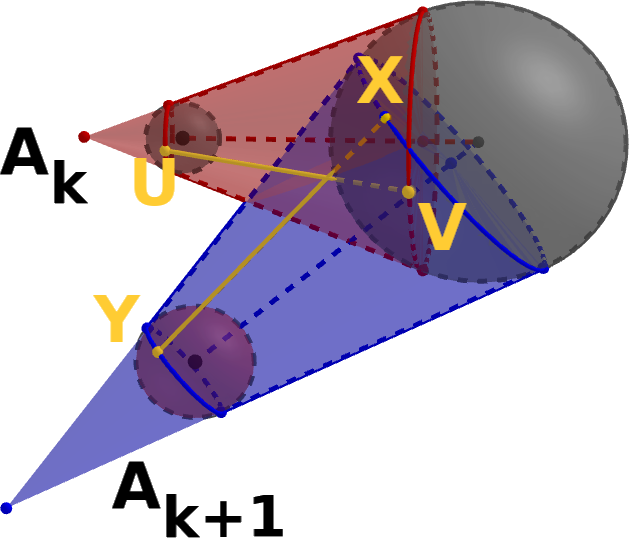}\\
        
        \includegraphics[width=0.15\textwidth]{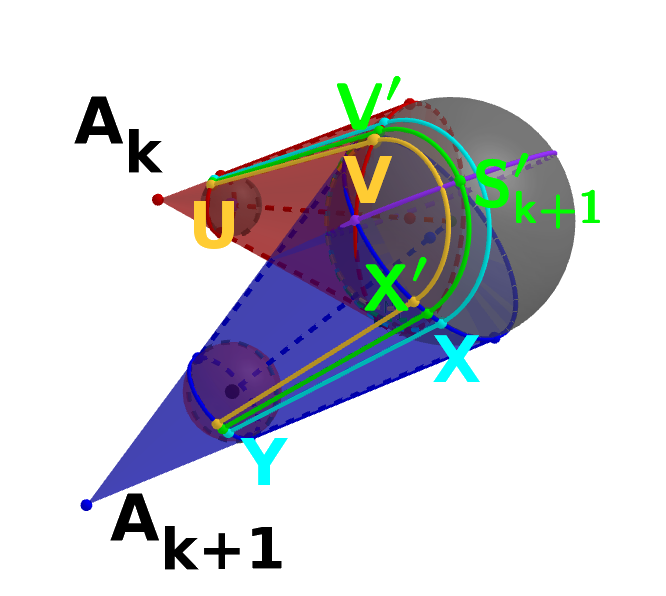}&
        \includegraphics[width=0.15\textwidth]{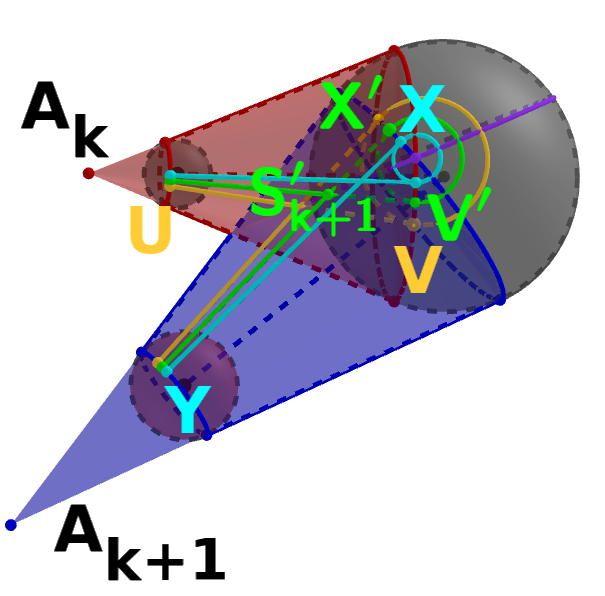}&
        \includegraphics[width=0.15\textwidth]{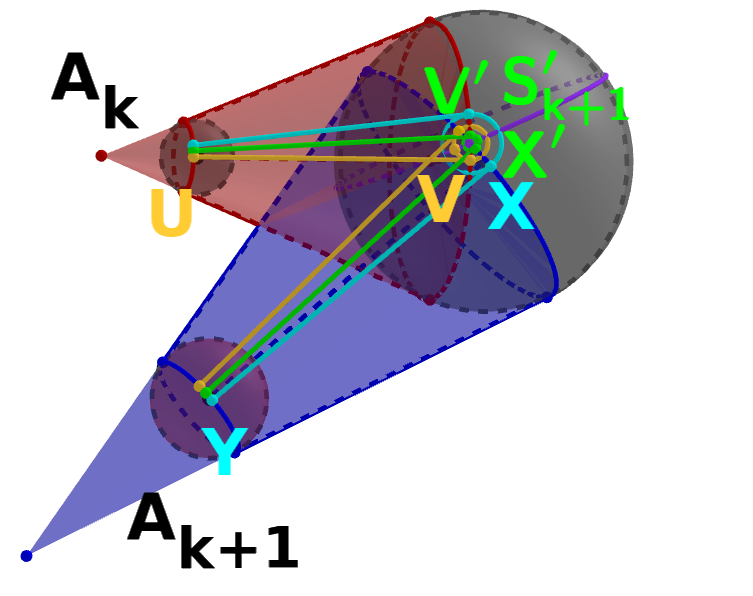}&
        \includegraphics[width=0.15\textwidth]{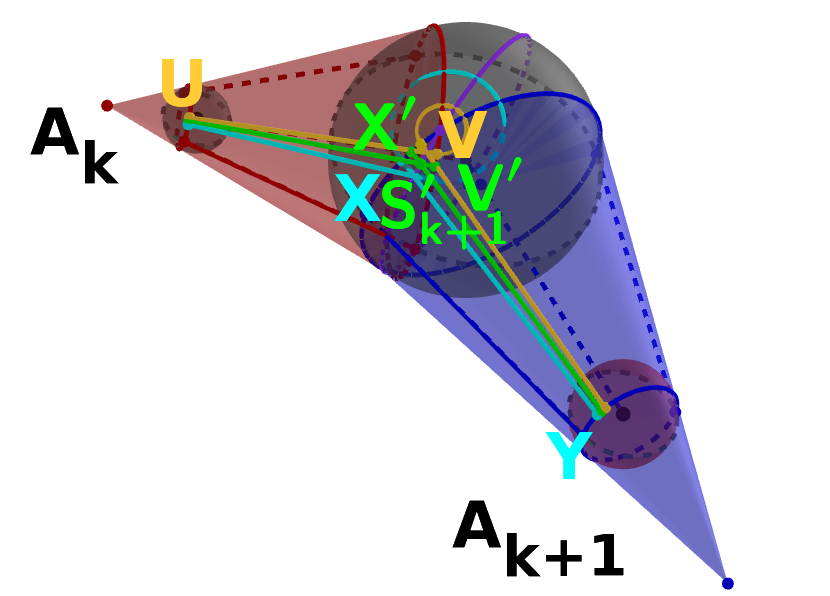}&
        \includegraphics[width=0.15\textwidth]{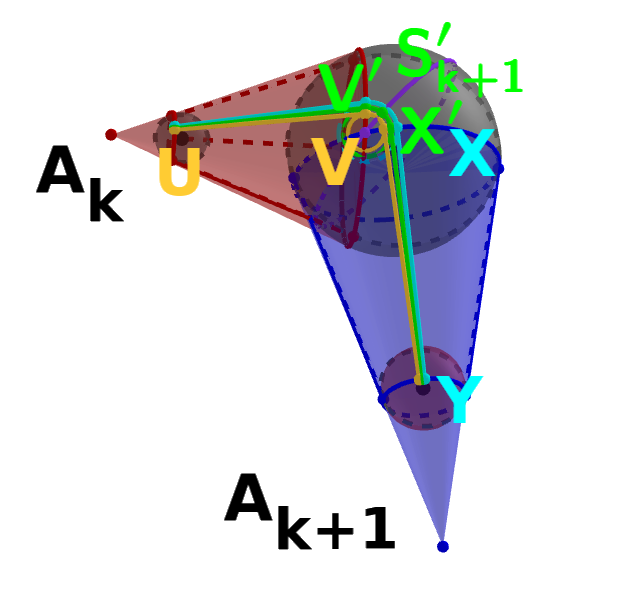}&
        \includegraphics[width=0.15\textwidth]{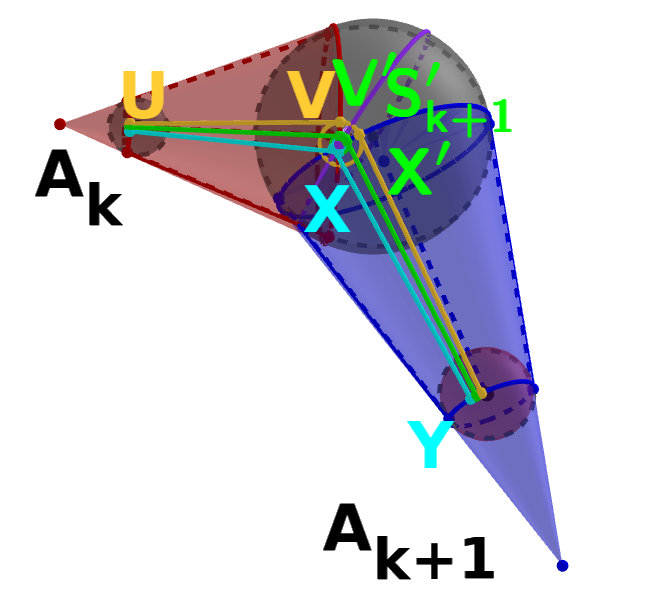}\\
        
        \multicolumn{1}{c}{\textbf{(a)}}& 
        \multicolumn{1}{c}{\textbf{(b)}}&
        \multicolumn{1}{c}{\textbf{(c)}}&
        \multicolumn{1}{c}{\textbf{(d)}}& 
        \multicolumn{1}{c}{\textbf{(e)}}&
        \multicolumn{1}{c}{\textbf{(f)}}
    \end{tabular}
    \caption{Different cases of baseline deformation following a bend. The first row shows baseline initial positions. The second row shows baseline deformations after bend rotation. The green curve shows the intersection of the plane containing point $V'$ and $X'$ on the bones.
    (a) initial baseline passing in the convex area of the joint, deformed baseline obtained by interpolation of two baselines in the convex area, (b) initial baseline passing in the convex area of the joint, deformed baseline obtained by interpolation of two baselines in the concave area, 
    (c) initial baseline passing in the convex area of the joint, deformed baseline obtained by interpolation of two baselines in the convex area and the concave area, (d) initial baseline passing in the concave area of the joint, deformed baseline obtained by interpolation of two baselines in the concave area, (e) initial baseline passing in the concave area of the joint, deformed baseline obtained by interpolation of two baselines in the convex area, (f) initial baseline passing in the concave area of the joint, deformed baseline obtained by interpolation of two baselines in the convex area and the concave area.}
    \label{fig:bend}
\end{figure*}

In Figure \ref{fig:intermediate_plane}, we illustrate how to find the relevant intermediate plane of the sheaf between a baseline passing in the convex area of the joint and a baseline passing in the concave area. Our approach works for all other situations that we illustrate in Figure \ref{fig:bend}. This Figure also illustrates that a baseline initially positioned in the convex part of a joint may evolve in the concave part after a bend, and vice versa.

%The target position $V'$ (resp. $X'$) corresponds to a target rotation angle $\theta_k^V$ (resp. $\theta_{k+1}^X$) for the plane of the sheaf including $V$ (resp. $X$).
The target position $V'$ (resp. $X'$) corresponds to a target rotation angle $\theta_k^V$ (resp. $\theta_{k+1}^X$) of $V$ (resp. $X$) around the axis of $B_k$ (resp. $B_{k+1}$).
%\ton{The target position $V'$ (resp. $X'$) corresponds to a target rotation angle $\theta_{bk}^V$ (resp. $\theta_{bk+1}^X$) that the angle is between two planes passing through the axis of the bone $B_k$ (resp. $B_k{k+1}$) and including $V$ and $V'$ (resp. $X$ and $X'$). }
Once these target rotation angles are determined, the points of the initial segment $UV$ (resp. $YX$) are rotated in turn around the axis of the cone by interpolating the rotation angle between $0$ and $\theta_{k}^V$ (resp. between $0$ and $\theta_{k+1}^X$) using a function of the normalized distance to $U$ (resp. $Y$) on the initial segment $UV$ (resp. $YX$).
%Once these target rotation angles are determined, the points of the initial segment $UV$ (resp. $YX$) are rotated in turn around $A_kA_{k+1}$ by interpolating the rotation angle between $0$ and $\theta_k^V$ (resp. between $0$ and $\theta_{k+1}^X$) using a function of the normalized distance to $U$ (resp. $Y$) on the initial segment $UV$ (resp. $YX$). 
As for the twist rotation, we propose to use a cubic interpolation of the angle, but once again it is possible to use another profile, possibly sketched by an artist, learned from experimental data, or resulting from physic-based criteria.
Thus, the point $q(d)$ with normalized distance $d$ to point $U$ on segment $UV$ is rotated by an angle $\theta_{k}(d)=-2\theta_{k}^Vd^3+3\theta_{k}^Vd^2$ around the axis of $B_k$. The same is done for the points of the initial segment $YX$, by using their normalized distance $d$ to $Y$ and by using it to interpolate the rotation angle $\theta_{k+1}(d)$ between $0$ and $\theta_{k+1}^X$. The resulting deformation of the segment is illustrated in Figure \ref{fig:baseline_after_trans}.

The curves $UV'$ and $YX'$ obtained by deforming the segments $UV$ and $YX$ are still included on the cone part of their respective bones. If they do not intersect, they can be connected by a circular arc since $V'$ and $X'$ belong to a common plane of the sheaf of planes passing through $A_kA_{k+1}$. If they intersect, we crop the parts of the curves that get inside the other cone, and we connect the truncated parts.

\paragraph{Deformation of a complete baseline on a chain of bones} 
The deformation of the segments composing a baseline is done by sequentially processing the bones composing a chain, as is always the case when positioning the skin of an articulated body. 
%A first pass treats the bend of the joints, and a second pass deals with the twist of the bones.

The bend of a joint distorts the segments on both its incident bones. If rotations occur at both joints bordering a bone, there is a combined effect on the common bone with a target position on each side for the endpoints of segments being deformed. However, what is said above can be adapted: we move the points of the segments by interpolating a rotation angle around the axis of the cone between two target angles (see Figure \ref{fig:baseline_3os_after}).
%the sequence of the two bends can be handled by as the sequence of a bend and a twist.

If a bone is also affected by a twist, there is a combined effect of the bend and the twist on the segments of the bone. The twist is simply handled by an additional rotation of one of the endpoint, corresponding to adding the angles $\tau_{max}$ to the actual target angle. Then, we proceed by cubic interpolation of the angle between the two target angles for rotating the segment inner points around the axis of the bone.

The deformed segments are completed with circular arcs in the spherical convex parts of the joints, and they are truncated where they intersect in the concave parts of the sphere.

If the transformations of the bones of the skeleton are pre-computed first, the relative displacement of the points above the bones can benefit from parallelism.

%If the bend of a joint is followed by the bend of the next joint, there is a combined effect on the common bone. 

%We start by moving the points of the initial segments to handle the first bend while keeping in memory the original value of the normalized abscissa $d$ for each point. We then add the effect of the second bend by adding the second displacement driven by the initial value of $d$ at each point. The same approach applies when applying the twist of a bone after a bend of an incident joint.}

\paragraph{Deformation of baselines at the junction between several chain of bones}

If a bone is bent at a common junction with other chains of bones, we propose to locally deform the baselines by establishing a correspondence between the baselines before and after skeletal deformation. This can be done by giving a normalized coordinate $c$ to each baseline of a given type, based on the proportion of angle crossed by the baseline on the pivot circular arc associated with the pair of its corresponding bones.
%\raphq{Sur l'image, identifier ainsi une baseline que l'on suit avant et après.}
\begin{figure}[ht]
    \centering
    \begin{subfigure}[ht]{0.22\textwidth}
        \centering
        \includegraphics[width=\textwidth]{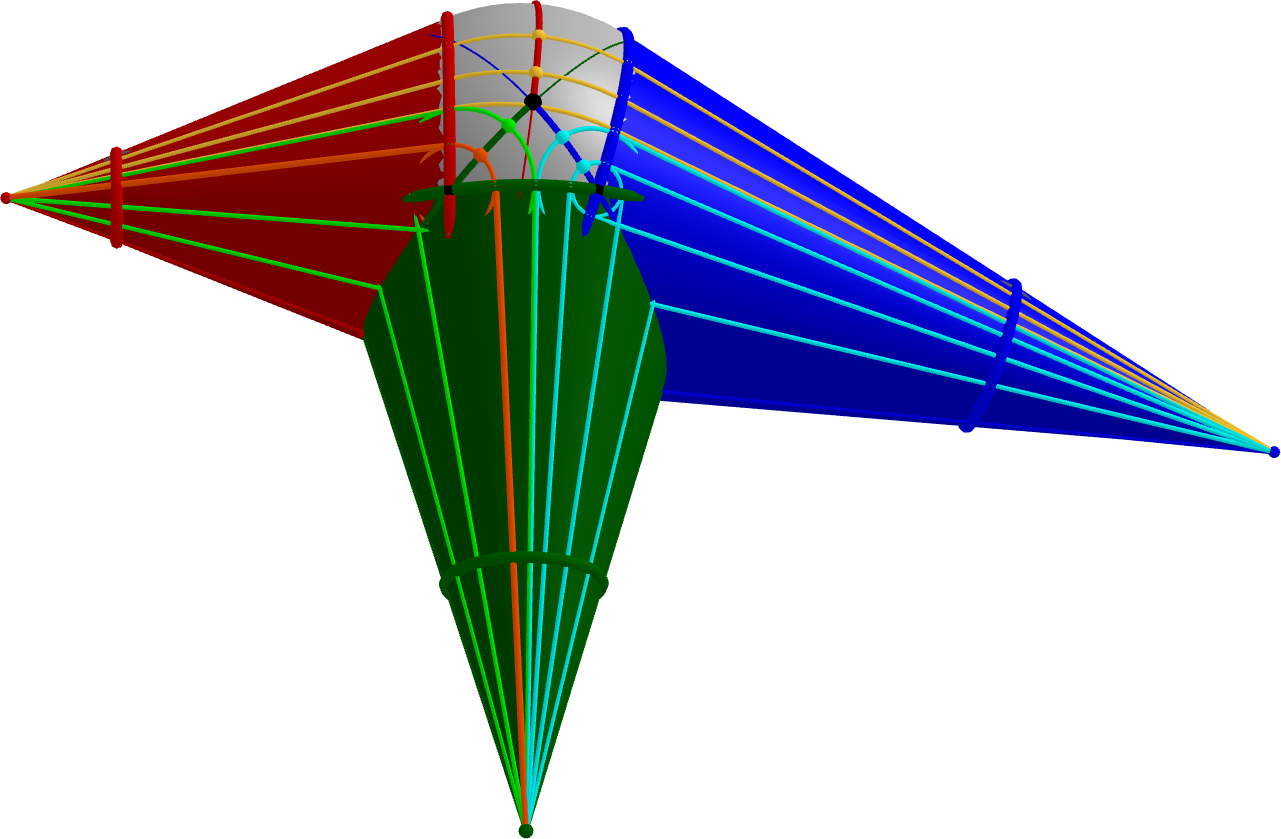}
        \caption{Bundle of baselines at a junction before bend rotation}
        \label{fig:jonction_before}
    \end{subfigure}
    ~
     \begin{subfigure}[ht]{0.16\textwidth}
        \centering
        \includegraphics[width=\textwidth]{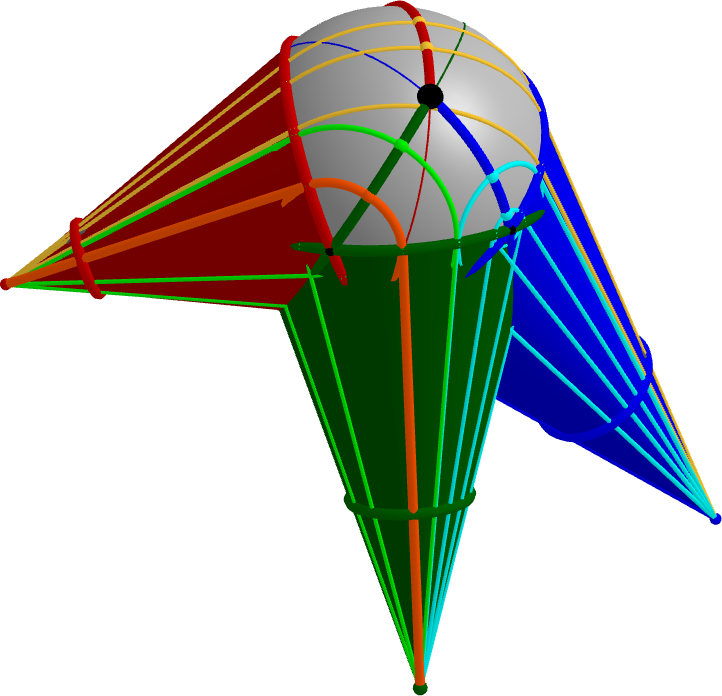}
        \caption{Bundle of baselines at a junction after bend rotation}
        \label{fig:jonction_after}
    \end{subfigure}
    \caption{The position of a bundle of baselines at junction. We show one baseline in orange to follow its position before and after a bend rotation.}
    \label{fig:jonction_deformation}
\end{figure}

\begin{figure}[ht]
    \centering
    \begin{subfigure}[ht]{0.25\textwidth}
        \centering
        \includegraphics[width=\textwidth]{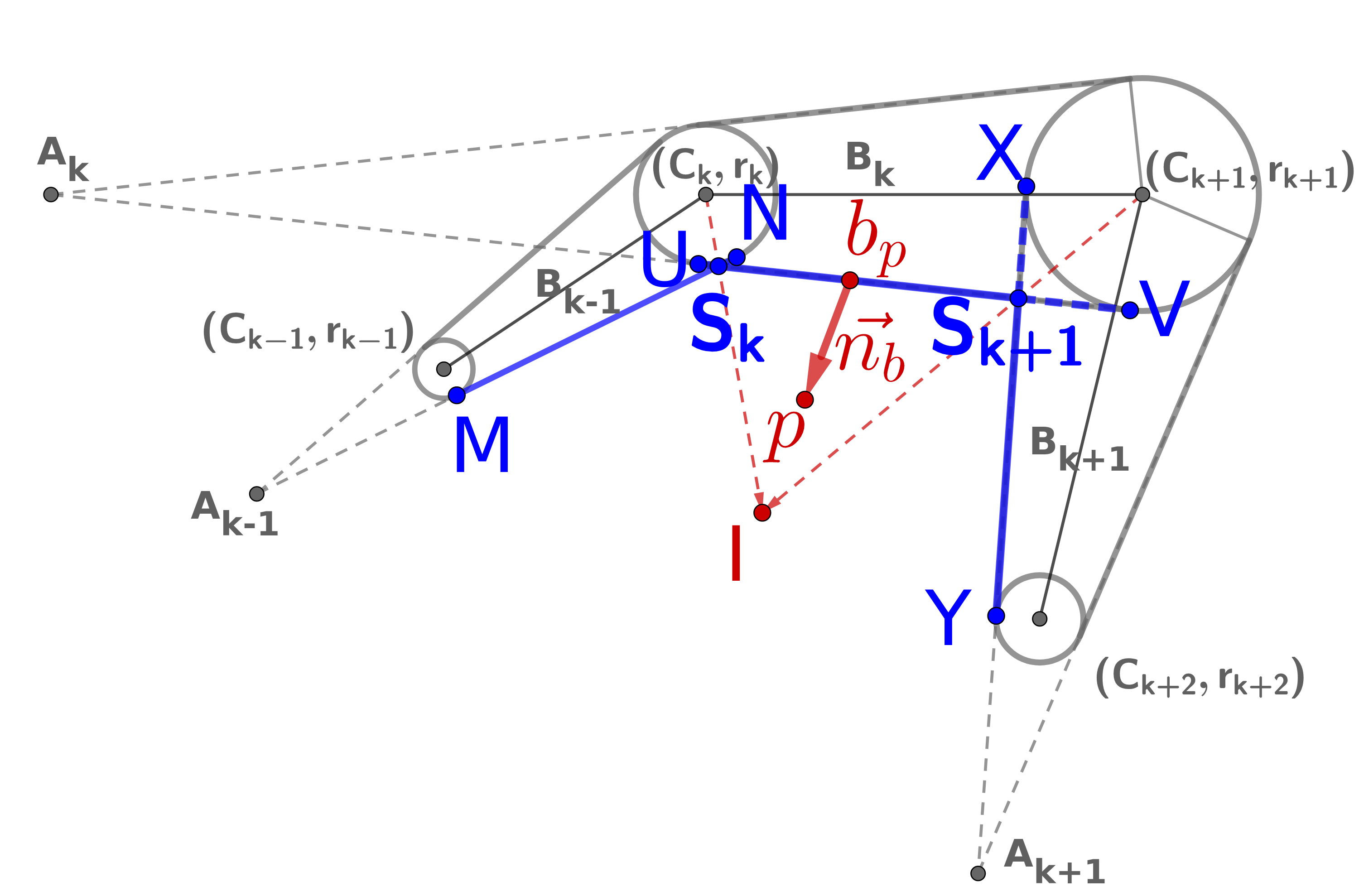}
        
        \caption{Initial baseline for point $p$}
        \label{fig:baseline_3os_0}
    \end{subfigure}
     \begin{subfigure}[ht]{0.21\textwidth}
        \centering
        \includegraphics[width=\textwidth]{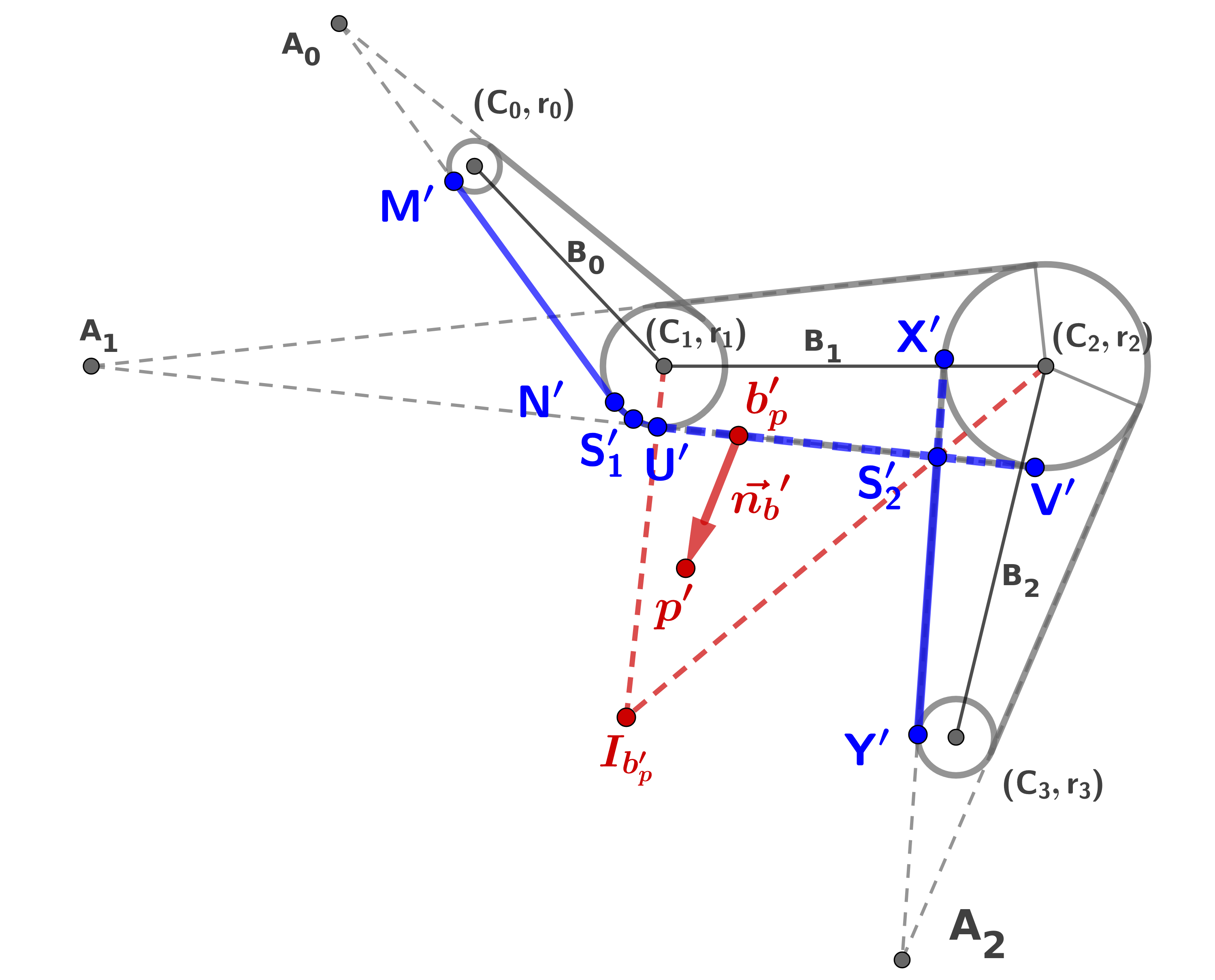}
        \caption{Baseline after transformation}
        \label{fig:baseline_3os_1}
    \end{subfigure}
    \caption{2D illustration of a baseline associate with point $p$ and the detail direction before and after a motion. In this figure the baseline has been represented in a 2D flat form to illustrate the determination of the detail direction above a segment. In practice, a baseline is flat by part at creation, and the segments are then deformed into non-planar curves. But the direction of detail is locally determined as if the deformed baseline were flat.}
    \label{fig:ex_baseline_3os}
\end{figure}

\begin{figure}[ht]
    \centering
        \includegraphics[width=0.1\textwidth]{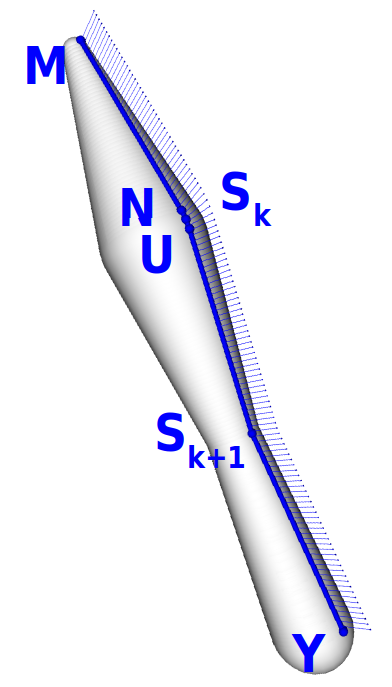}
        \includegraphics[width=0.12\textwidth]{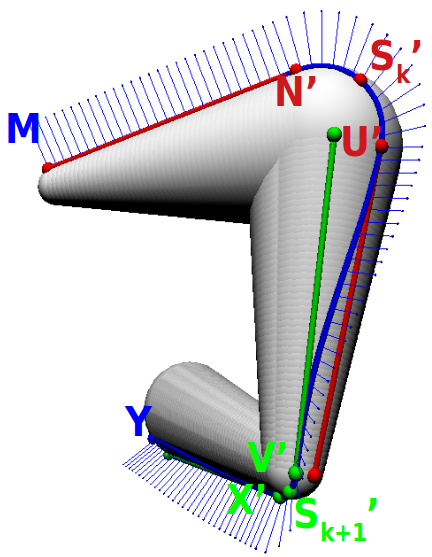}
        \includegraphics[width=0.12\textwidth]{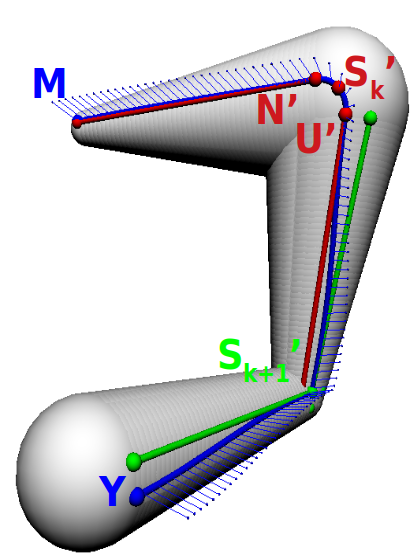}
        \includegraphics[width=0.12\textwidth]{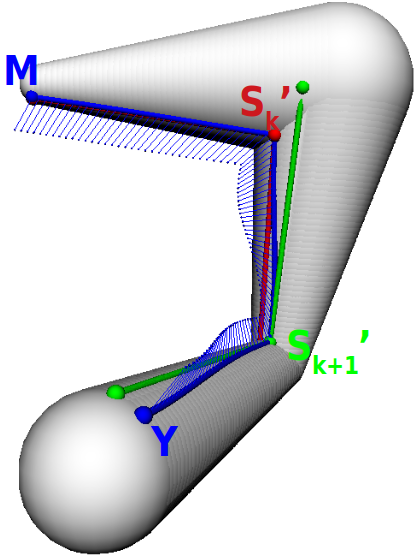}
    \caption{Baseline and detail direction field after a bend. The entire baselines are shown in blue. %Red curves are new baselines given by the first two bones. Green curves are new baselines given by the last two bones. 
    The first figure shows the initial position of the baseline and its detail direction field. From the second to the fourth figures shows the baseline curve after the bend with consistent direction of detail contains: two circular arcs, one circular arc and no circular arc. The interpolation angle driving the evolution of the baseline is cubic. The detail directions are not determined by using the deformed baselines, but by using the construction of a new bundle of baselines. 
    %\raphq{VIRER LA DERNIERE COLONNE puisqu'on ne se sert plus de cette approche.}
    %The interpolation for the baseline evolution is linear and the interpolation for the detail direction is cubic in the third column.
    }
    \label{fig:bend3os}
\end{figure}

\paragraph{Evolution of the detail direction field over the deformed baselines}
After the movement of the skeleton, the detail directions above a deformed baseline can be constructed without knowledge of the original baseline. It depends only on the position of the points on the surface of the sphere-mesh.
This amounts to compute portions of new baselines as they would be constructed with the new position of the skeleton and to construct the direction field above the segments and circular arcs of this new bundle of baselines. This is illustrated in 2D in Figure \ref{fig:ex_baseline_3os} and in Figure \ref{fig:bend3os}.

 %\raphq{Peut-être la figure suivante peut-elle être mise dans cette partie? Je viens de la référencer dans le texte.}

%(cela revient à tourner la direction du détail du même angle que le point du segment de baseline de même abscisse t, avec éventuellement un changement de l'inclinaison $\beta(t)$ au dessus de la baseline en fonction de ce qu'il se passe dans les parties concave.
%2) On peut vouloir injecter un peu de "retard" sur l'évolution des détails. A savoir que l'on peut utiliser une interpolation de l'angle de rotation d'ordre supérieur. Par exemple interpolation linéaire de l'angle pour la base-line et interpolation linéraire cubique du détail.... Ou bien profil d'angle donné par l'utilisateur. \ton{oir figure \ref{fig:bend3os}.

%The definition of $\alpha_d(t)$ is similar as $\alpha_p(t)$. We can use linear interpolation, quadratic interpolation, cubic interpolation or other type of the interpolation that a user can choose. If $\alpha_d(t)$ = $\alpha_p(t)$, the detail direction $\overrightarrow{n_b}'$ of one point $b_p'$ is in the same plan which consists of the axe of the bone and point $b_p'$. But it is not necessary that $\alpha_p(t)$ and $\alpha_d(t)$ are identical. $\overrightarrow{n_u}$ and $\overrightarrow{n_v}$ are the direction of details on point $S_1'$ (or $U_1'$) and $S_2'$ ($V_2'$). $\overrightarrow{n_u}$ is in the direction $\overrightarrow{C_kS_k'}$ (or $C_kU_1'$) and $\overrightarrow{n_v}$ is in the direction $\overrightarrow{C_{k+1}S_{k+1}'}$ (or $C_{k+1}S_{k+1}'$). We illustrates in Figure \ref{fig:baseline_3os_1}. After an interpolation:
%$$\overrightarrow{n_u}' = Rotation(\overrightarrow{C_kC_{k+1}},\alpha_d(t))(\overrightarrow{n_u})$$
%$$\overrightarrow{n_v}' = Rotation(\overrightarrow{C_{k+1}C_k},(1-\alpha_d(t))(\overrightarrow{n_v})$$
%The direction of detail of point $b_p'$ is defined by:
%$$\overrightarrow{n_b}' = (1-t)*\overrightarrow{n_u'} + t*\overrightarrow{n_v'}$$

\subsection{Base-points and points after transformation}
\label{sec:interpolation}
\begin{figure}[ht]
    \centering
        \begin{subfigure}{0.07\textwidth}
        \includegraphics[width=\textwidth]{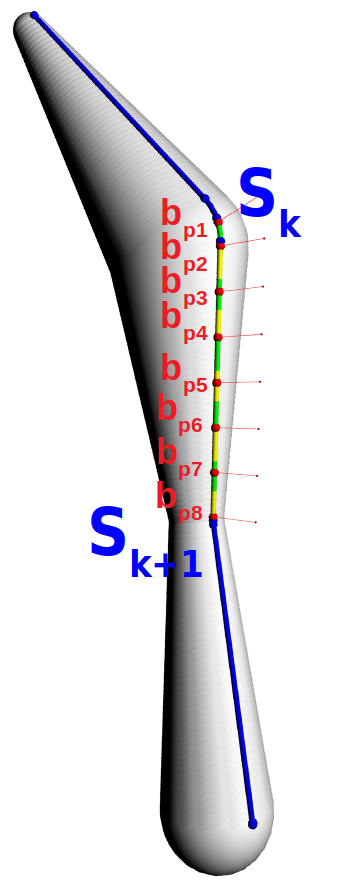}
        \caption{}
        \end{subfigure}
        \begin{subfigure}{0.18\textwidth}
        \includegraphics[width=\textwidth]{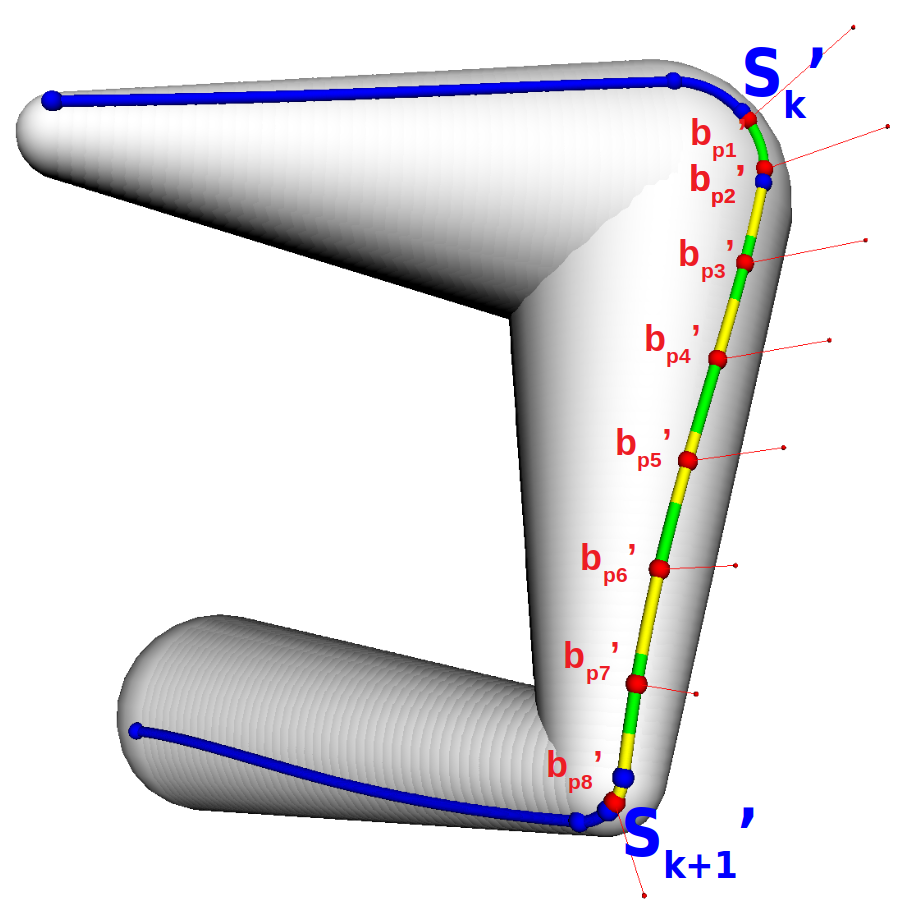}
        \caption{}
        \end{subfigure}
        \begin{subfigure}{0.17\textwidth}
        \includegraphics[width=\textwidth]{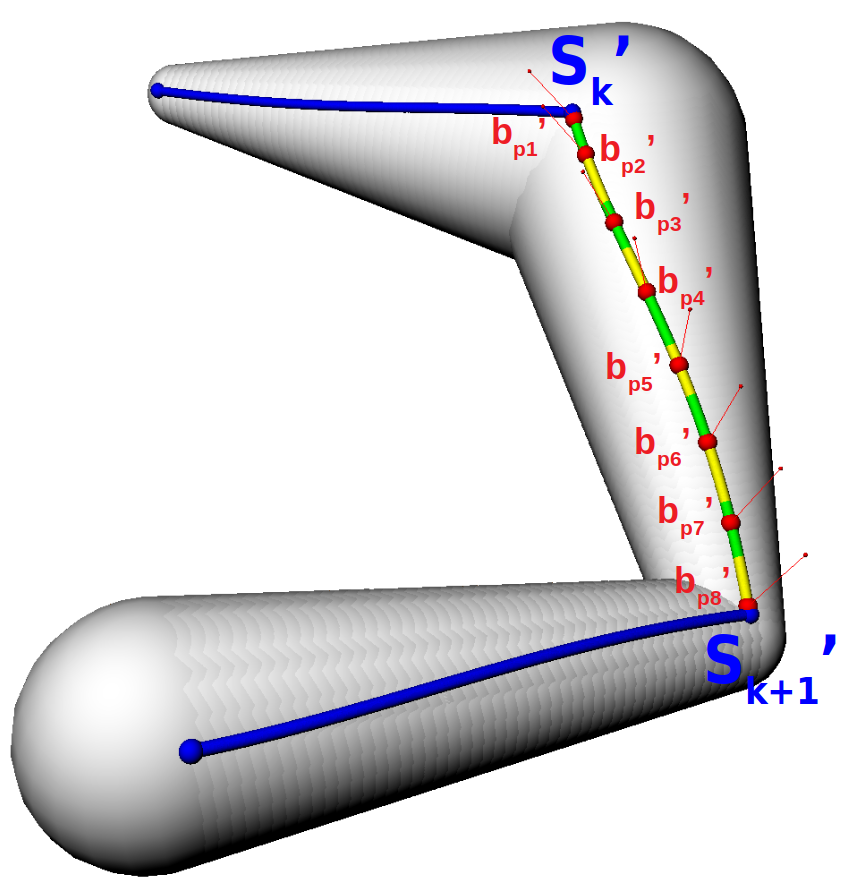}
        \caption{}
        \end{subfigure}
    \caption{Displacement of several sampled base-points between two anchor points. The scale is shown in green and yellow between two anchor points. (a) Initial position, (b) base-points displacement over a stretched baseline after deformation , (c) base-points displacement over a shrank baseline after deformation. }
    \label{fig:basepoint}
\end{figure}

\paragraph{Base-points displacement}
When the skeleton moves, the baseline of a point $p$ moves along with the bones. The base-point $b_p$ moves along with this baseline, but it also displaces along the deformed baseline to deal with the enlargement or shrinkage of the baseline parts between two consecutive joints. More precisely, a baseline is divided into sections by anchor points ($S_k$ and $S_{k+1}$ in Figure \ref{fig:basepoint}) that were placed at the intersection with the separator planes (on the circular arcs of the baseline or at the intersection between two baseline segments). After movement of the skeleton, anchor points are still fixed at the intersection of the deformed baselines and the separator planes, and base-points are distributed over the baseline sections in such a way as to preserve the curvilinear distance ratios $t$ to the bordering anchor points. The interest of our approach is that it guarantees a good distribution of details in the areas of the surface which undergo a stretching or a shrinking.
%\raphq{Figure qui illustre l'élongation ou le racourcissement d'un tronçon de base line entre deux anchor point avant et après déformation + montrer des points qui se retrouvent à bouger 1) au sein d'un segment 2) d'un segment sur un arc de cercle 3) d'un arc de cercle sur un segment. }
\paragraph{Point set deformation}
In our skinning process, the new position of each point $p$ of the shape is encoded with respect to its new base-point $b_p'$ through a displacement $h$ using the corresponding direction of detail $\overrightarrow{n_b}'$. 
In practice, for each point $p$, we only need to identify the corresponding portion of the baseline on the original shape and the relative abscissa of its base-point $b_p$ with respect to bordering anchor points on the baseline. This portion of the baseline is further deformed with the skeleton, and the new position of $b_p'$ is computed before reporting the detail in the direction corresponding to $b_p'$.

\paragraph{Point set skinning algorithm}
The full process of point set skinning is summarized in Algorithm \ref{algo:baseline}.

\begin{algorithm}
%\footnotesize % \small, \footnotesize, \scriptsize, or \tiny
\caption{Pose and anatomy change using baseline skinning}
\label{algo:baseline}
\begin{algorithmic}[1]

 \REQUIRE Sphere-mesh model registered to the input point set $P$. \\
          Target pose and anatomy. 
 \ENSURE  Point set $P^{\prime}$ corresponding to the target changes.
  \STATE Define for each point $p\in P$ its base point $b_p$, baseline $b$ on the registered sphere-mesh and compute $h$ and $t$ (ratio of curvilinear distance to bordering anchor points)
  \STATE Deform the baseline $b$ into $b'$ using the target pose and anatomy
  \STATE Determine $b_p'$ and its corresponding  $\overrightarrow{n_b}'$ by reporting $t$ on baseline $b'$
  \STATE Lift point $b_p'$ to the skin surface using $h$ and $\overrightarrow{n_b}'$
\end{algorithmic}
\end{algorithm}

%For a point $p_k$ attached to bone $B_k$, \ju{we need to find the baseline spreading over} its predecessor bone $B_{k-1}$, its attached bone $B_k$ and its successor bone $B_{k+1}$. To do so, we consider each time a pair of bones. Using the notations of Figure \ref{fig:baseline_3os_0}, for $B_0$ and $B_1$, the intersection consists of segments $MS_1$ and $S_1V$. We get key points $M$, $N$, $U$, $V$ and $S_1$. They are all in the plane $A_0A_1p_b$. Similarly, we can get key points $U$, $V$, $X$, $Y$ and $S_2$ for $B_1$ and $B_2$, which are all in the plane $A_1A_2p_b$. Segment $UV$ is on the intersection line of plane $A_0A_1p_b$ and plane $A_1A_2p_b$. In this example, the baseline of point $p$ is composed by segments $MS_1$, $S_1S_2$ and $S_2Y$. The complete baseline is determined by 8 points separating segments or circular arcs: $M$, $N$, $U$, $V$, $X$, $Y$, $S_1$ and $S_2$.

\subsubsection{Modulation of detail amplitude $h$}
\begin{figure}[ht]
    \centering
    \begin{subfigure}[ht]{0.24\textwidth}
        \centering
        \includegraphics[width=\textwidth]{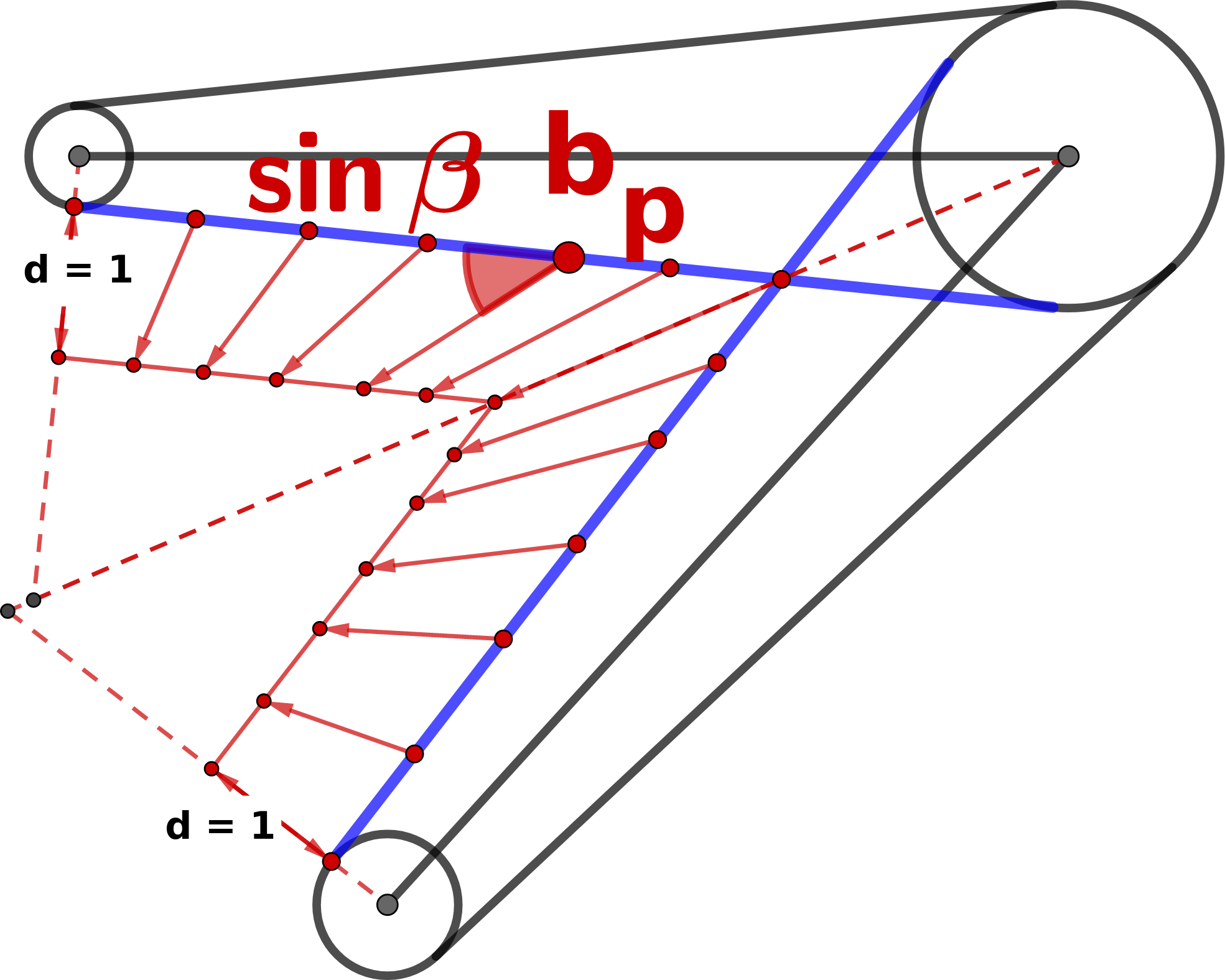}
        
        \caption{Surface defined as a sphere-mesh offset before a movement ($d=1$ everywhere)}
        \label{fig:module_avant}
    \end{subfigure}
    ~
     \begin{subfigure}[ht]{0.20\textwidth}
        \centering
        \includegraphics[width=\textwidth]{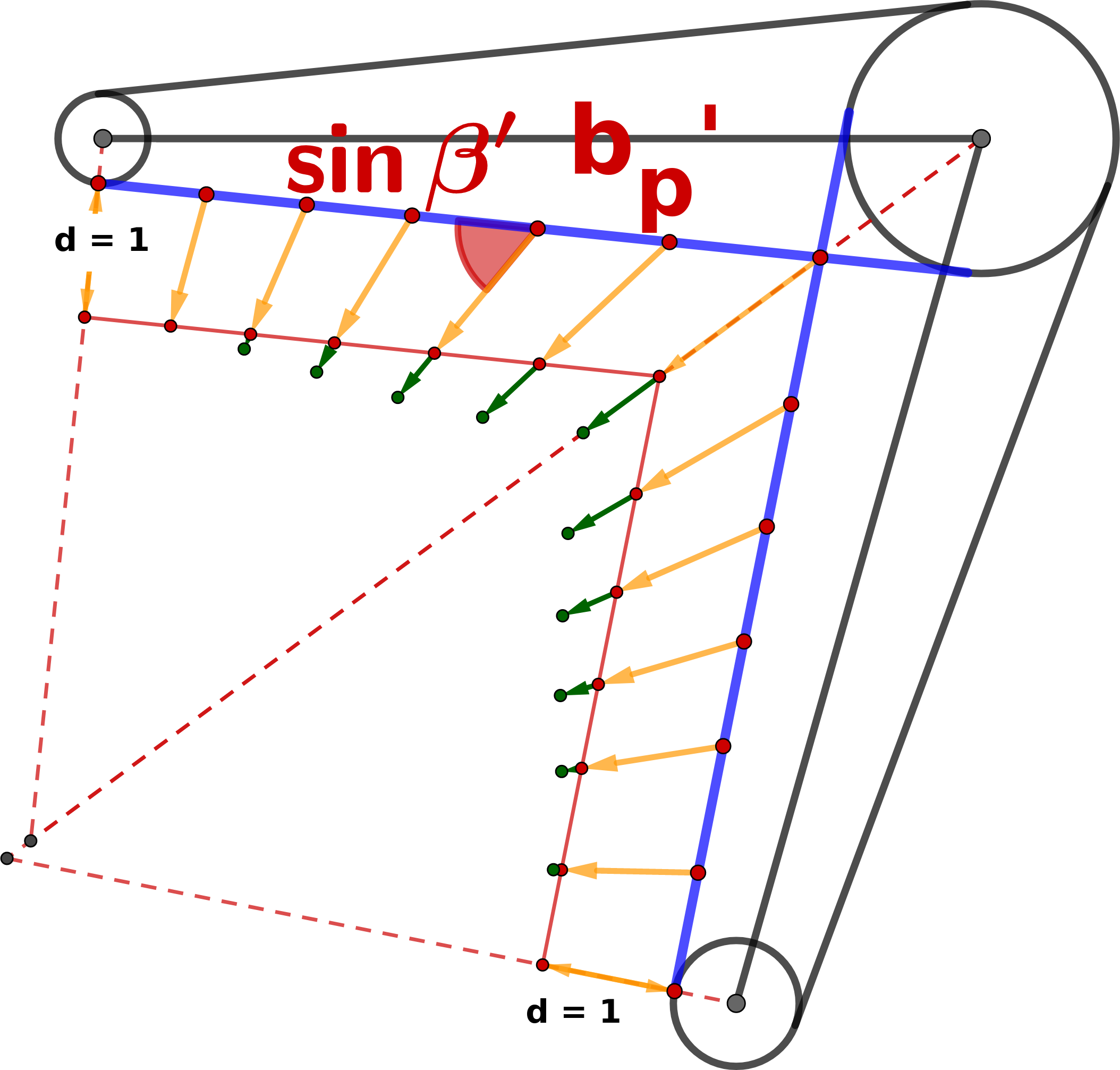}
        \caption{Angle-based modulation of the detail to preserve the offset.}
        \label{fig:module_apres}
    \end{subfigure}
    \caption{Angle-based modulation of the detail for a surface defined as a sphere-mesh offset $1$. The surface is sketched in red and some points are sampled on the surface. Yellow vectors represent the modulated details and green vectors represent the details without modulation.}
    \label{fig:modulation}
\end{figure}

If the shape we consider is an offset surface above the sphere-mesh ($h$ constant everywhere), we would like it to remain an offset surface after deformation. This amounts to a local principle of volume preservation which was not addressed in our approach (see green details in Figure \ref{fig:module_apres}.
%\raphq{+ Préciser les couleurs dans la légende}). 
To introduce this principle it is sufficient to modulate the amplitude of the detail $h$ with a function of the angle between the direction of detail $\overrightarrow{n_b}$ and the normal to the sphere-mesh (see red details in Figure \ref{fig:module_apres}):
The normalized value $h'$ is computed by:
$$h' = \frac{\sin_{\beta}}{\sin_{\beta}'}h$$
where $\sin\beta$ (or $\sin\beta'$) represents the angles of the detail direction $\overrightarrow{n_b}$ (or $\overrightarrow{n_b}'$)
%\raphq{MODIFIER Beta1 en Beta Beta2 en Beta' dans la légende.}.
It is this modulated detail that is reported over the base-point after movement of the skeleton. $\sin\beta = 1$ when a base-point is on the sphere part or on the convex part of the sphere-mesh model.

%In Figure \ref{fig:modulation}, we illustrate a 2D example of a surface (shown as a red line) over the sphere-mesh model with a \ju{constant offset equal to $1$} before and after a motion. We sample some points from this surface and show \ju{their detail amplitudes} as red vectors (Figure \ref{fig:module_avant}). We show the \ju{modulated detail amplitude for each point with modulation as green vectors and the amplitude without modulation as red vectors}. We can observe in Figure \ref{fig:module_apres} that the surface offset property is not preserved after bending an articulation without a modulation of displacement value of $h$.    

\subsubsection{Missing data when unfolding a joint}
\begin{figure}[ht]
    \centering
    \begin{subfigure}[ht]{0.15\textwidth}
        \centering
        \includegraphics[width=\textwidth]{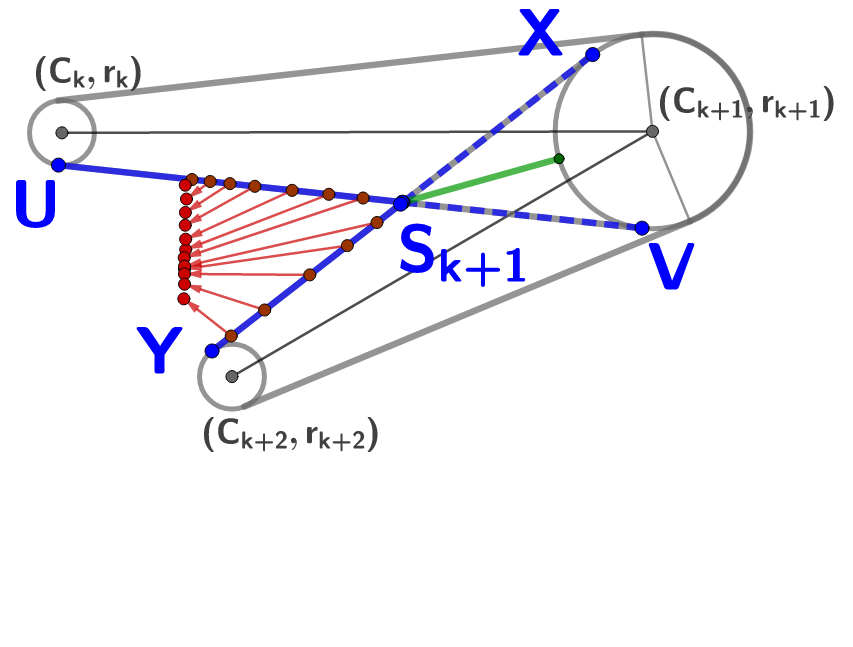}
        \caption{Our model with a fold in initial position.}
        \label{fig:ip_0}
    \end{subfigure}
     \begin{subfigure}[ht]{0.15\textwidth}
        \centering
        \includegraphics[width=\textwidth]{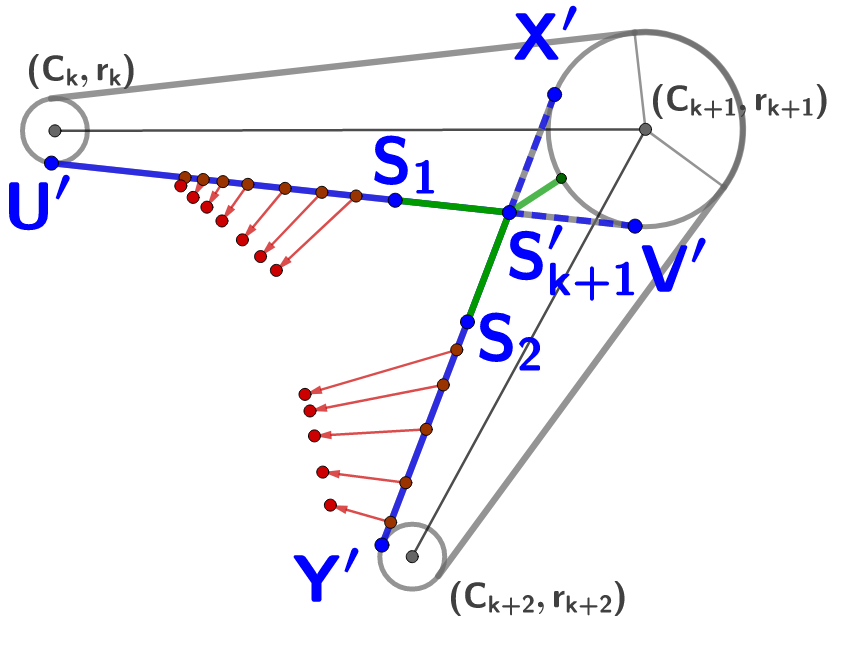}
        \caption{Unfold the joint with modified baseline skinning taking the fold into account.}
        \label{fig:ip_1}
    \end{subfigure}
     \begin{subfigure}[ht]{0.15\textwidth}
        \centering
        \includegraphics[width=\textwidth]{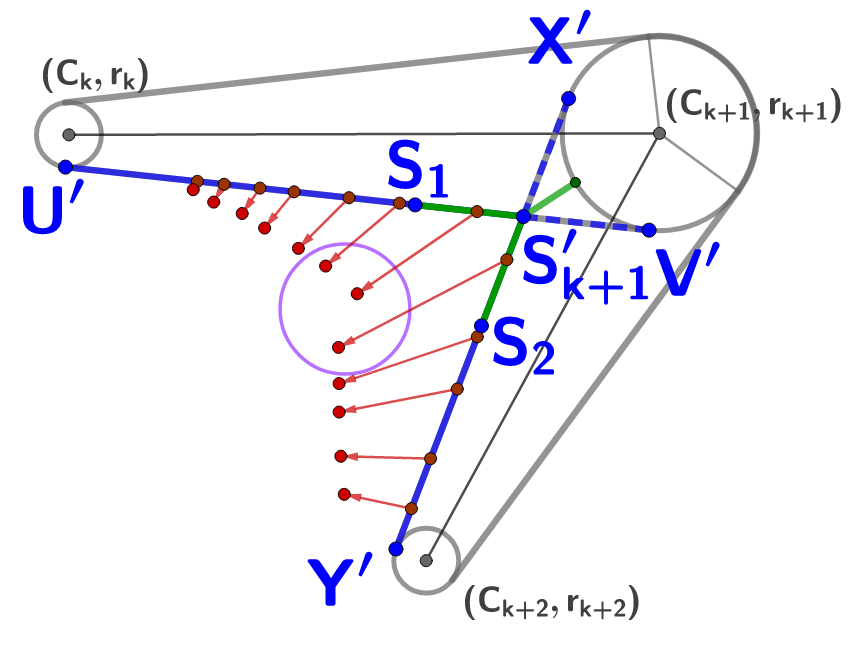}
        \caption{Unfold the joint with default baseline skinning that ensures a rough inpainting.}
        \label{fig:ip_3}
    \end{subfigure}
    \caption{Inpainting problem raised by unfolding a joint. }
    \label{fig:inp}
\end{figure}

\begin{figure}[ht]
    \centering
      \includegraphics[width=0.45\textwidth]{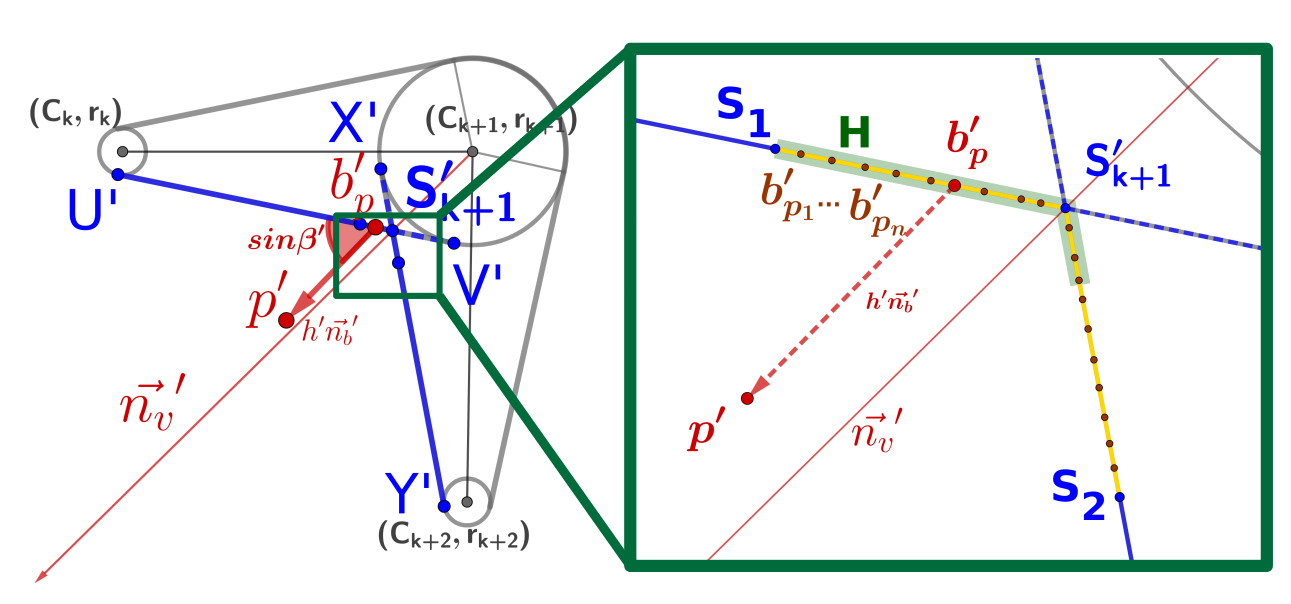}
    \caption{The illustration of the smooth zone}
    \label{fig:lissage}
\end{figure}

%\raphq{Par ailleurs, quand une articulation est pliée et qu'on souhaite la déplier, nous manquons parfois d'information. Il faut faire un schéma 2D pour illustrer le problème (cf slack): quand une articulation est pliée, il arrive que de l'information soit masquée dans le pli. Après dépliement, notre modèle se retrouve à attacher des zones qui sont parfois très éloignées sur la surface et on se retrouve face à un problème d'inpainting que nous listons dans notre future work. Cependant, en pratique, lors d'un dépliement, on étend les deux parties de la baseline sur la zone à inpainter en suivant notre méthode, et un lissage progressif gaussien de cette zone 1D permet de réduire le petit artefact visuel qui se produit lorsque la position de départ était trop pliée et que l'on se retrouve à attacher entre elles des zones trop distantes (Voir figure).}
When a joint is initially folded like in Figure \ref{fig:ip_0} and we want to unfold it, we sometimes lack information because it was hidden in the fold (green area in Figure \ref{fig:ip_0}). Figure \ref{fig:ip_1} shows the position of points on the baseline if we unfold the joint by taking into account the hidden area:
a new zone (segment $S_1S_{k+1}'$ and segment $S_{k+1}'S_2$) appears on the baseline after unfolding, for which we lack detail information. This phenomenon appears only in the concave part of a joint being unfolded. This can be solved by an inpainting process, which is a difficult task for surfaces. When taking into account only the visible part of the baseline (as it was done in previous subsections), we adopt a simpler strategy by not taking the hidden fold into account: we simply extend the two visible parts of the baseline over the area to be inpainted. The problem is that the junction may be unpleasant after reporting the detail since we stitch together details that may correspond to remote areas on the surface of the object (Figure \ref{fig:ip_3} in purple zone). Therefore, we propose a gradual Gaussian smoothing of this 1D area to reduce the small visual artifact which occurs when the starting position was too bent.
%\raphq{on applique bien le baseline skinning dans ce cas, mais en ajoutant la longueur du pli que tu n'as pas représenté sur la figure (cf image du slack). Il faut donc dessiner le pli dans (a), il est encore présent dans (b) mais plus petit et la différence de longueur se reporte sur la nouvelle zone sans info qui apparait.}: 
%\raphq{Ajouter une illustration de la longueur de la zone 1D à lisser, éventuellement avec des couleurs qui peuvent permettre d'exprimer des reports de distances}. 
We apply this smoothing process only in the context where we have a severe fold that possibly hides information. 
%We smooth the detail amplitude $h(b_p)$ of base-point $b_p$ by a Gaussian-weighted average of its neighbor points detail amplitude. 
For example, in Figure \ref{fig:lissage}, we gradually apply this smoothing process only for base-points on segment $S_1S'$ and segment $S'S_2$. $S_1$ and $S_2$ represent the positions of the initial point $S$ on the left bone and the right bone before unfolding the joint. The resulting height amplitude of $p$ is thus:
$$\bar{h}(b_p') = \frac1{cst}\sum_{b_{p_i}' \in H}exp(-\frac{\|b_p'-b_{p_i}'\|^2}{2\delta(b_p')^2}) h(b_{p_i}')$$
with $cst$ a weight normalizing factor. $H$ is a subset of $b_p'$'s neighbors' base points shown in green zone in Figure \ref{fig:lissage}. The value of $\delta(b_p')$ depends on the position of point $b_p'$: it becomes smaller when $b_p'$ is close to the boundary ($S_1$ or $S_2$ in our example) of the smooth zone:
$\delta(b_p') = \frac{\|S_1b_p'\|}{3}$ if $b_p'$ is on the segment $S_1S_{k+1}'$ or $\delta(b_p') = \frac{\|S_2b_p'\|}{3}$ if $b_p'$ is on the segment $S_{k+1}'S_2$. 

Let note that in contrast, when we bend a joint, we choose the approach which consists in leaving the detail visible and pushing it in its entirety outside the fold. But we could very well set thresholds beyond which we would choose to hide part of the detail inside the fold.

\section{Results}
In this section, we show the performance of our baseline skinning method on synthetic and real point sets.
We implemented our algorithm in C++, using OpenMP for computing baselines in parallel. All experiments are run on an Intel Core i7-4790K CPU @ 4.00GHz.

\begin{figure}[ht]
    \centering
        \begin{subfigure}{0.45\textwidth}
        \centering
        \includegraphics[width=0.3\textwidth]{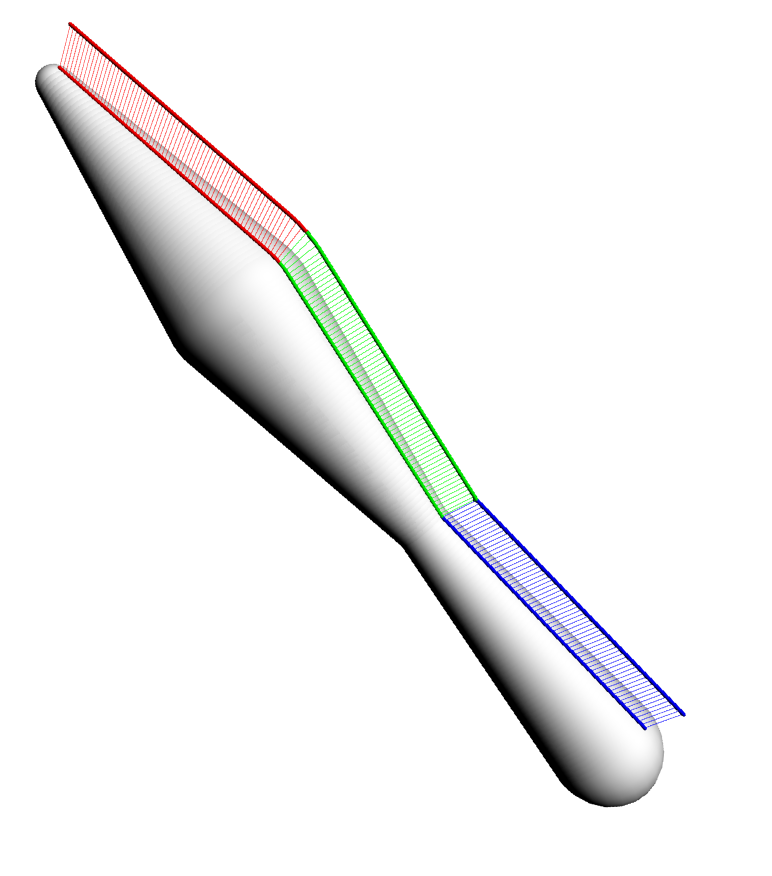}
        \includegraphics[width=0.3\textwidth]{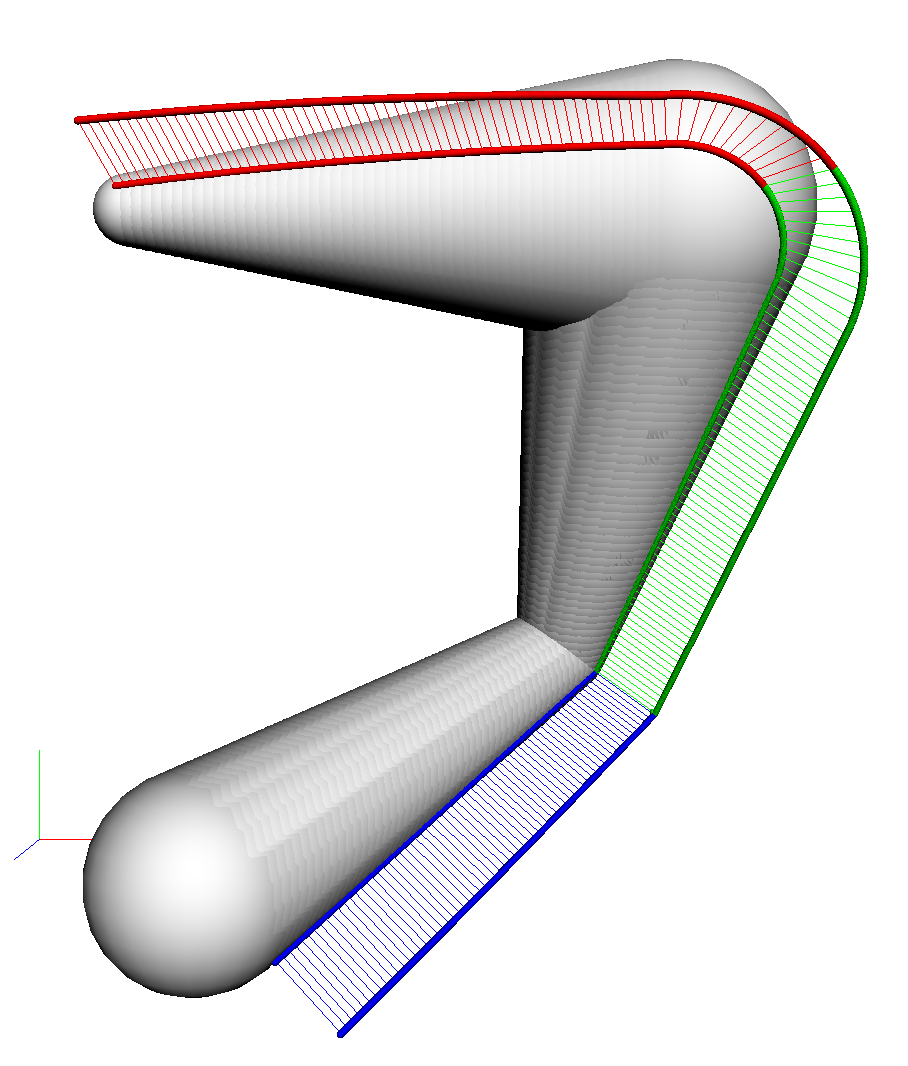}
        \includegraphics[width=0.3\textwidth]{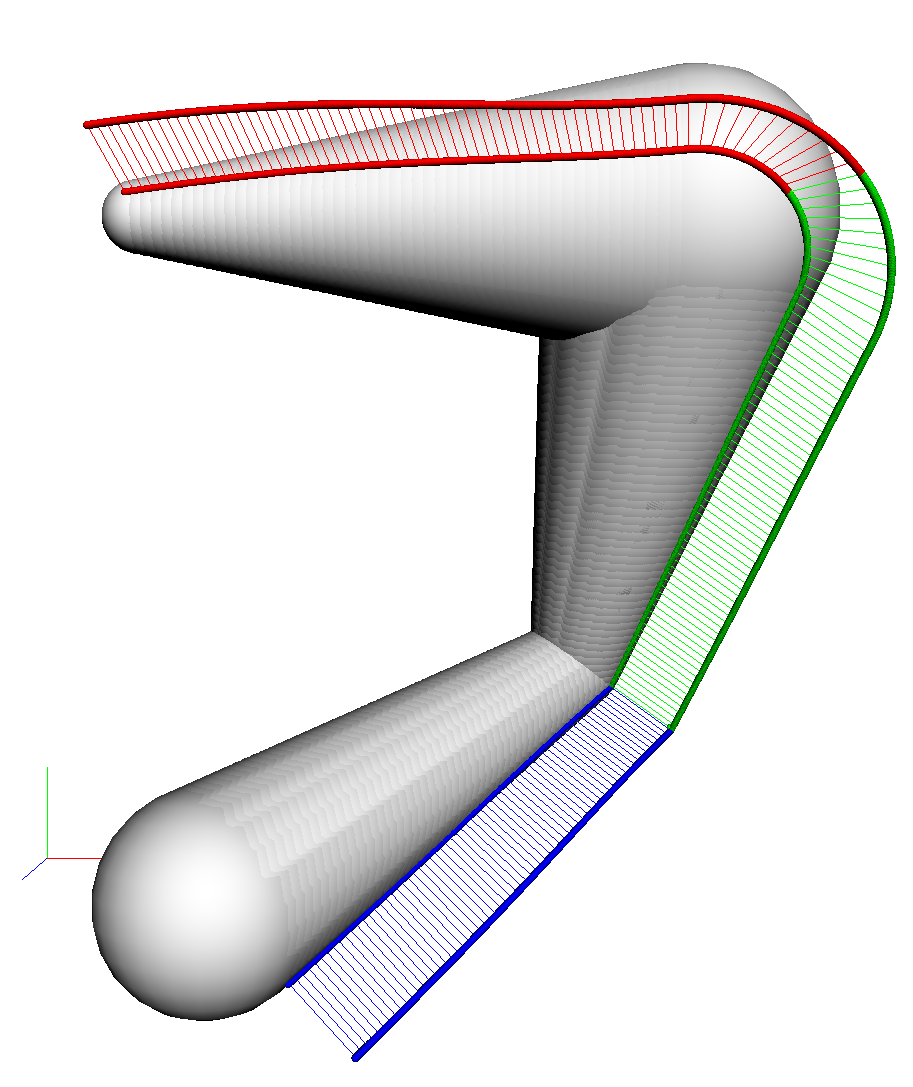}
        \caption{A bending motion for a baseline in a convex part after deformation.}
        \end{subfigure}
        
        \begin{subfigure}{0.45\textwidth}
        \centering
        \includegraphics[width=0.3\textwidth]{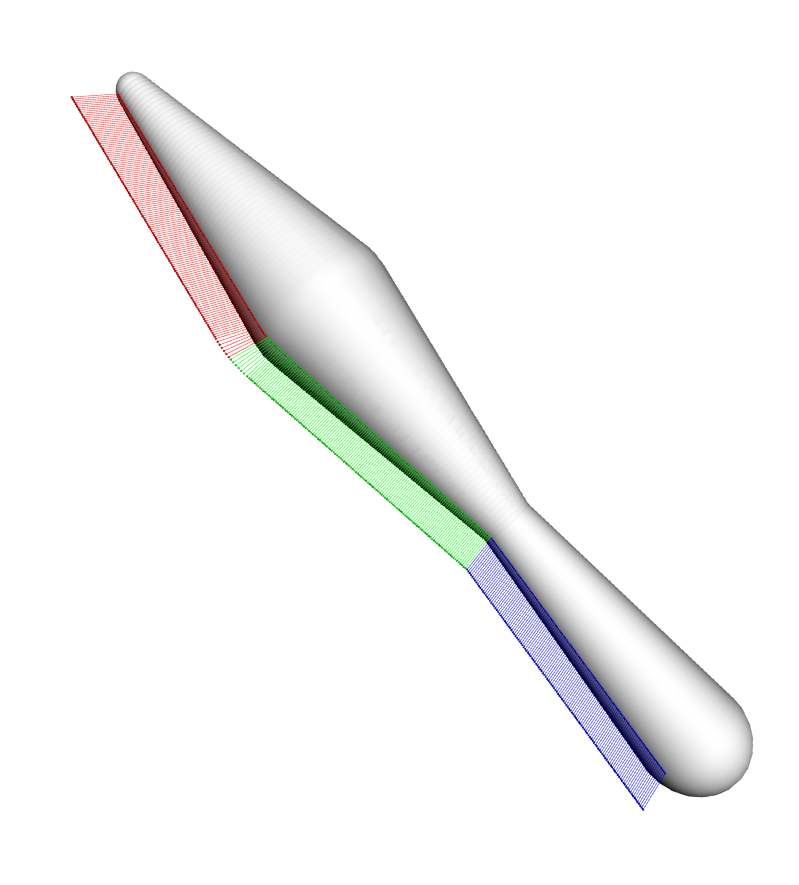}
        \includegraphics[width=0.3\textwidth]{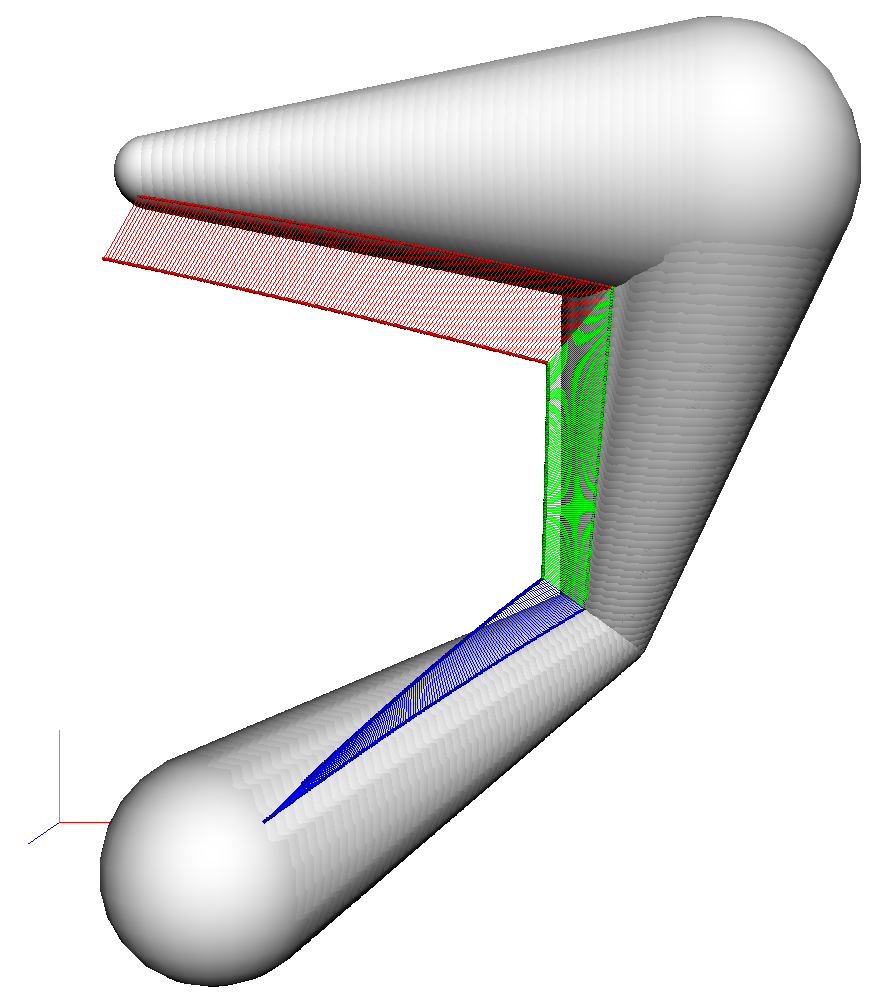}
        \includegraphics[width=0.3\textwidth]{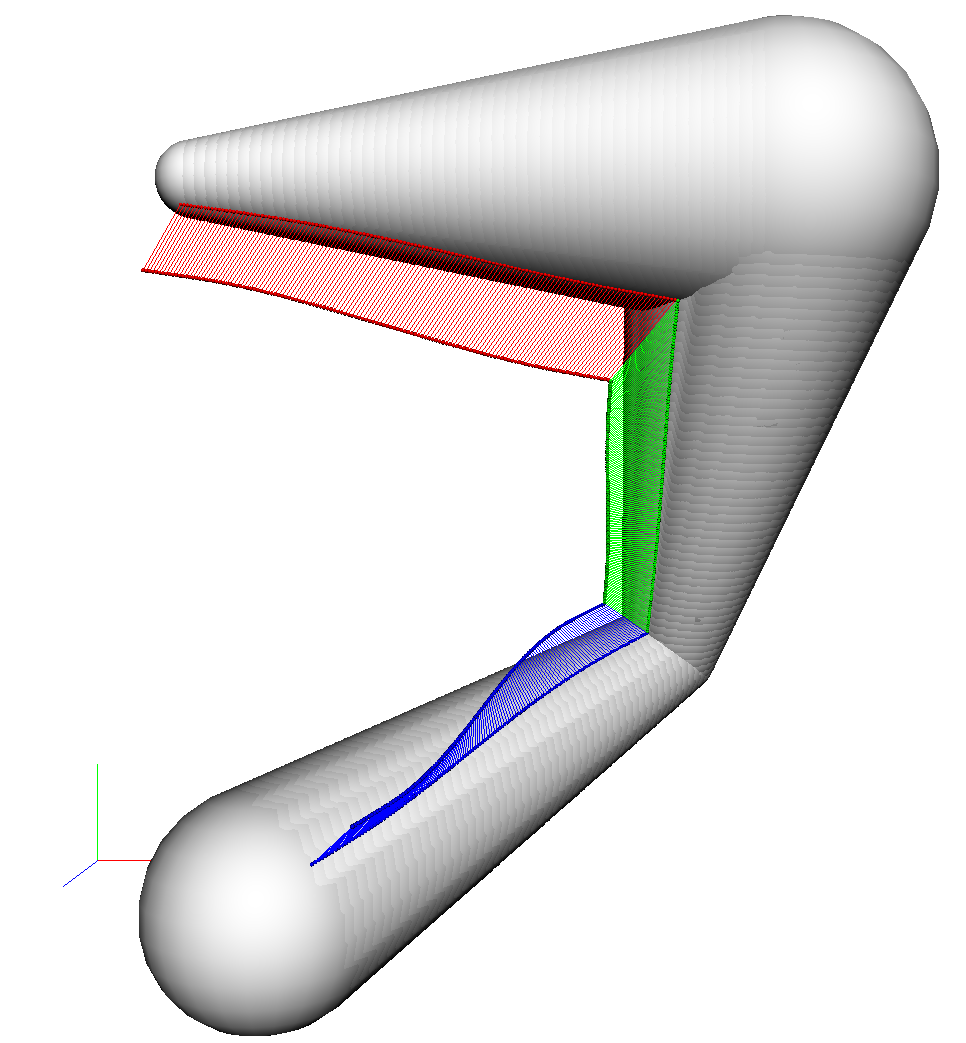}
        \caption{A bending motion for a baseline in a concave part after deformation.}
        \end{subfigure}
        
        \begin{subfigure}{0.45\textwidth}
        \centering
        \includegraphics[width=0.3\textwidth]{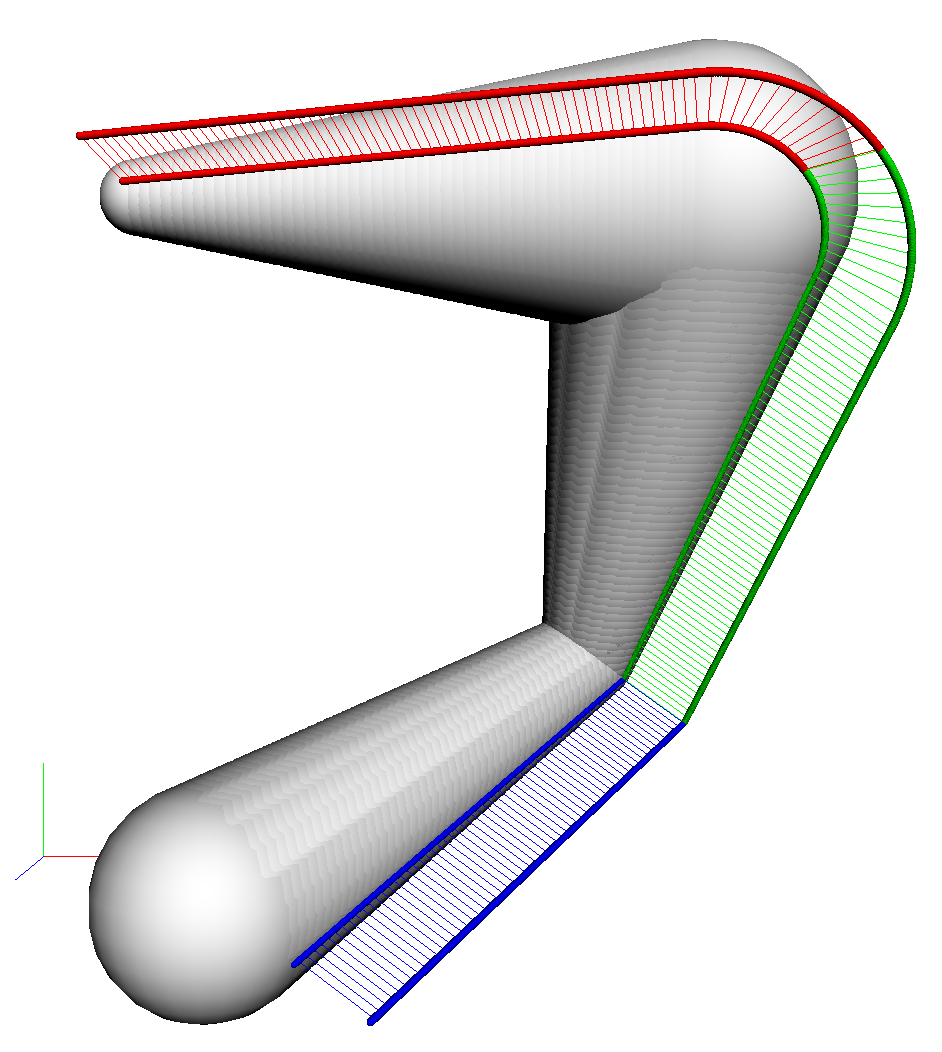}
        \includegraphics[width=0.3\textwidth]{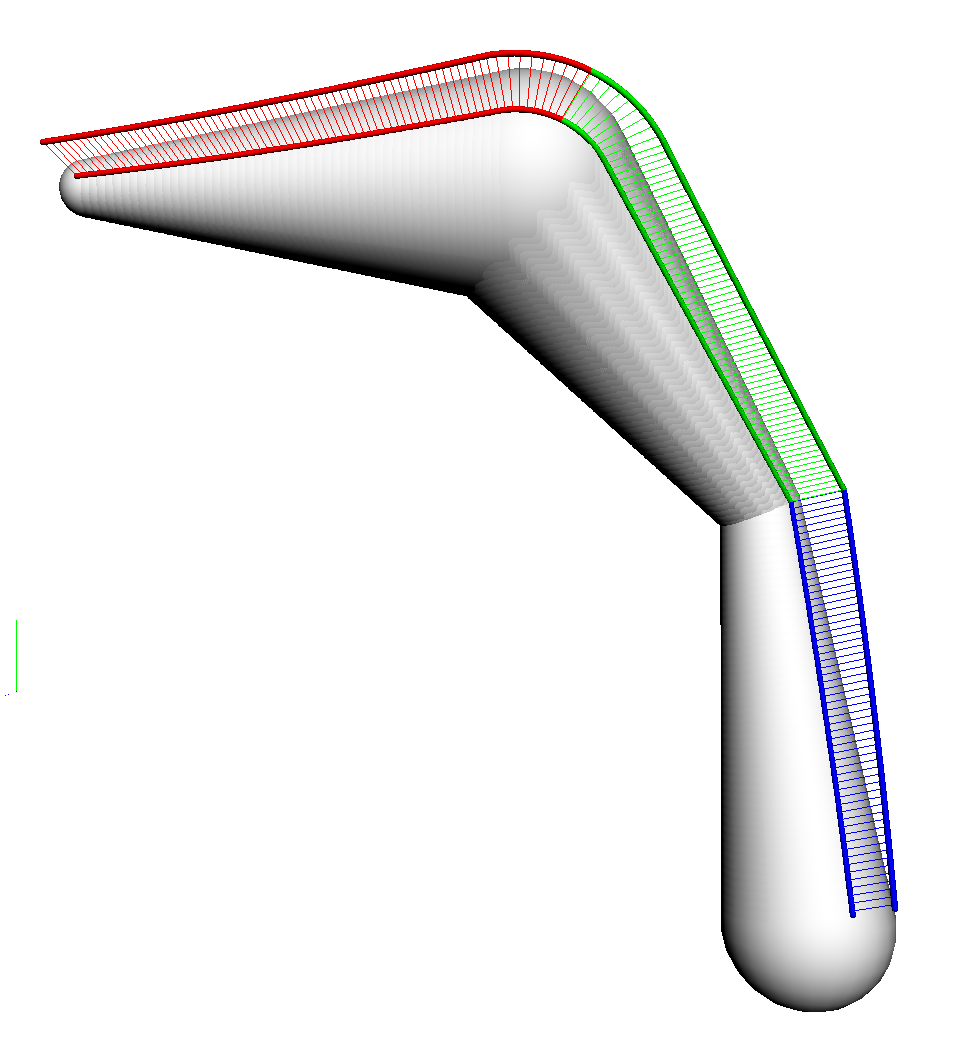}
        \includegraphics[width=0.3\textwidth]{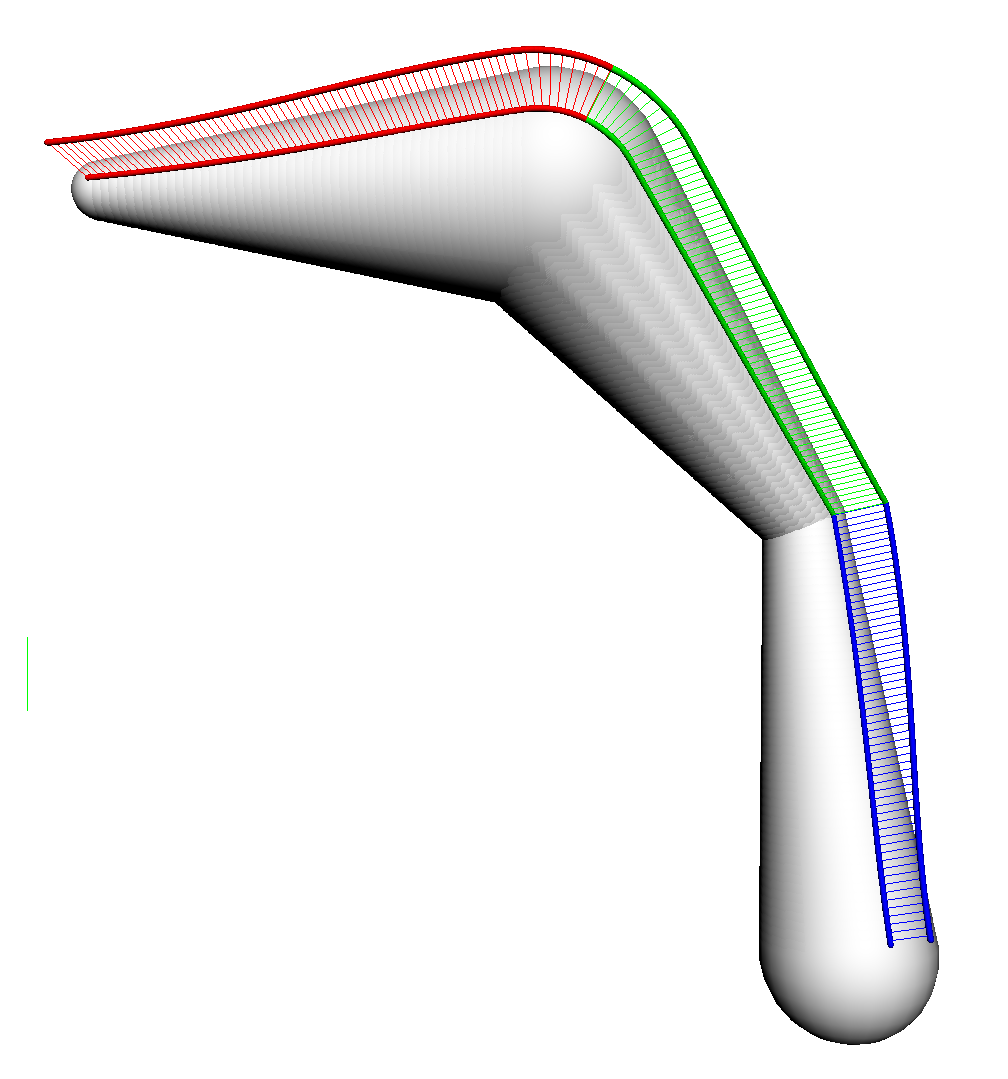}
        \caption{An unfolding motion for a baseline in a convex part after deformation.}
        \end{subfigure}
        
        \begin{subfigure}{0.45\textwidth}
        \centering
        \includegraphics[width=0.3\textwidth]{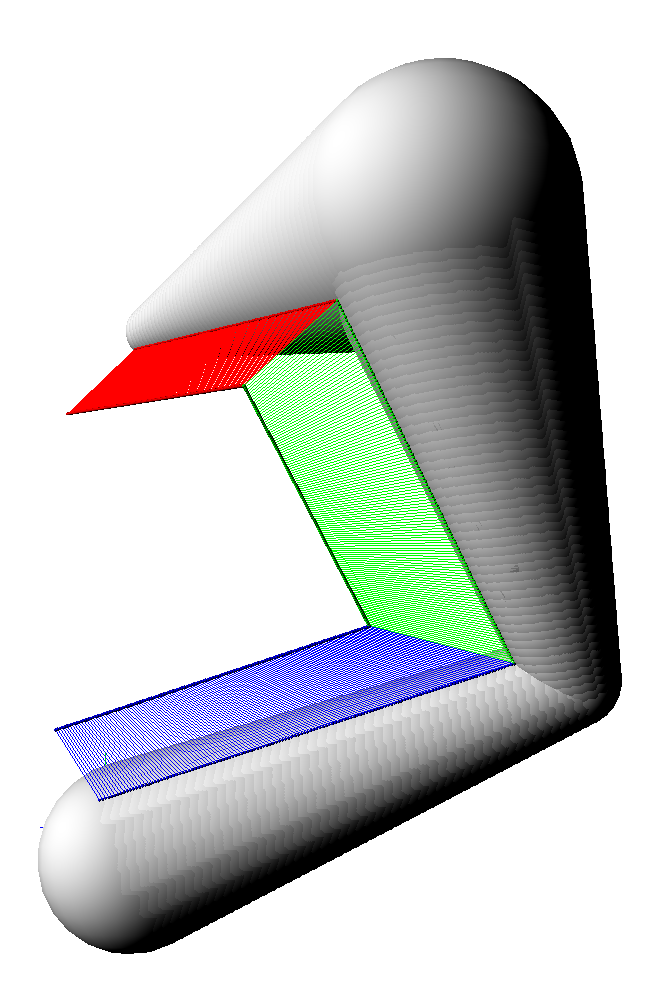}
        \includegraphics[width=0.3\textwidth]{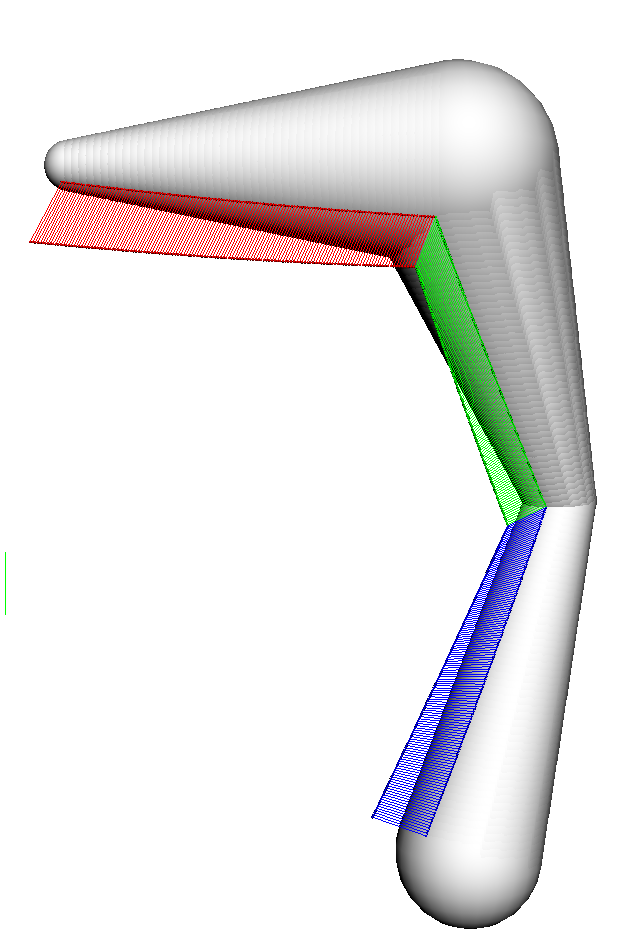}
        \includegraphics[width=0.3\textwidth]{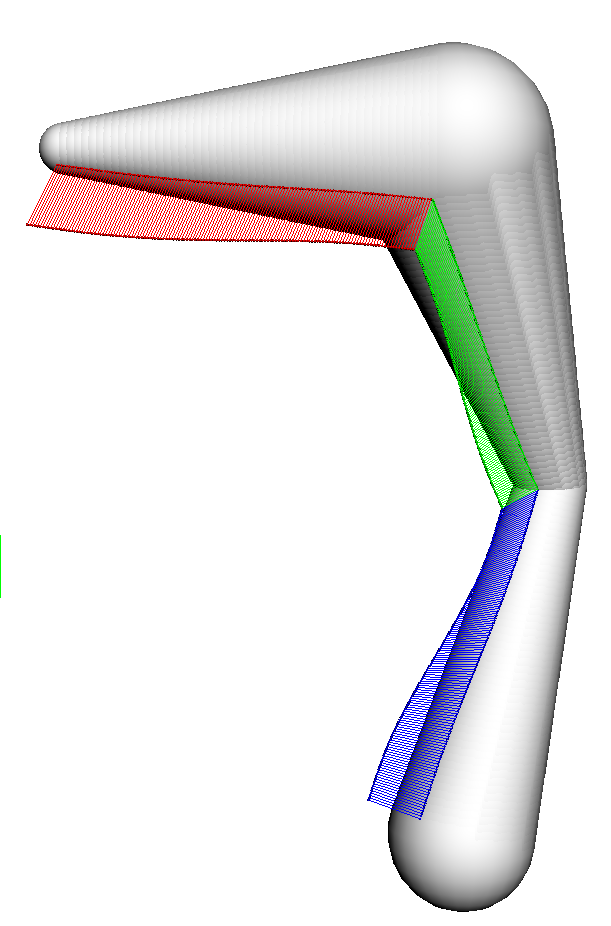}
        \caption{An unfolding motion for a baseline in a concave part after deformation.}
        \end{subfigure}
    \caption{Baseline skinning on synthetic data for point on a single baseline. First column: initial position; second column: linear interpolation for $\alpha_p$ and $\alpha_d$; third column: cubic interpolation for $\alpha_p$ and $\alpha_d$.}
    \label{fig:results_syn}
\end{figure}

We first test our algorithm on synthetic data. We generated a set of points lying on the same baseline and used a chain of three bones. The attachment of points to bones is shown in different colors: red, green and blue. 
In Figure \ref{fig:results_syn}, we show original points, their base points and detail directions.
For different bone motion, we show the initial position (first column), our baseline skinning results by linear interpolation (2nd column) and by cubic interpolation (3rd column) for both $\alpha_p$ and $\alpha_d$.
We show the results for a folding motion (first and second row), and an unfolding motion (third and fourth rows), for baselines situated in convex or concave parts after deformation. In all these synthetic problems our baseline skinning method performs well. The skinning result can also be further adjusted by changing interpolation schemes. Since our detail direction is not orthogonal to the bones in concave parts, we avoid the detail direction intersecting near the joints. Then we compare our method with LBS, dual quaternion skinning and skinning method presented in~\cite{h.20201290} on a stripy synthetic surface \ref{fig:fig:results_plie_deplie}. We show the results of twisting, bending and re-bending to initial position. Our method gives a good performance for all these movements. 
 
\begin{figure*}[ht]
    \centering
        \begin{subfigure}{0.23\textwidth}
        \centering
        \includegraphics[width=\textwidth]{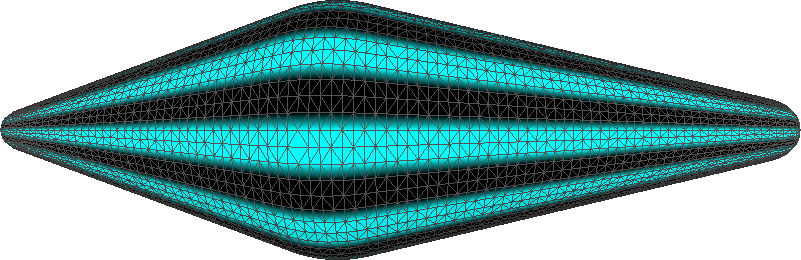}\\
        \includegraphics[width=\textwidth]{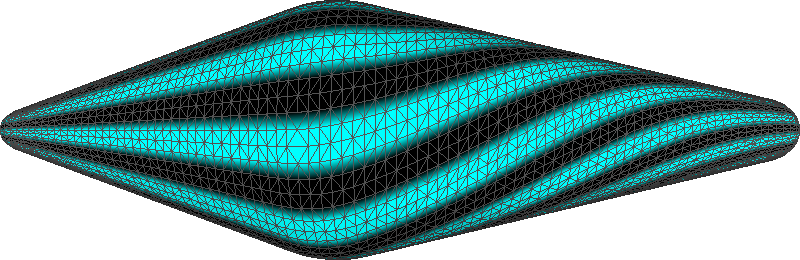}\\
        \includegraphics[height = 3cm]{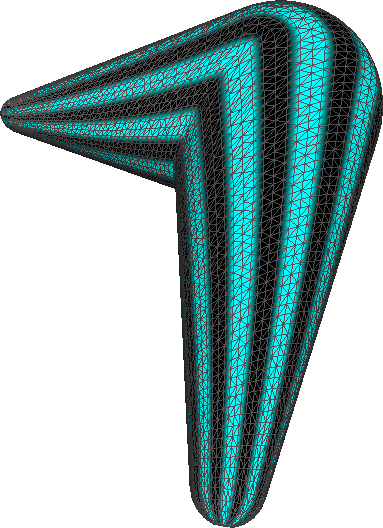}\\
        \includegraphics[width=\textwidth]{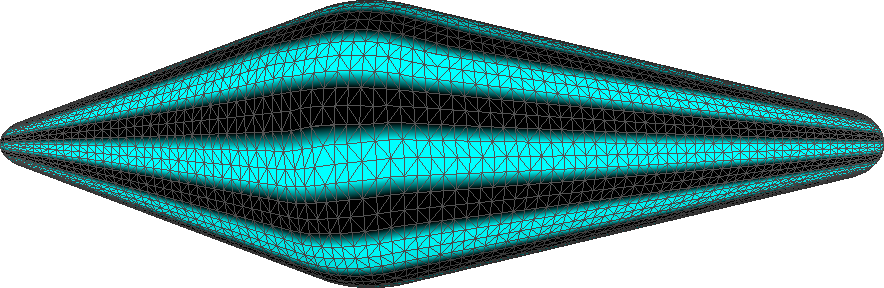}
        \caption{Our method}
        \end{subfigure}
        \begin{subfigure}{0.23\textwidth}
        \centering     
        \includegraphics[width=\textwidth]{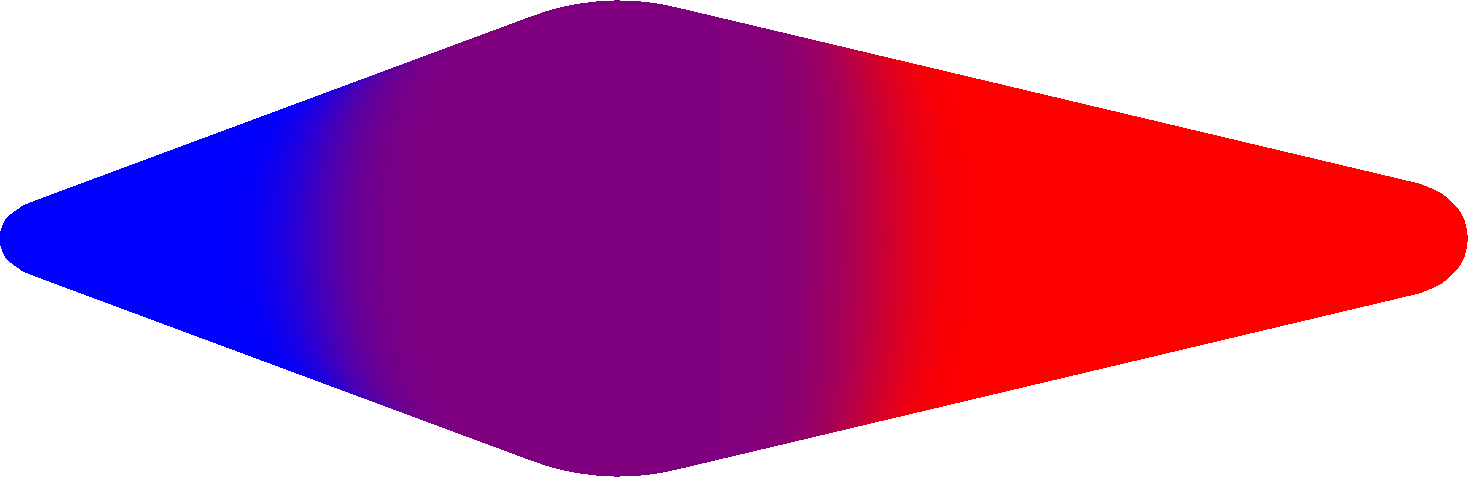}\\
        \includegraphics[width=\textwidth]{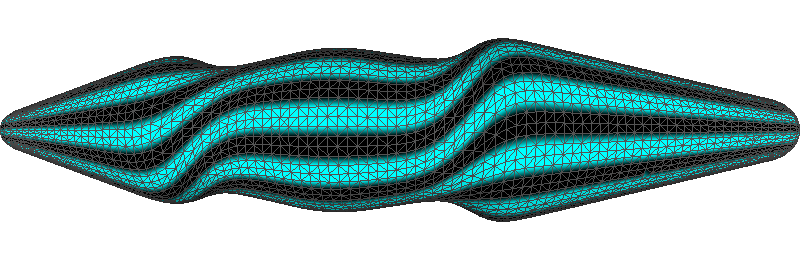}\\
        \includegraphics[height = 3cm]{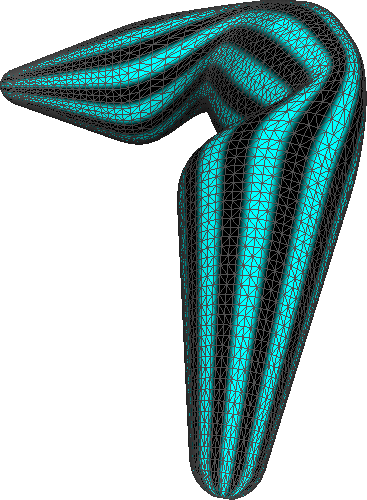}\\
        \includegraphics[width=\textwidth]{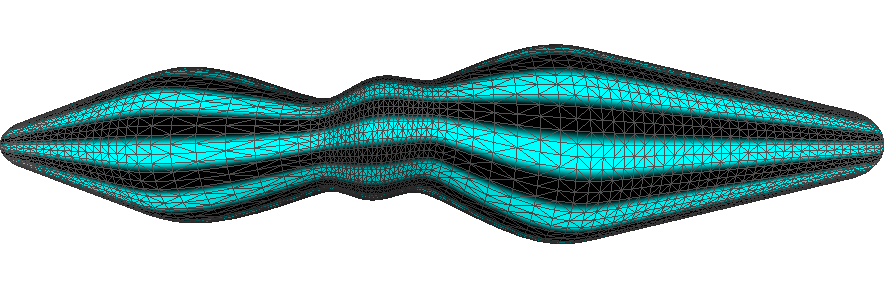}
        \caption{Linear Blend skinning}
        \end{subfigure}
        \begin{subfigure}{0.23\textwidth}
        \centering
        \includegraphics[width=\textwidth]{poids_lbs_dq00.png}\\
        \includegraphics[width=\textwidth]{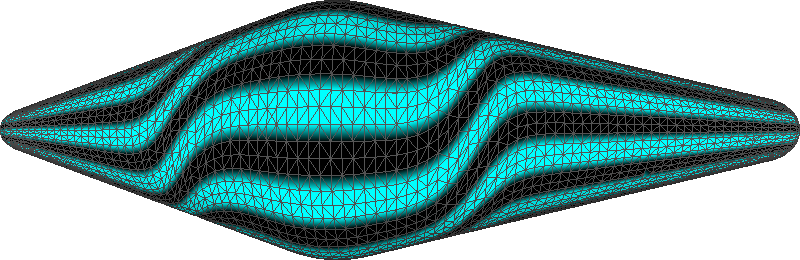}\\
        \includegraphics[height = 3cm]{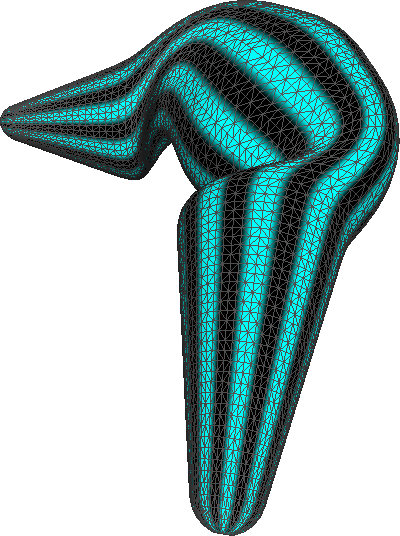}\\
        \includegraphics[width=\textwidth]{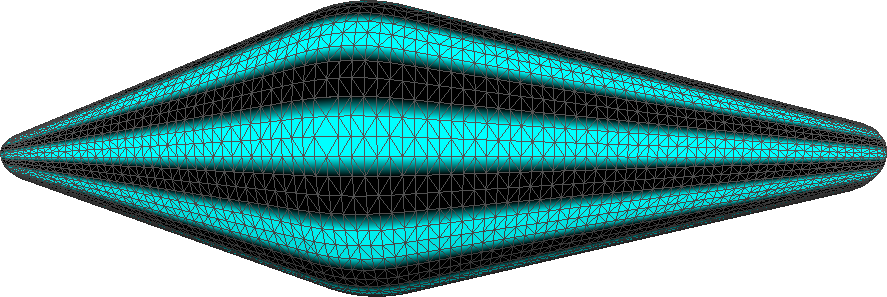}
        \caption{Dual quaternion skinning}
        \end{subfigure}
        \begin{subfigure}{0.23\textwidth}
        \centering   
        \includegraphics[width=\textwidth]{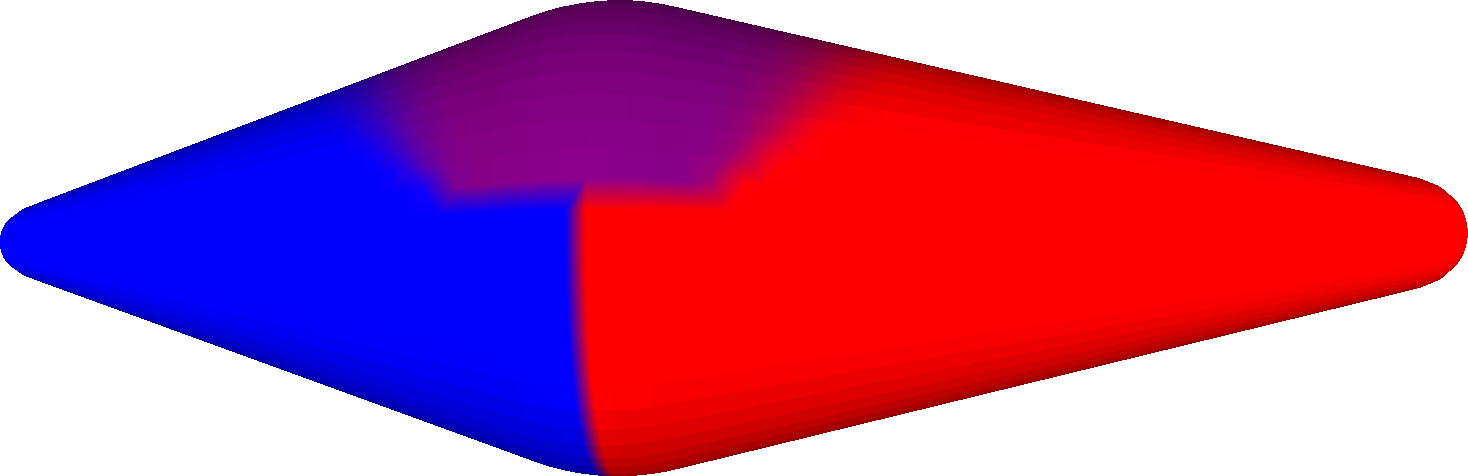}\\
        \includegraphics[width=\textwidth]{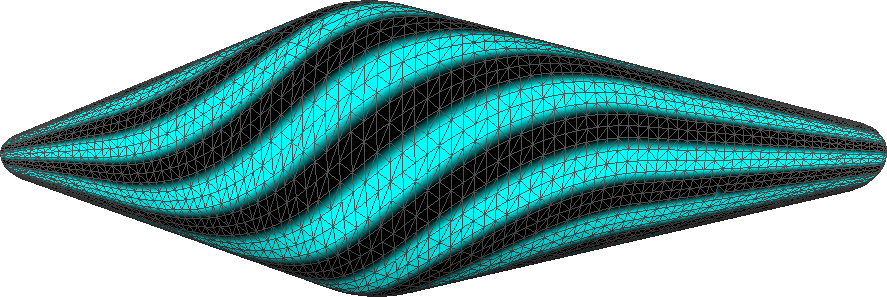}\\
        \includegraphics[height = 3cm]{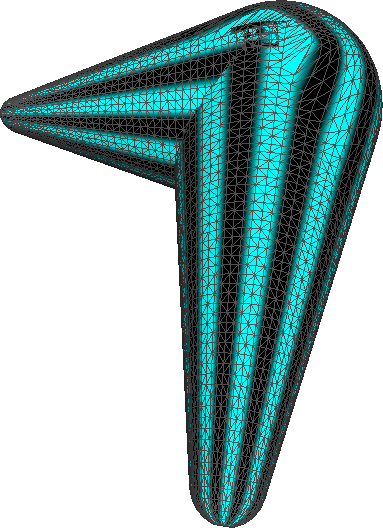}\\
        \includegraphics[width=\textwidth]{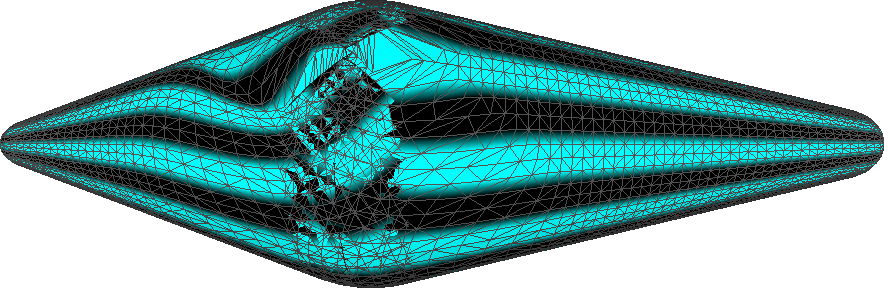}
        \caption{Fu et al.\cite{h.20201290}}
        \end{subfigure}
    \caption{In each sub-figure from up to down: initial position, twisting result, bending result, re-bending to initial position result. We show the weight distribution for the right bone in the first line in sub-figure (b), (c) and (d).  We used the same Gaussian weight computation introduced in \cite{h.20201290} for (b), (c) and (d). The support of the Gaussian weight is anisotropically reduced in (d). }
    \label{fig:results_plie_deplie}
\end{figure*}

\begin{figure}[ht]
    \centering
    \begin{subfigure}[ht]{0.3\textwidth}
        \centering
        \includegraphics[height=3cm]{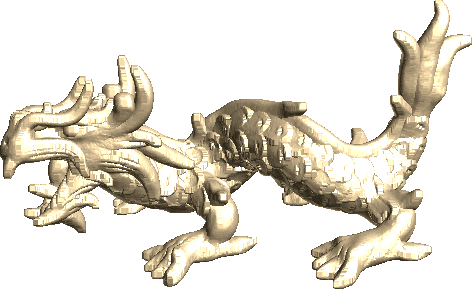}
        \caption{Dragon}
    \end{subfigure}
    \begin{subfigure}[ht]{0.15\textwidth}
        \centering
        \includegraphics[height=3cm]{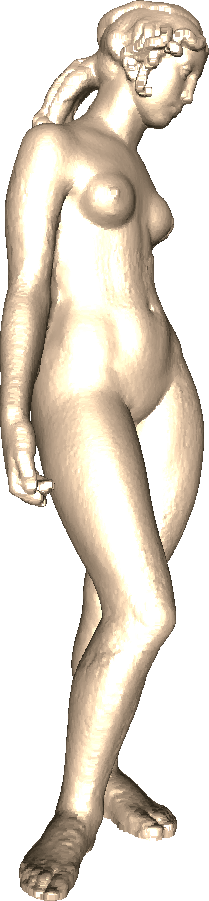}
        \caption{Aphrodite}
    \end{subfigure}
    
    \begin{subfigure}[ht]{0.15\textwidth}
        \centering
        \includegraphics[height=3cm]{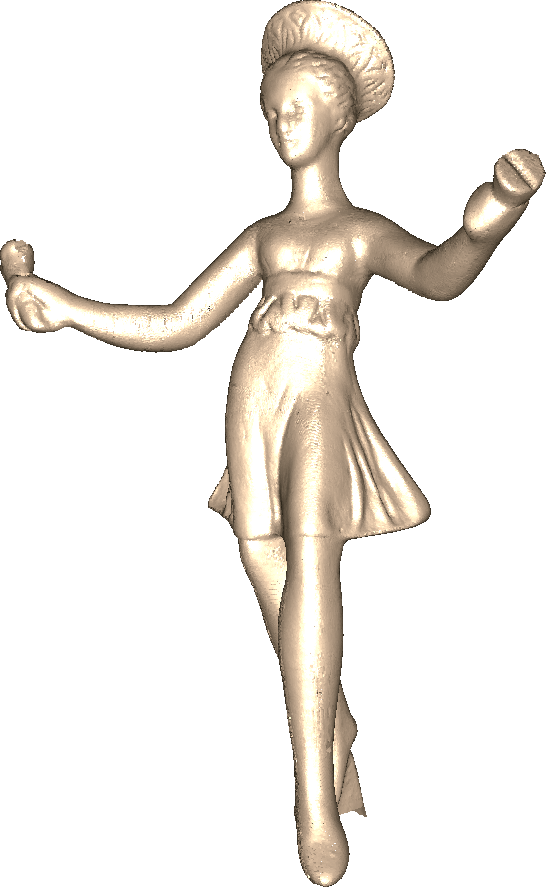}
        \caption{Dancer with Crotales}
    \end{subfigure}
    \begin{subfigure}[ht]{0.15\textwidth}
        \centering
        \includegraphics[height=3cm]{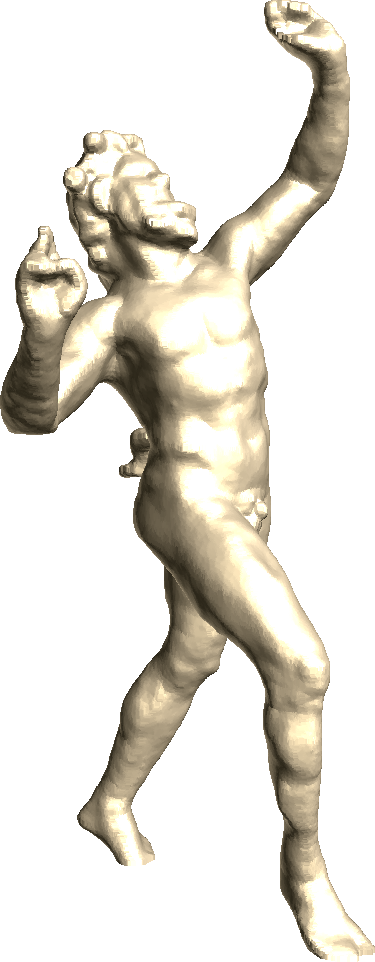}
        \caption{Dancing Faun}
    \end{subfigure}
    \begin{subfigure}[ht]{0.15\textwidth}
        \centering
        \includegraphics[height=3cm]{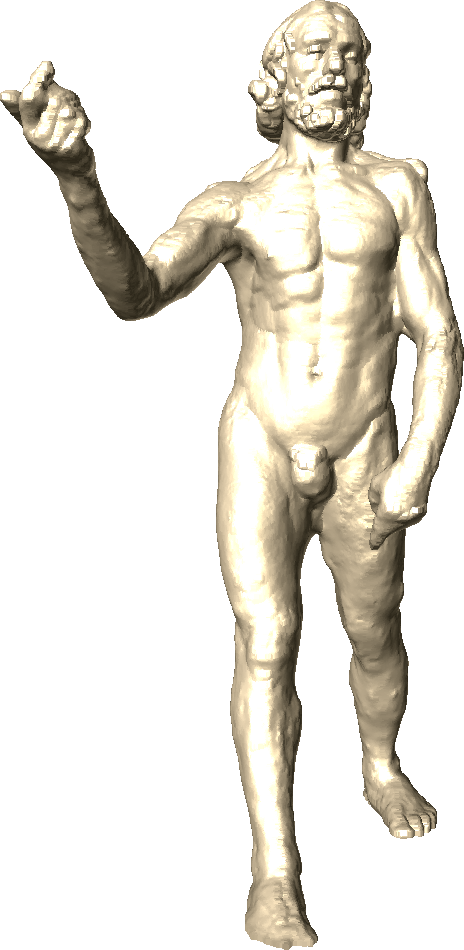}
        \caption{Saint John the Baptist}
    \end{subfigure}
    \caption{The original point sets of statues.}
    \label{fig:original}
\end{figure}

We selected a set of statues from various sources:
\begin{enumerate}
\item Dragon, Scan data of a dragon sculpture
\item Dancer with Crotales, Louvre Museum%, Paris, France.
\item Aphrodite, Thorvaldsens Museum%, Copenhagen, Denmark.
\item Dancing Faun, Pompei excavations %\raph{(complete model)}%, Italy.
\item Saint John the Baptist, Ny Carlsberg Glyptotek %\raphq{St Jean Baptiste A vérifier sur internet???}%, Copenhagen, Denmark
\end{enumerate}
While the 'Dancer with crotales' is a raw point set \cite{ipol.2011.dalmm_ps}. The other models are point sets sampled on meshes extracted from the Sketchfab website. We show original point set of statues in Figure\ref{fig:original}.

\begin{table}[ht]
    \centering
    \begin{tabular}{|p{1.8cm}|p{1.4cm}|p{0.7cm}|p{0.7cm}|p{0.7cm}|p{0.7cm}|}
         \hline
         model & number of points & Ours & Fu et al. & DQ & LBS \\
         \hline
         Dragon & 532 067& 0.48s & 0.62s & 0.34s & 0.49s \\
         \hline
         Dancer with Crotales & 539 136 & 1.32s & 1.4s & 0.21s & 0.27s\\
         \hline
         Aphrodite & 461 688 & 0.31s & 0.82s & 0.22s & 0.31s\\
         \hline
         Dancing Faun & 518 037 & 0.63s & 1.38s & 0.22s & 0.28s\\
         \hline
         Saint John the Baptist & 519 483 & 0.48s & 1.68s & 0.25s & 0.31s\\
         \hline
    \end{tabular}
     \caption{Execution times of different methods for the point sets. }
     \label{tab:time}
 \end{table}
     
\begin{figure}[ht]
    \centering
        \begin{subfigure}[ht]{0.11\textwidth}
        \centering
            \includegraphics[height=3cm]{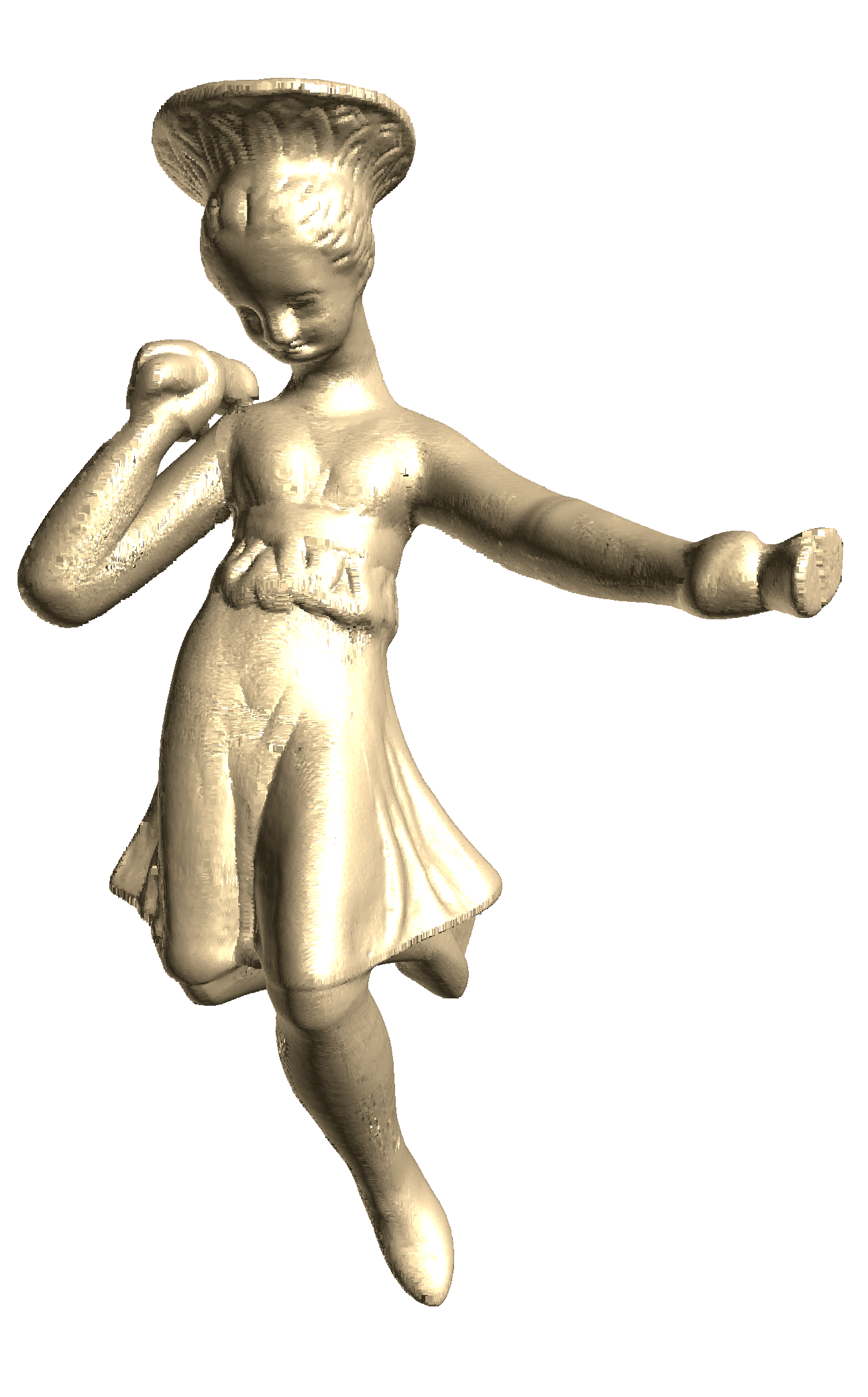}\\
            \includegraphics[height=2cm]{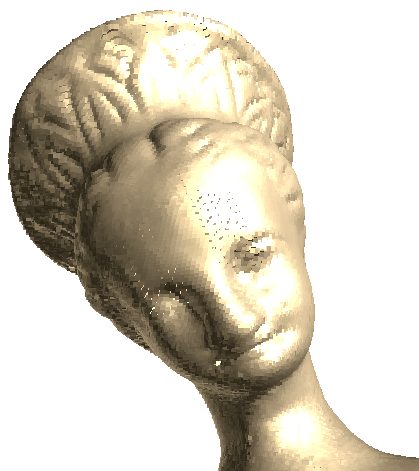}\\
            \includegraphics[height=2cm]{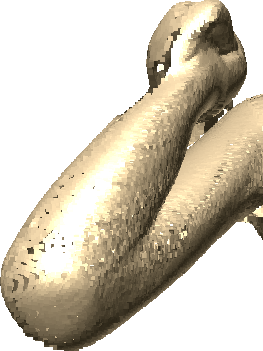}\\
            \includegraphics[height=0.8cm]{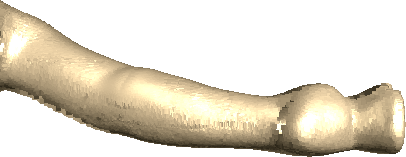}\\
        \caption{}
        \label{fig:danseuse_b}
        \end{subfigure}
        \begin{subfigure}[ht]{0.11\textwidth}
        \centering
            \includegraphics[height=3cm]{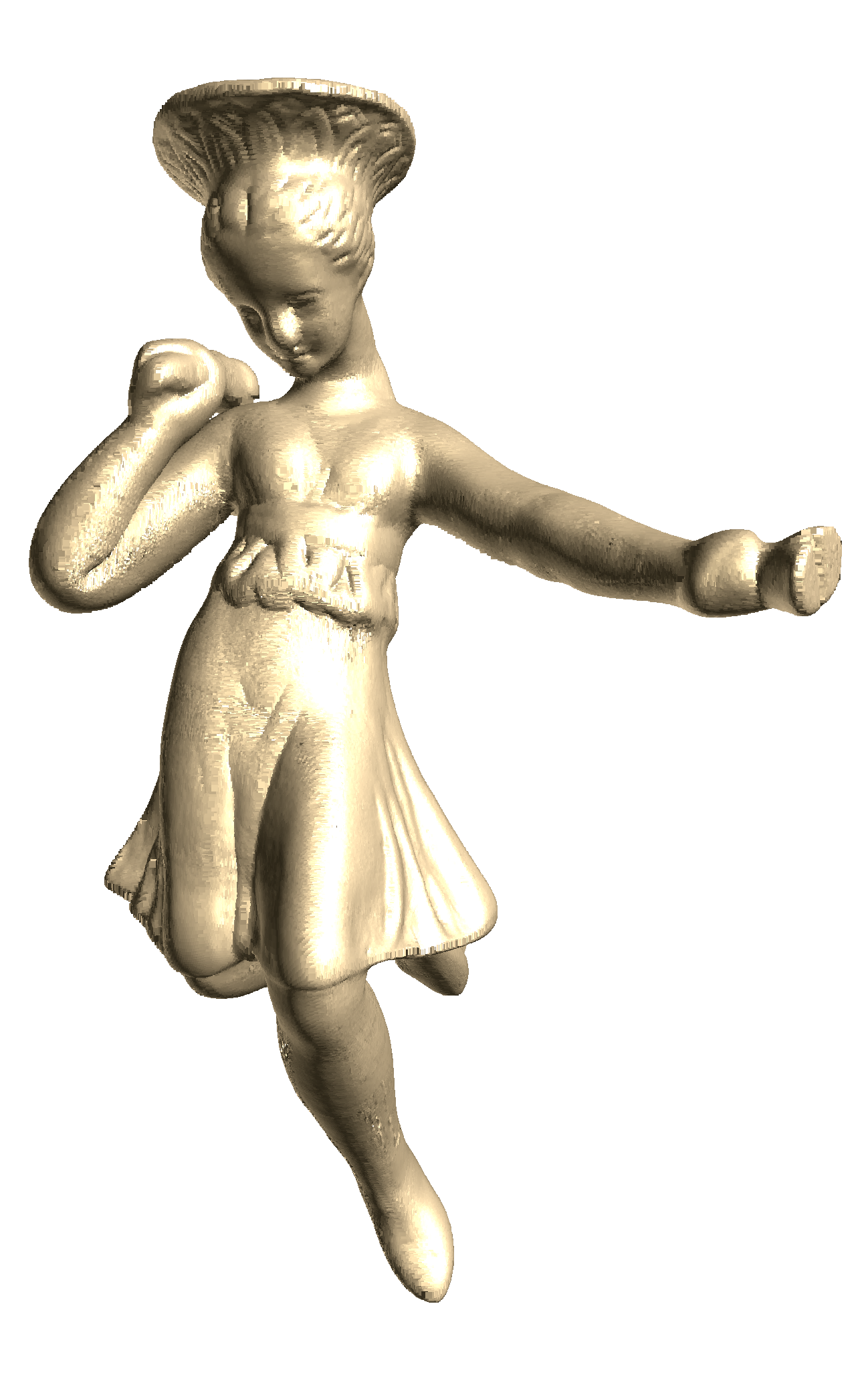}\\
            \includegraphics[height=2cm]{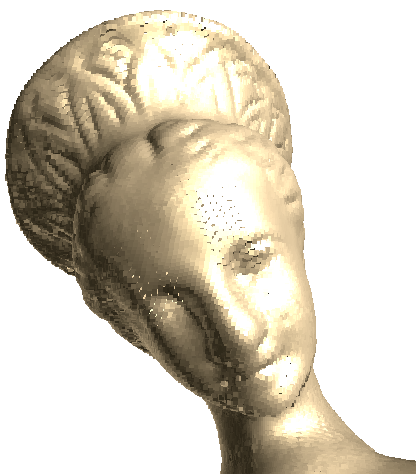}\\
            \includegraphics[height=2cm]{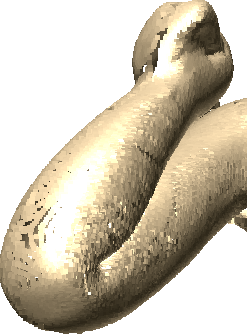}\\
            \includegraphics[height=0.8cm]{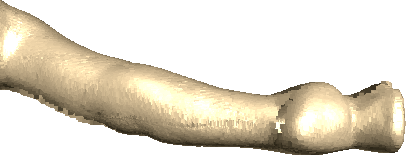}
        \caption{}
        \end{subfigure}
        \begin{subfigure}[ht]{0.11\textwidth}
        \centering
            \includegraphics[height=3cm]{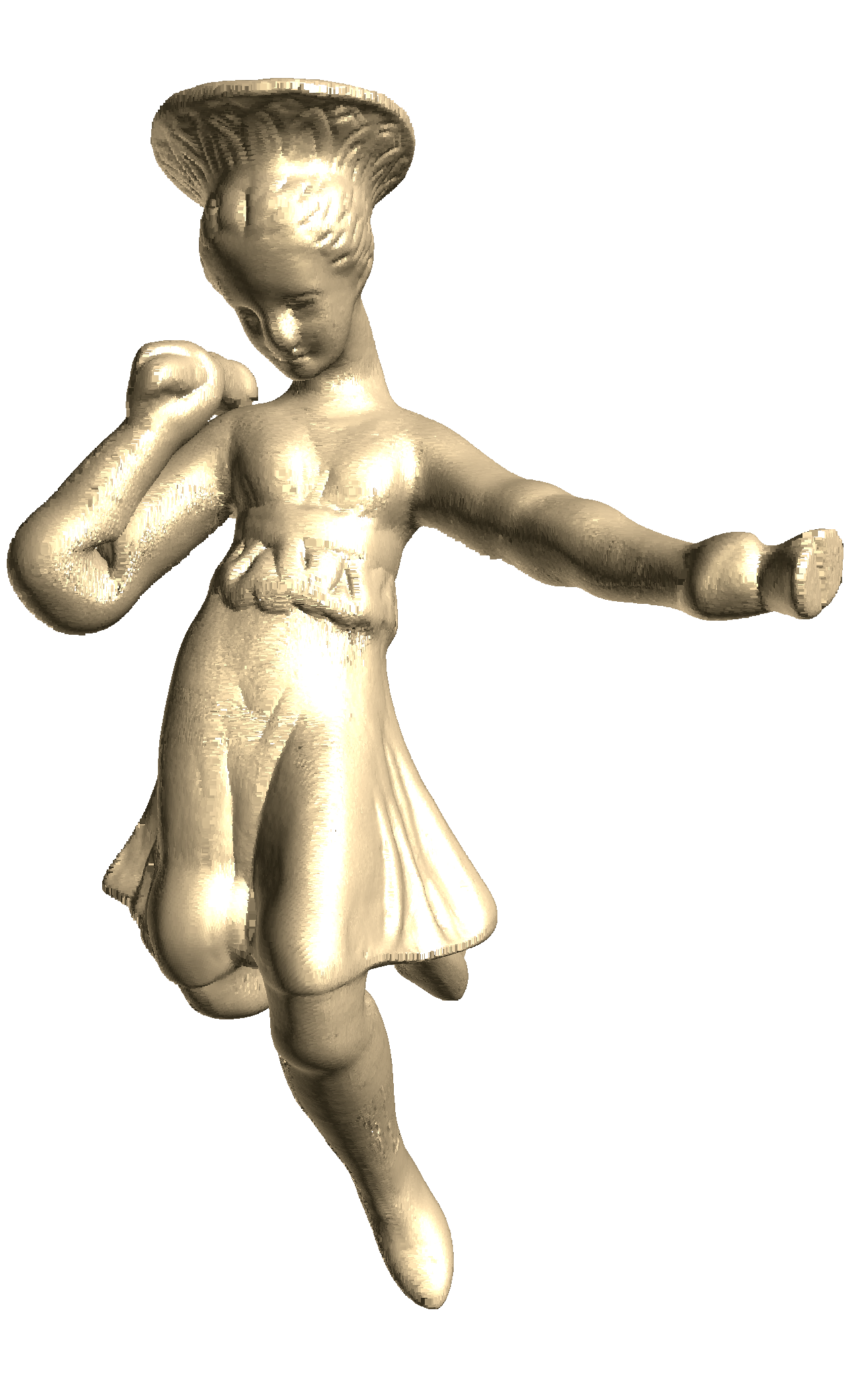}\\
            \includegraphics[height=2cm]{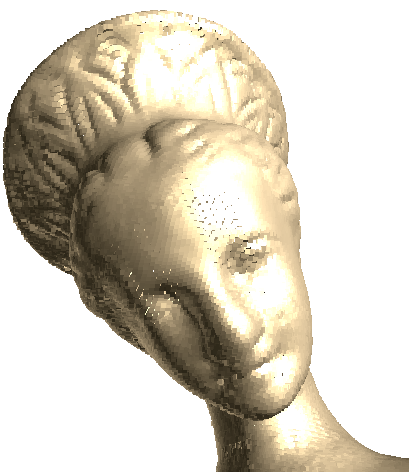}\\
            \includegraphics[height=2cm]{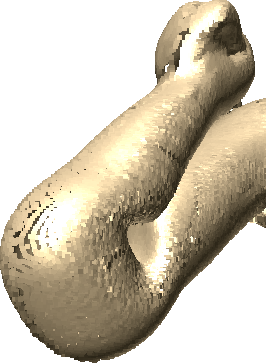}\\
            \includegraphics[height=0.8cm]{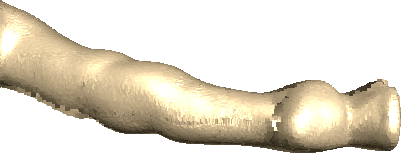}
        \caption{}
        \end{subfigure}
        \begin{subfigure}[ht]{0.11\textwidth}
        \centering
            \includegraphics[height=3cm]{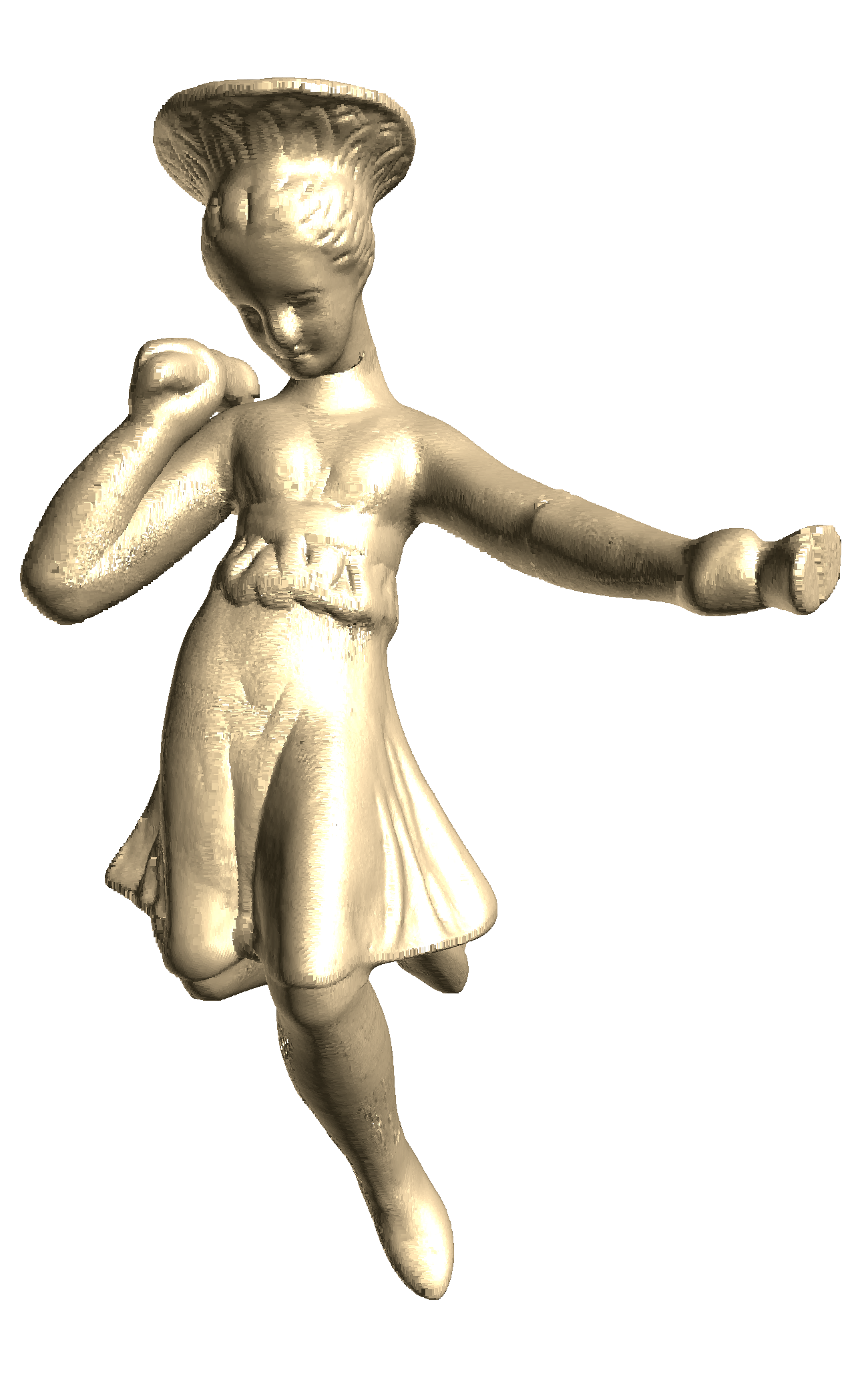}\\
            \includegraphics[height=2cm]{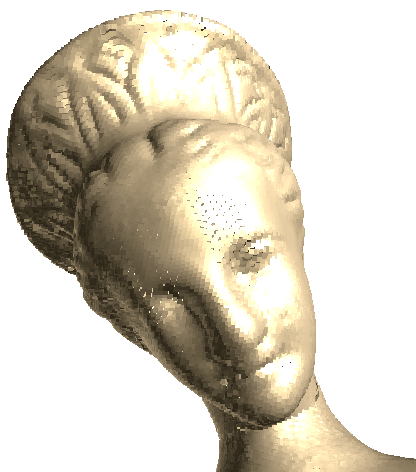}\\
            \includegraphics[height=2cm]{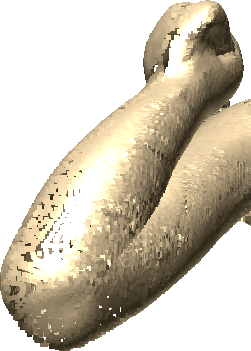}\\
            \includegraphics[height=0.8cm]{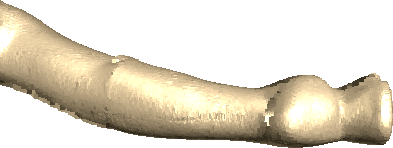}
        \caption{}
        \end{subfigure}
    \caption{Comparison of different skinning methods on the Dancer with Crotales point set. We show close-ups from the second to the fourth row. (a) Pose change with our baseline skinning method, (b) Pose change with Linear blend skinning, (c) Pose change with dual quaternion skinning, (d) Pose change with the method of \cite{h.20201290}.} %\raphq{Dans la thèse se référer à la partie précédente et pas à l'article.}
    \label{fig:result_danseuse}
\end{figure}

\begin{figure}[ht]
    \centering
        \begin{subfigure}[ht]{0.11\textwidth}
            \centering
            \includegraphics[height=3cm]{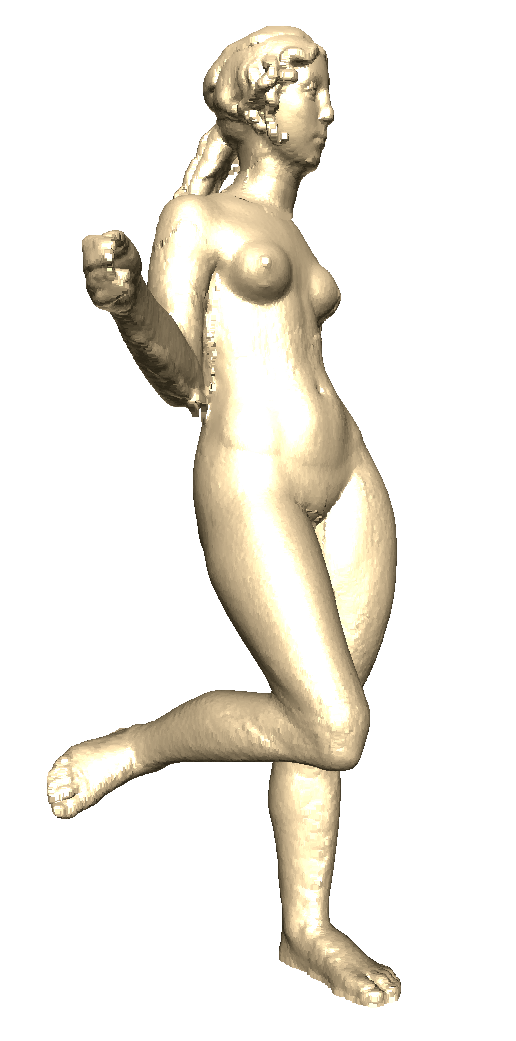}\\
            \includegraphics[height=2cm]{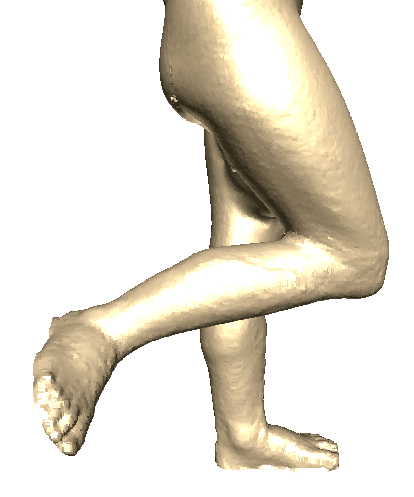}\\
            \includegraphics[height=2cm]{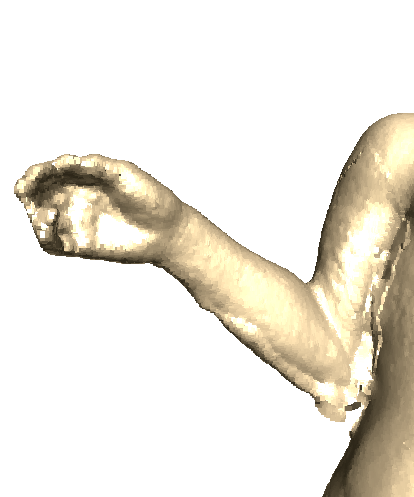}
        \caption{}
        \label{fig:stjean_b}
        \end{subfigure}
        \begin{subfigure}[ht]{0.11\textwidth}
        \centering
            \includegraphics[height=3cm]{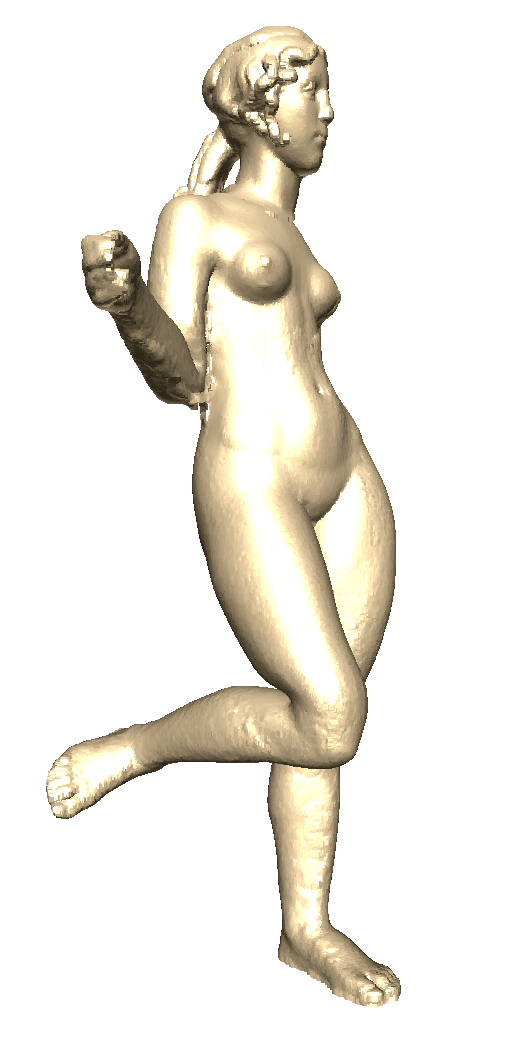}\\
            \includegraphics[height=2cm]{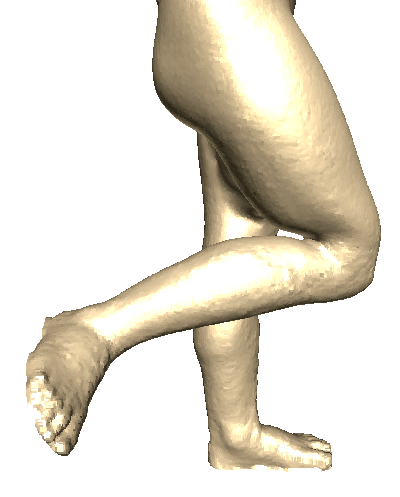}\\
            \includegraphics[height=2cm]{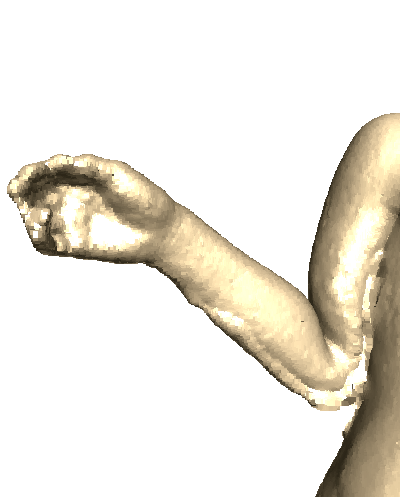}
        \caption{}
        \end{subfigure}  
        \begin{subfigure}[ht]{0.11\textwidth}
        \centering
            \includegraphics[height=3cm]{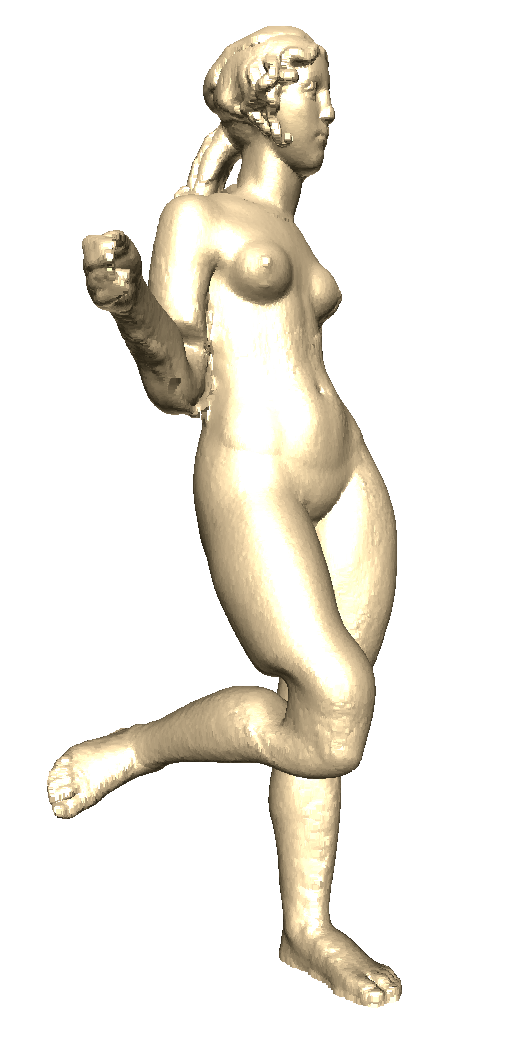}\\
            \includegraphics[height=2cm]{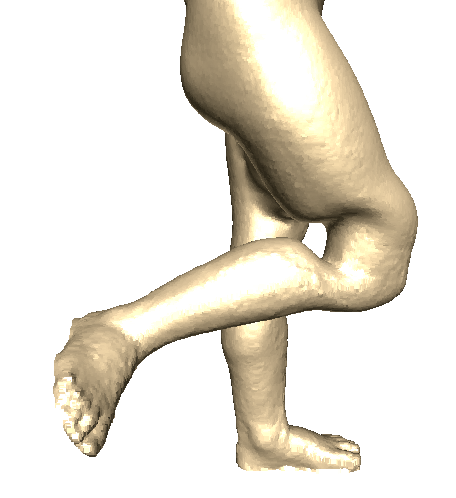}\\
            \includegraphics[height=2cm]{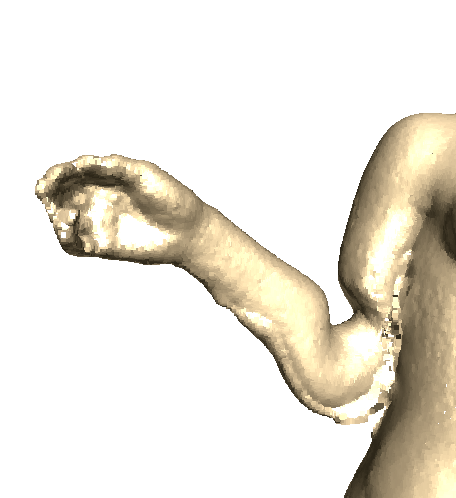}
        \caption{}
        \end{subfigure}  
        \begin{subfigure}[ht]{0.11\textwidth}
        \centering
            \includegraphics[height=3cm]{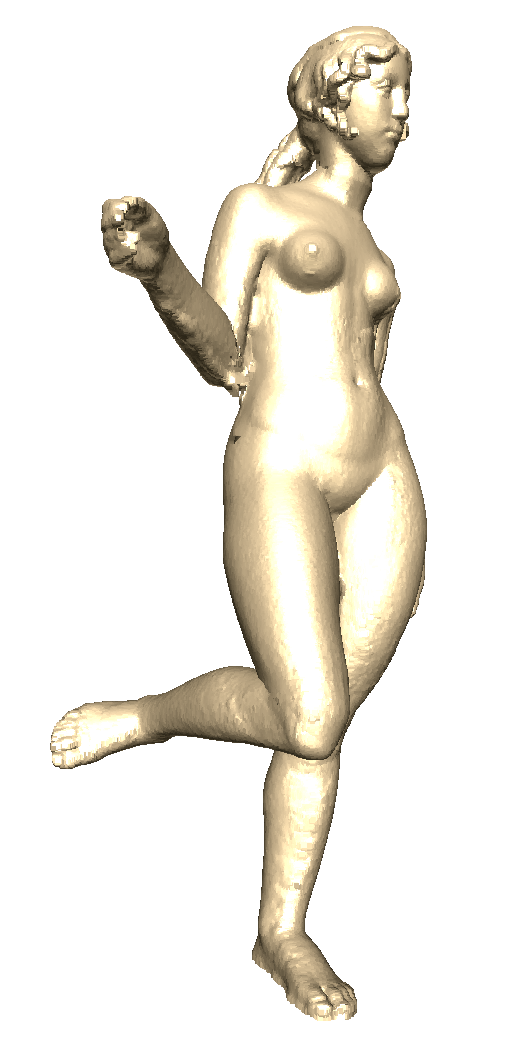}\\
            \includegraphics[height=2cm]{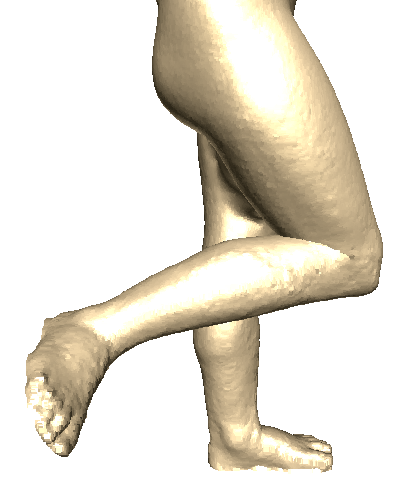}\\
            \includegraphics[height=2cm]{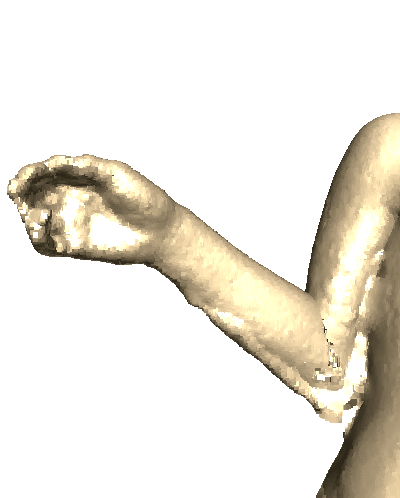}
        \caption{}
        \end{subfigure}
    \caption{Comparison of different skinning methods on the Aphrodite point set. We show close-ups from the second to the third row. (a) Pose change with our baseline skinning method, (b) Pose change with Linear blend skinning, (c) Pose change with dual quaternion skinning, (d) Pose change with the method of \cite{h.20201290}.}% \raphq{Dans la thèse se référer à la partie précédente et pas à l'article.}
    \label{fig:result_aph}
\end{figure}

\begin{figure}[ht]
    \centering
        \begin{subfigure}[ht]{0.11\textwidth}
            \centering
            \includegraphics[height=3cm]{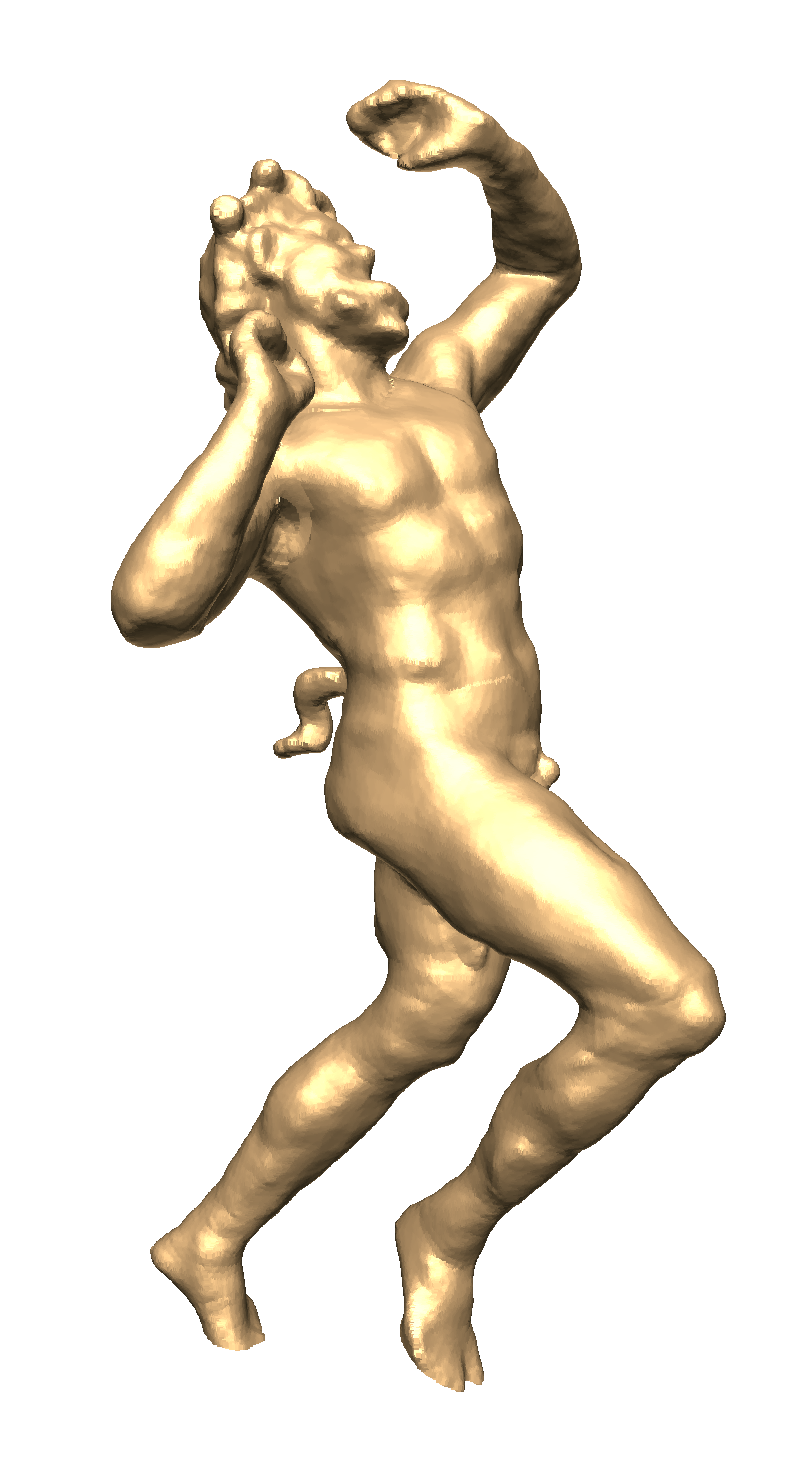}\\
            \includegraphics[height=2cm]{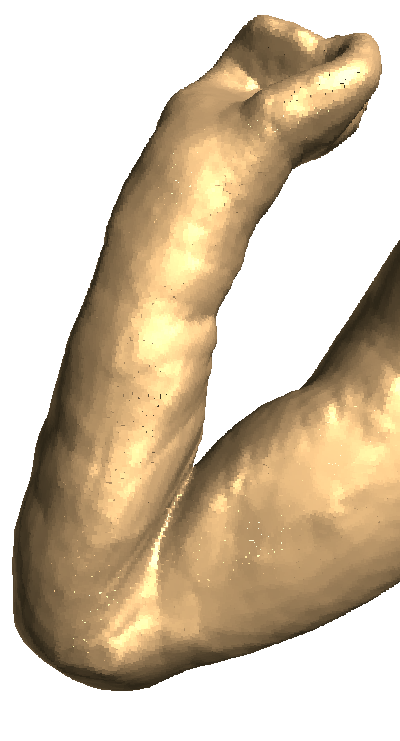}\\
            \includegraphics[height=2cm]{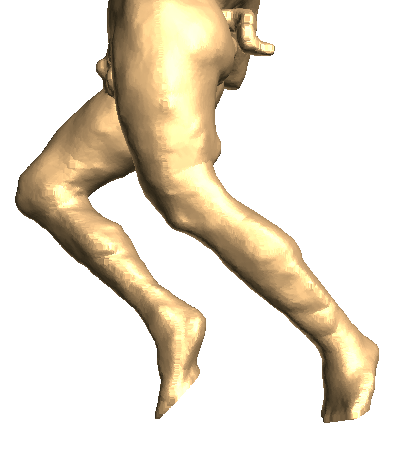}\\
            \includegraphics[height=2cm]{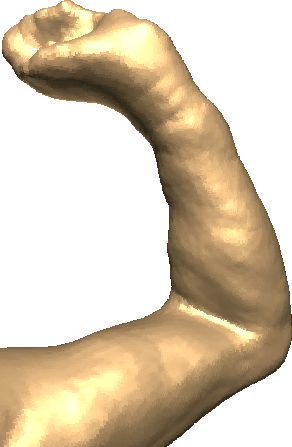}
        \caption{}
        \end{subfigure}
        \begin{subfigure}[ht]{0.11\textwidth}
        \centering
            \includegraphics[height=3cm]{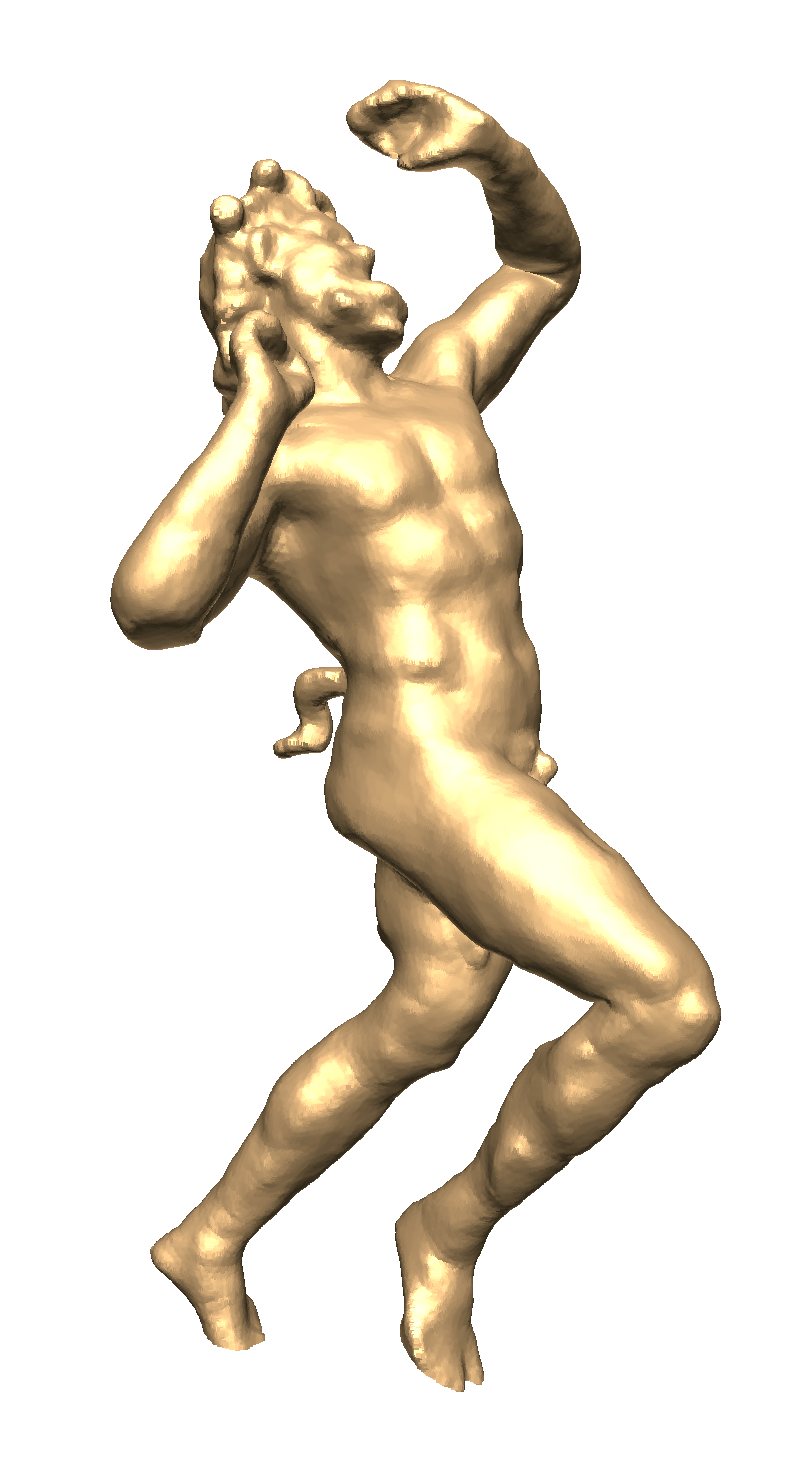}\\
            \includegraphics[height=2cm]{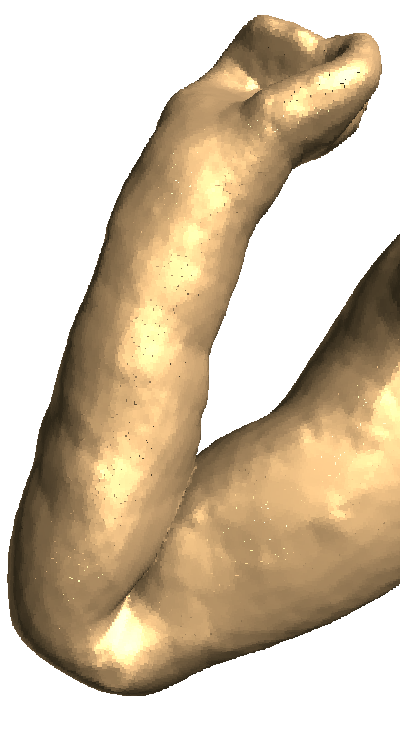}\\
            \includegraphics[height=2cm]{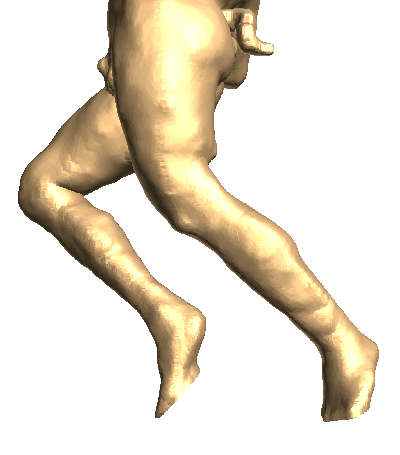}\\
            \includegraphics[height=2cm]{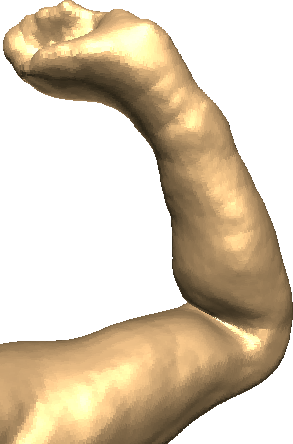}
        \caption{}
        \end{subfigure}  
        \begin{subfigure}[ht]{0.11\textwidth}
        \centering
            \includegraphics[height=3cm]{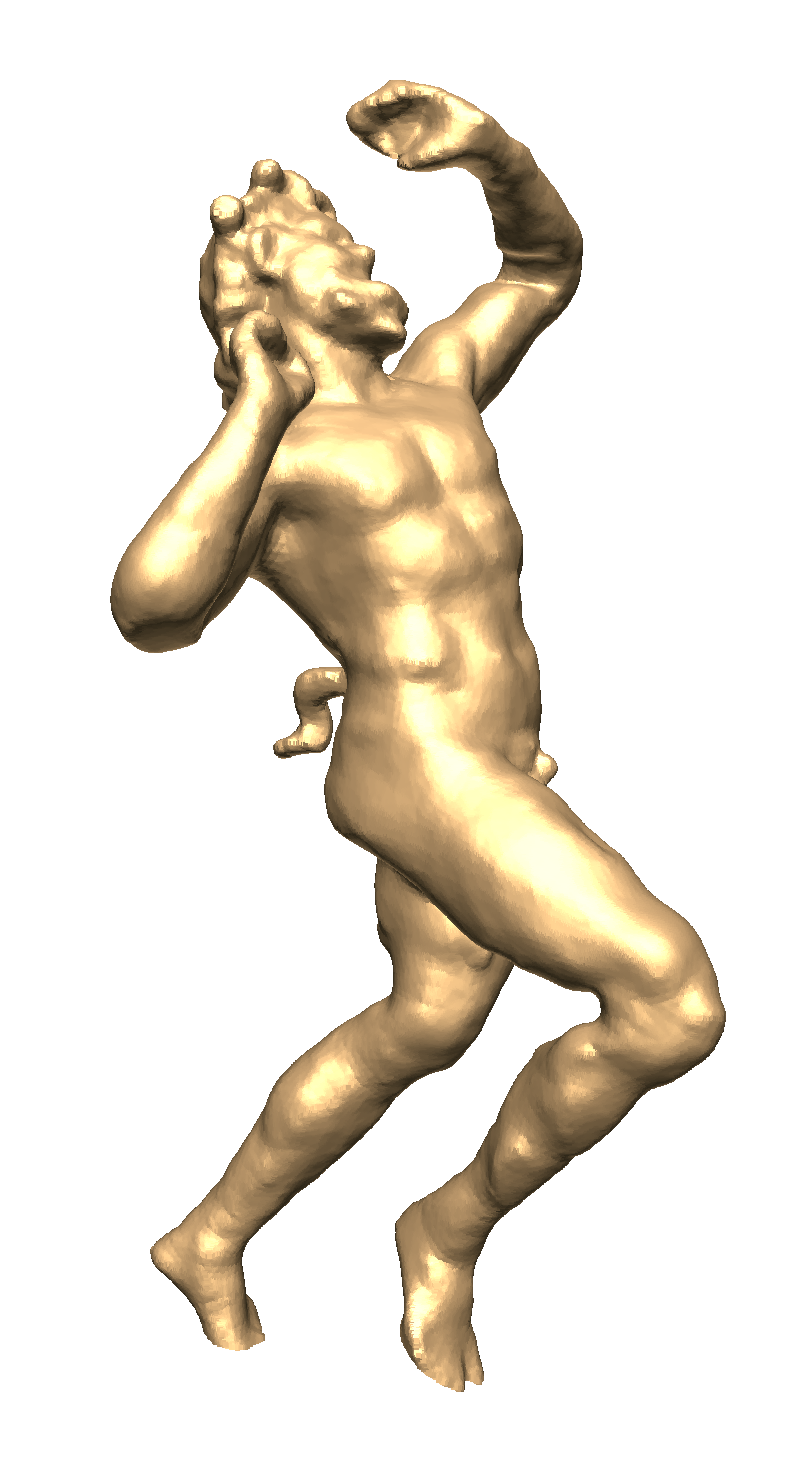}\\
            \includegraphics[height=2cm]{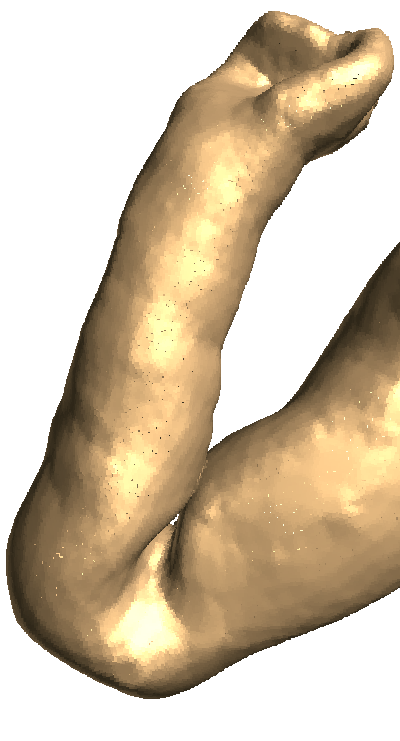}\\
            \includegraphics[height=2cm]{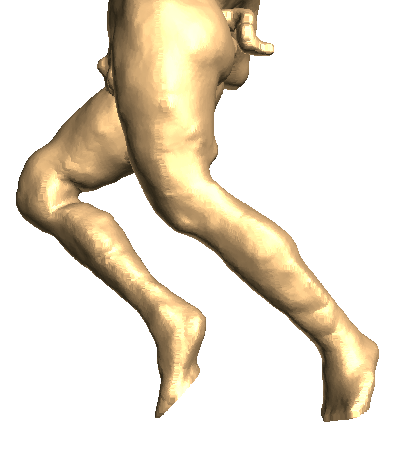}\\
            \includegraphics[height=2cm]{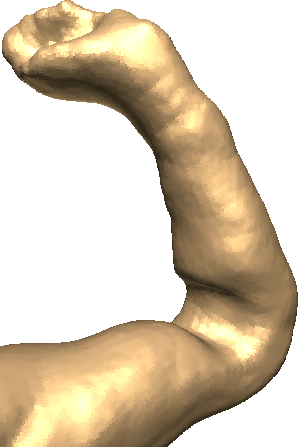}
        \caption{}
        \end{subfigure}  
        \begin{subfigure}[ht]{0.11\textwidth}
        \centering
            \includegraphics[height=3cm]{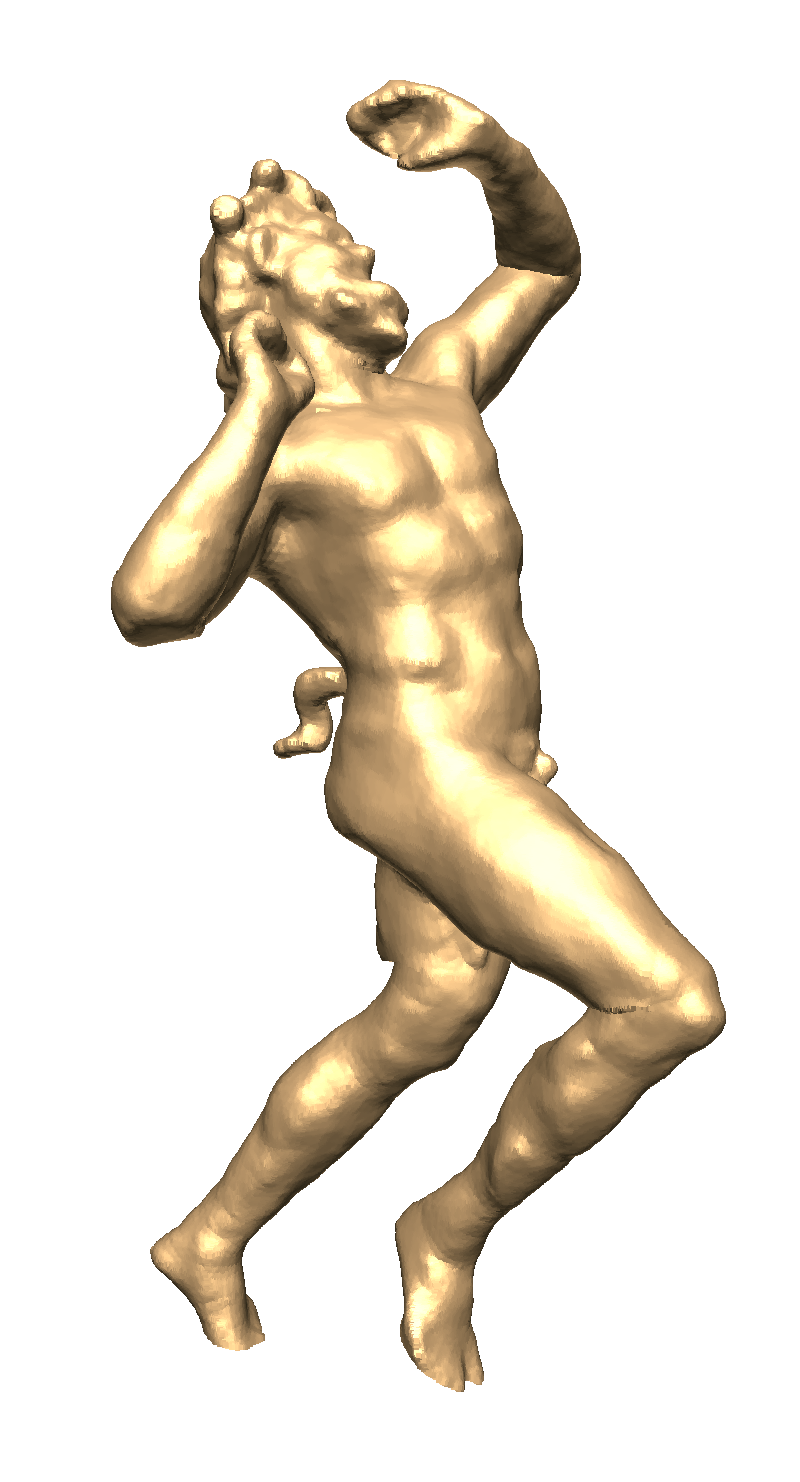}\\
            \includegraphics[height=2cm]{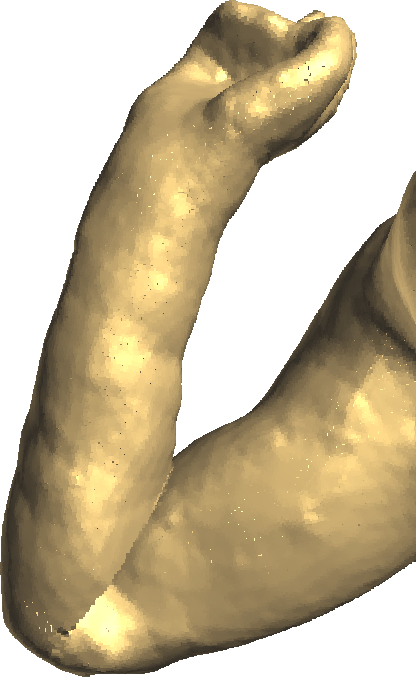}\\
            \includegraphics[height=2cm]{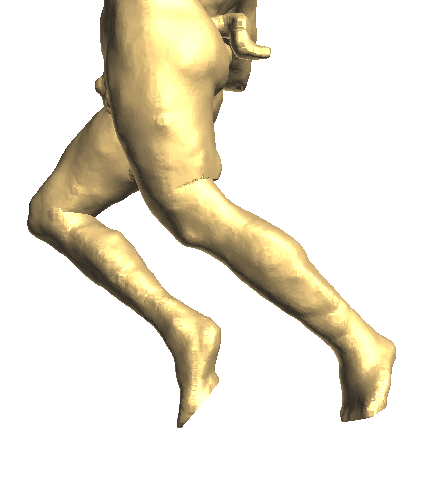}\\
            \includegraphics[height=2cm]{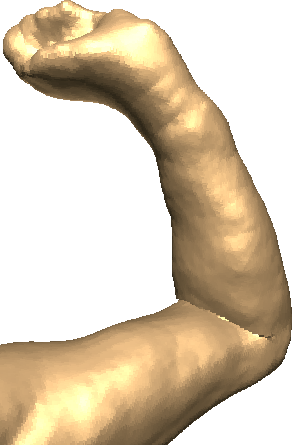}
        \caption{}
        \end{subfigure}
    \caption{Comparison of different skinning methods on the Dancing Faun point set. We show close-ups from the second to the fourth row. (a) Pose change with our baseline skinning method, (b) Pose change with Linear blend skinning, (c) Pose change with dual quaternion skinning, (d) Pose change with the method of \cite{h.20201290}.}%\raphq{Dans la thèse se référer à la partie précédente et pas à l'article.}
    \label{fig:result_faun}
\end{figure}

\begin{figure}[ht]
    \centering
        \begin{subfigure}[ht]{0.11\textwidth}
            \centering
            \includegraphics[height=3.5cm]{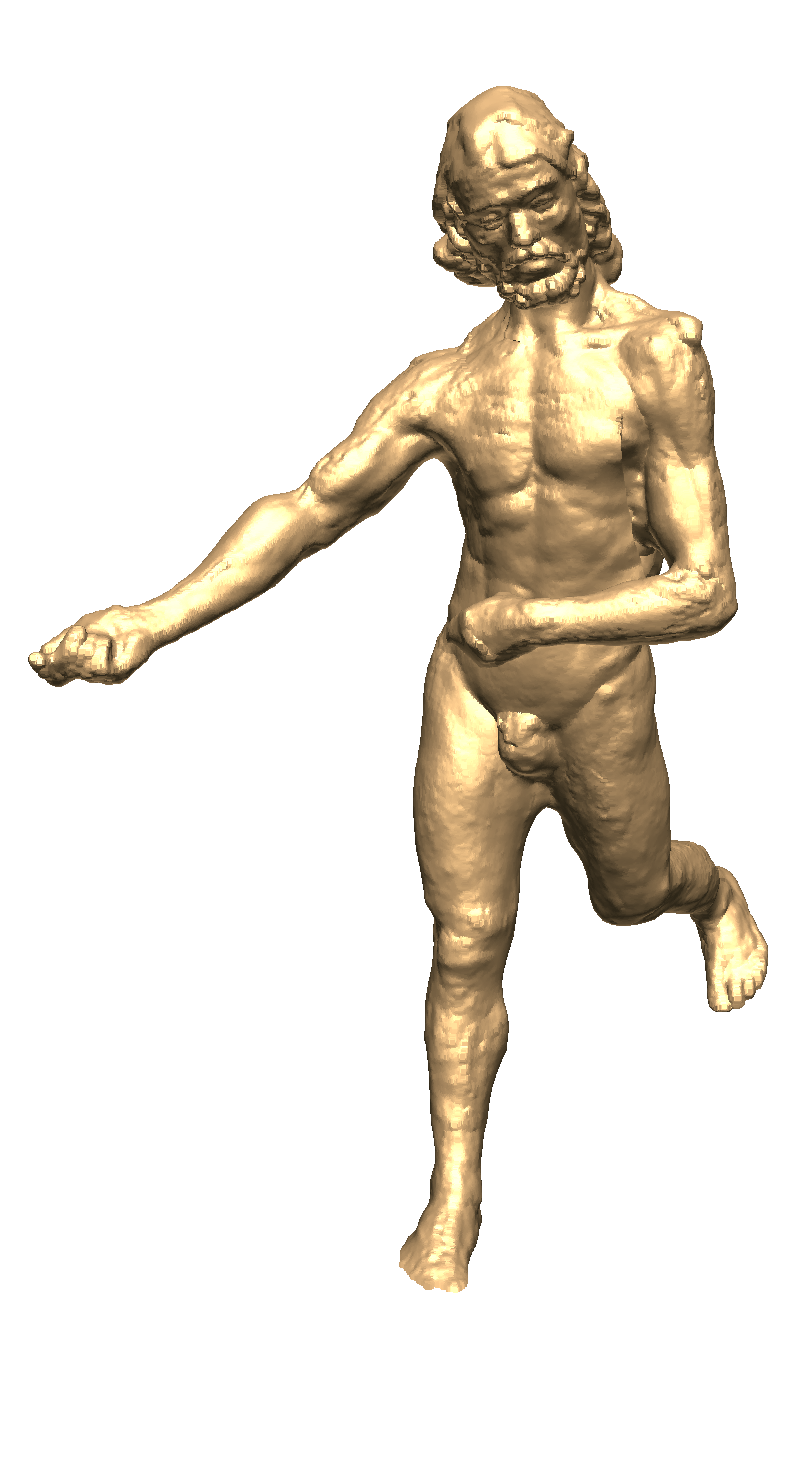}\\
            \includegraphics[height=2cm]{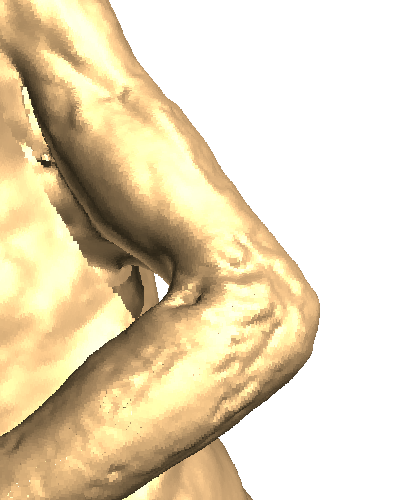}\\
            \includegraphics[height=2cm]{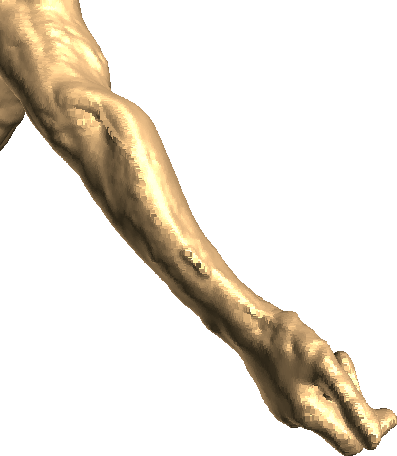}\\
            \includegraphics[height=2cm]{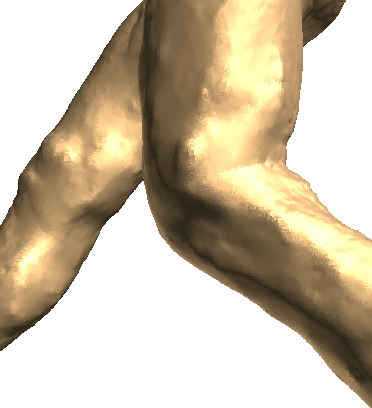}
        \caption{}
        \end{subfigure}
        \begin{subfigure}[ht]{0.11\textwidth}
        \centering
            \includegraphics[height=3.5cm]{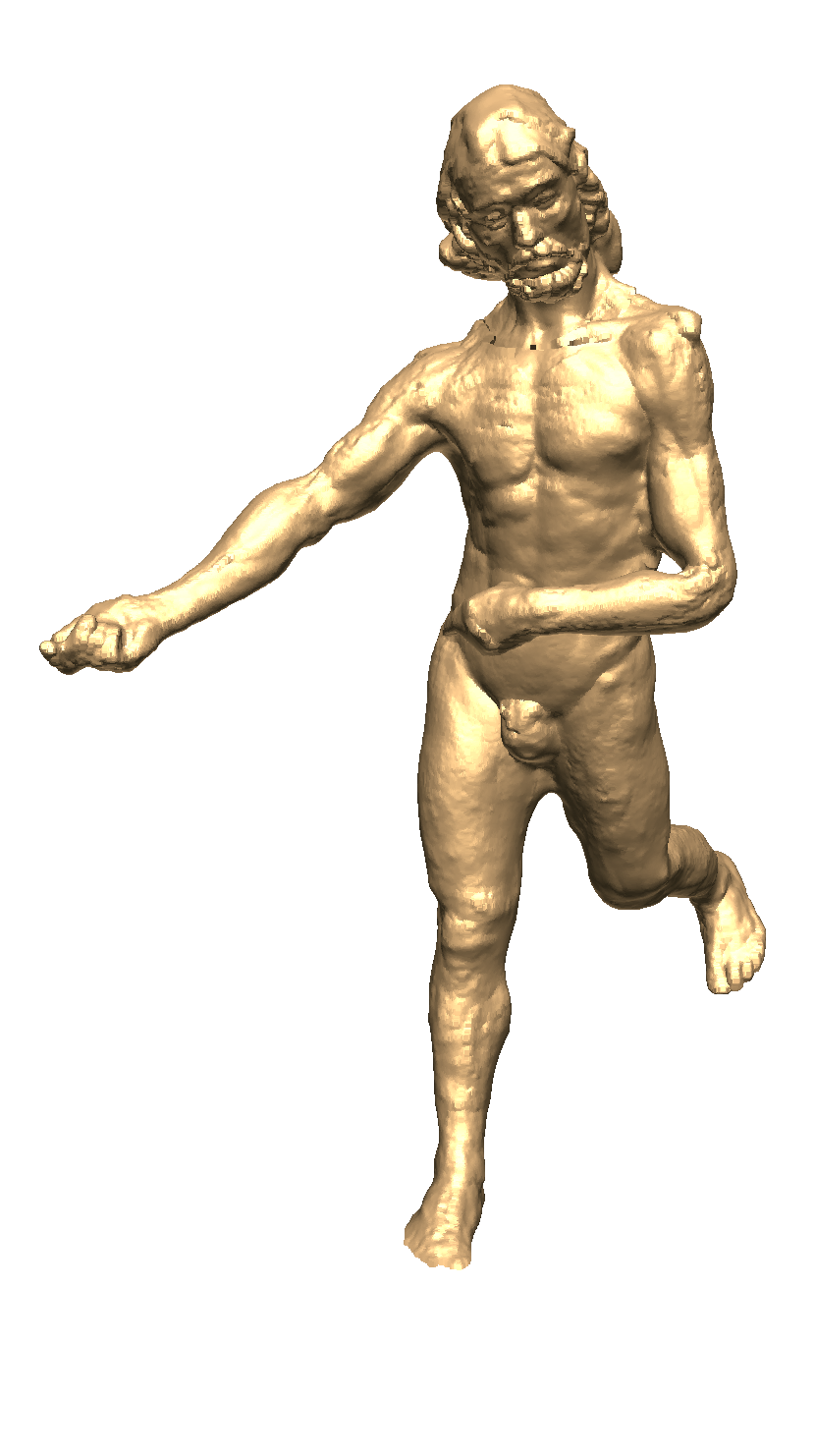}\\ 
            \includegraphics[height=2cm]{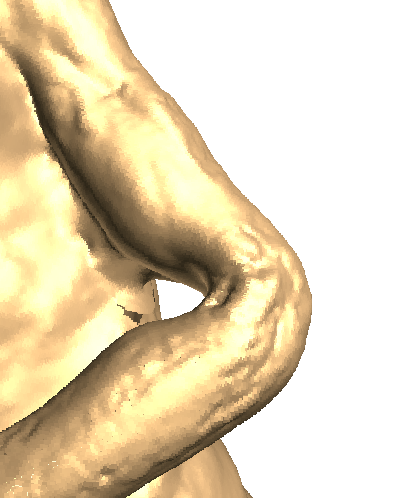}\\
            \includegraphics[height=2cm]{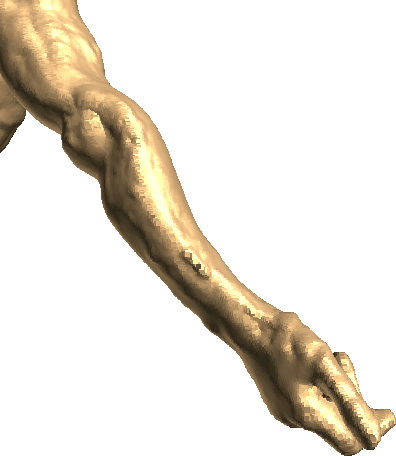}\\
            \includegraphics[height=2cm]{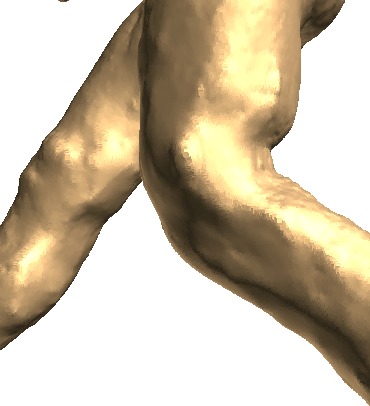}
        \caption{}
        \end{subfigure}  
        \begin{subfigure}[ht]{0.11\textwidth}
        \centering
            \includegraphics[height=3.5cm]{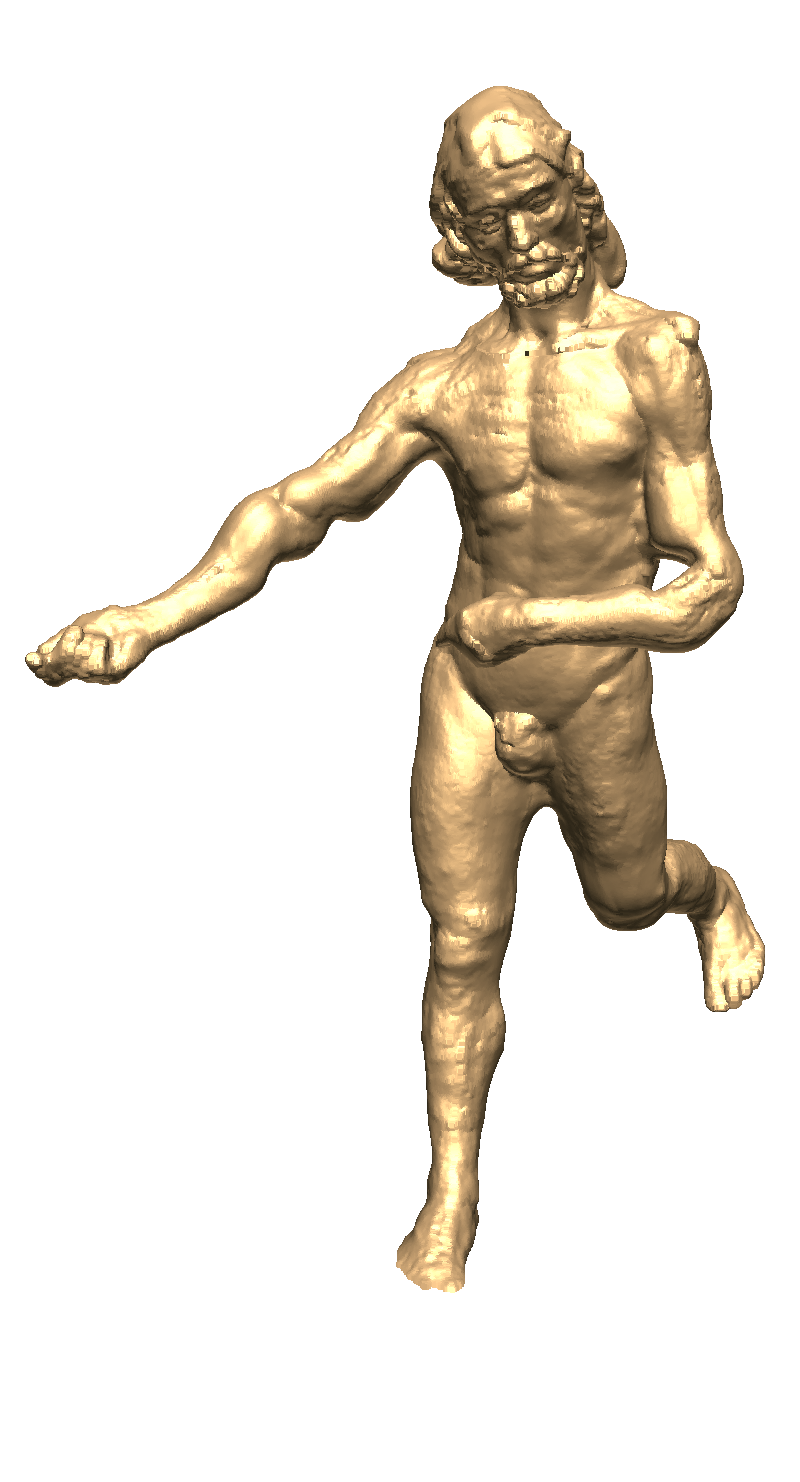}\\
            \includegraphics[height=2cm]{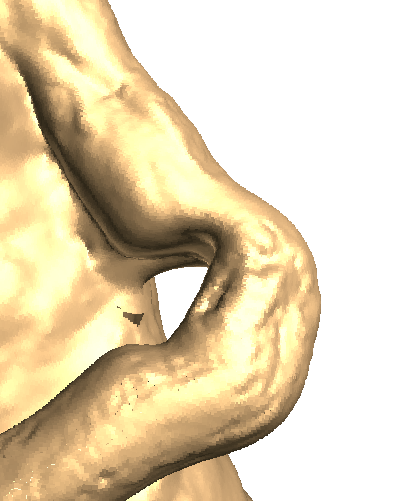}\\
            \includegraphics[height=2cm]{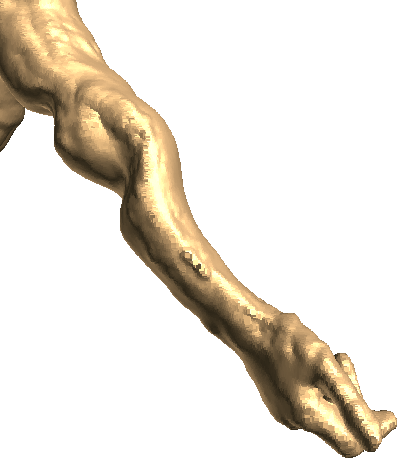}\\
            \includegraphics[height=2cm]{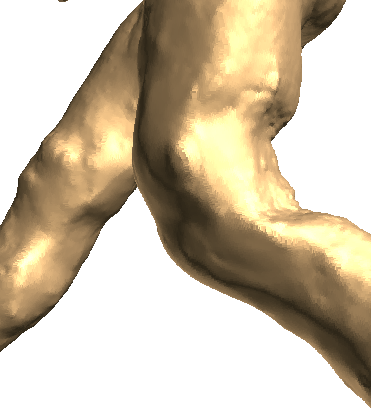}
        \caption{}
        \end{subfigure}  
        \begin{subfigure}[ht]{0.11\textwidth}
        \centering
            \includegraphics[height=3.5cm]{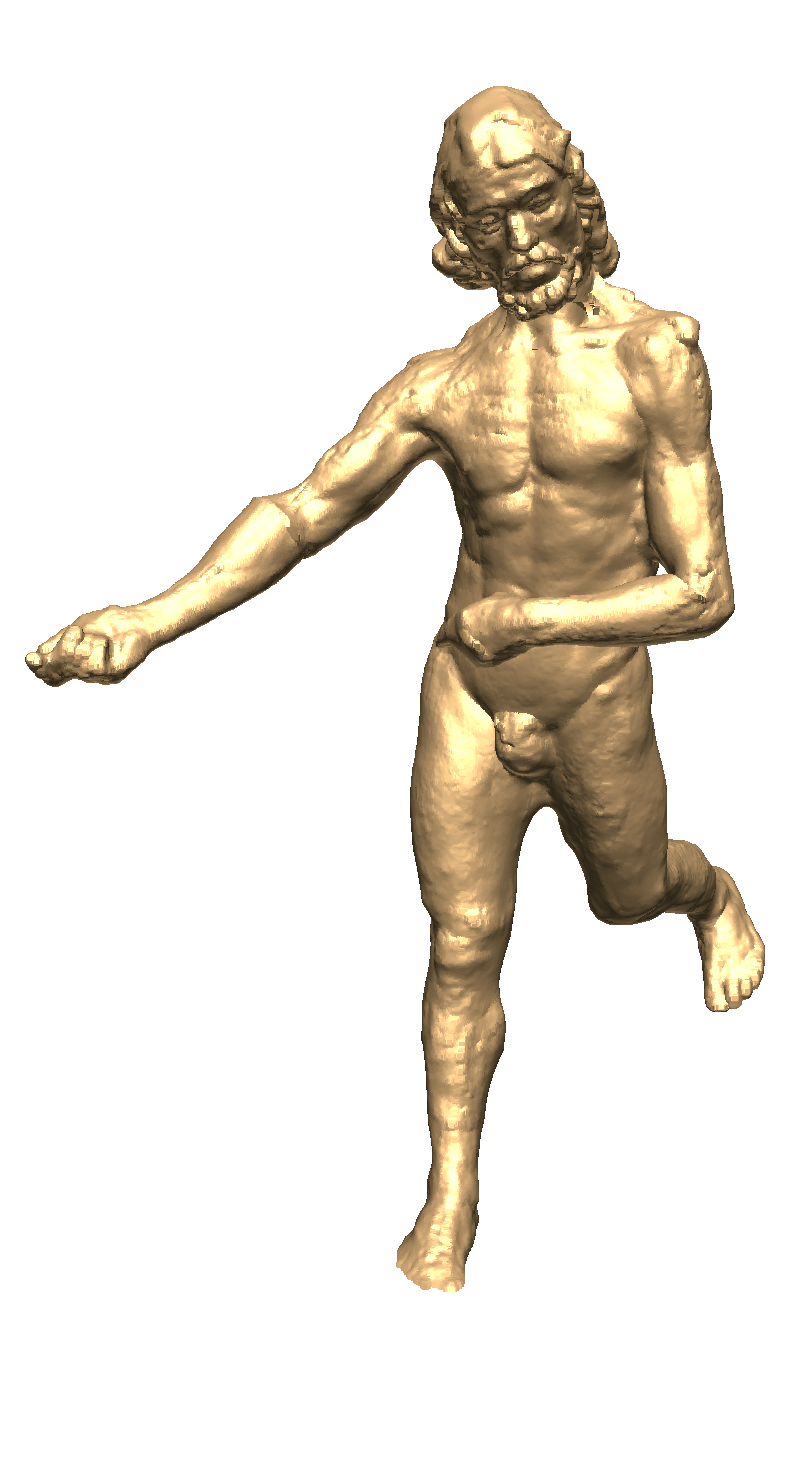}\\
            \includegraphics[height=2cm]{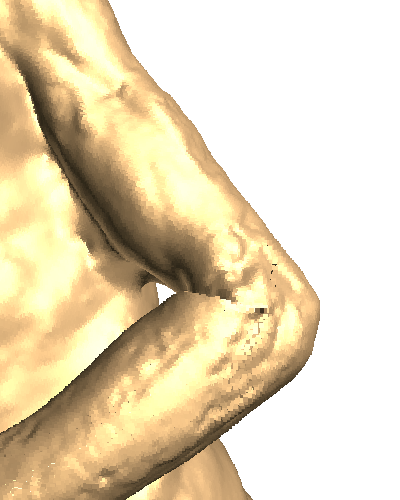}\\
            \includegraphics[height=2cm]{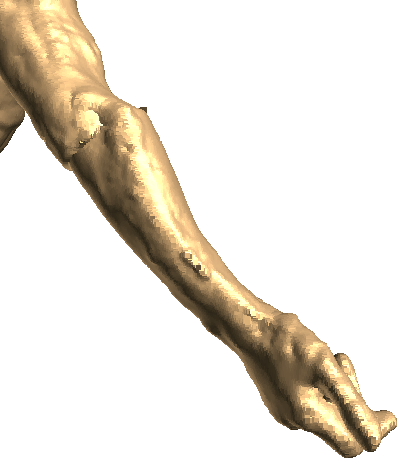}\\
            \includegraphics[height=2cm]{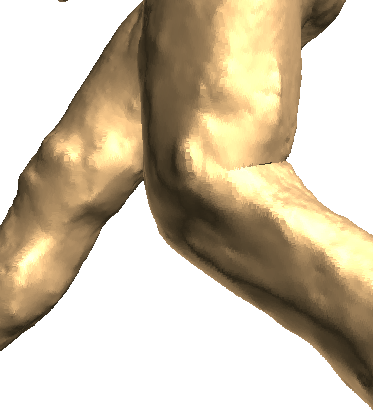}
        \caption{}
        \end{subfigure}
    \caption{Comparison of different skinning methods on the Saint John the Baptist point set. We show close-ups from the second to the fourth row. (a) Pose change with our baseline skinning method, (b) Pose change with Linear blend skinning, (c) Pose change with dual quaternion skinning, (d) Pose change with the method of \cite{h.20201290}.} %\raphq{Dans la thèse se référer à la partie précédente et pas à l'article.}
    \label{fig:result_stjean}
\end{figure}

\begin{figure}[ht]
    \centering
        \begin{subfigure}[ht]{0.225\textwidth}
            \centering
            \includegraphics[width=0.48\textwidth]{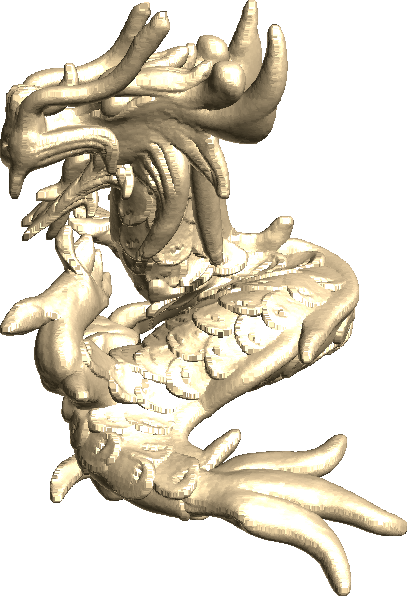}
            \includegraphics[width=0.48\textwidth]{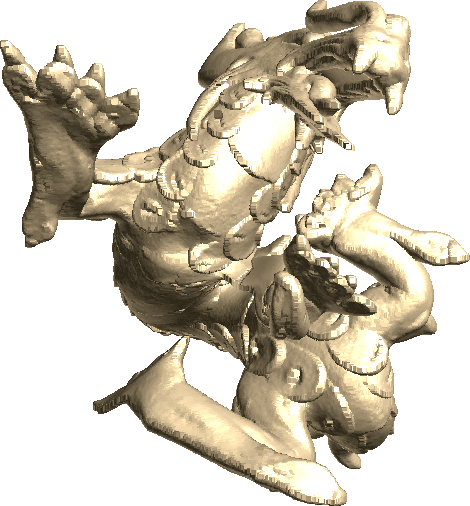}
        \caption{Baseline skinning method}
        \end{subfigure}
        \begin{subfigure}[ht]{0.225\textwidth}
        \centering
            \includegraphics[width=0.48\textwidth]{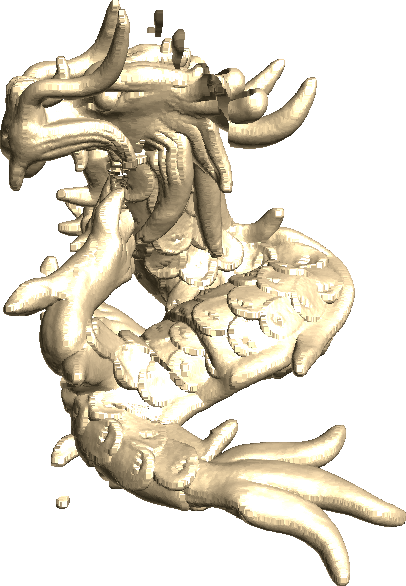}
            \includegraphics[width=0.48\textwidth]{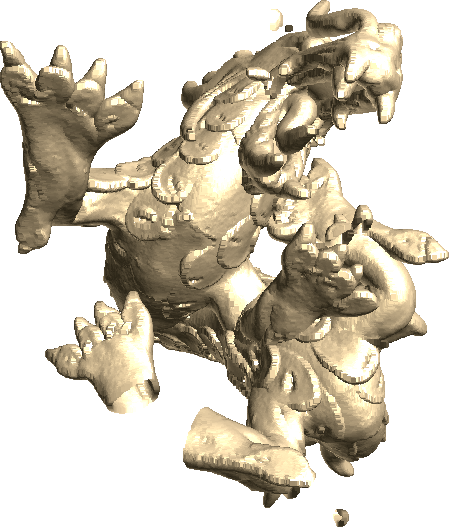}
        \caption{Linear blend skinning}
        \end{subfigure}
        
        \begin{subfigure}[ht]{0.25\textwidth}
            \centering
            \includegraphics[width=0.32\textwidth]{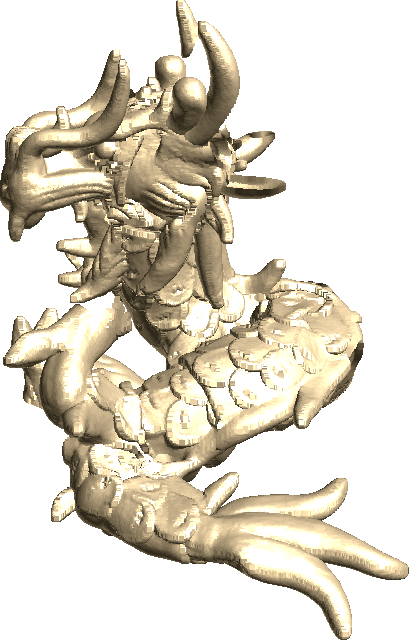}
            \includegraphics[width=0.32\textwidth]{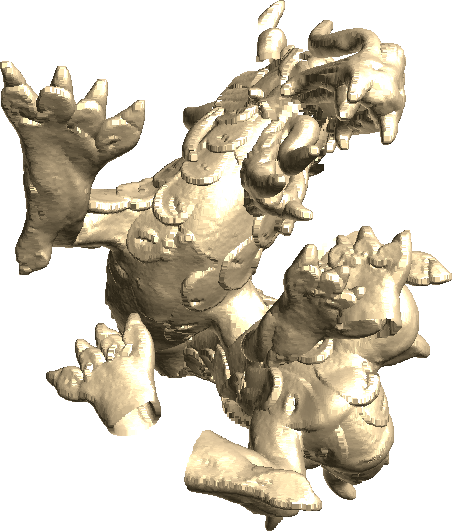}
            \includegraphics[width=0.32\textwidth]{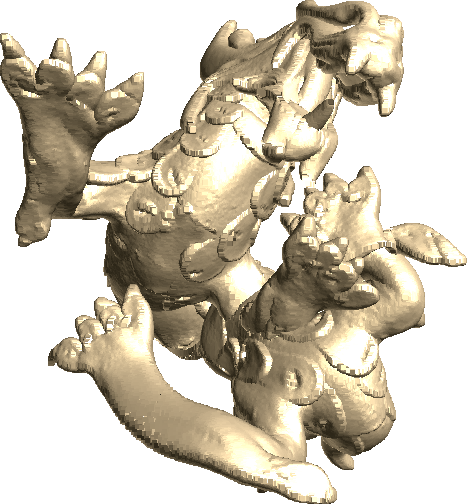}
        \caption{Dual quaternion skinning}
        \label{fig:dragon_dq}
        \end{subfigure}
        \begin{subfigure}[ht]{0.2\textwidth}
        \centering
            \includegraphics[width=0.48\textwidth]{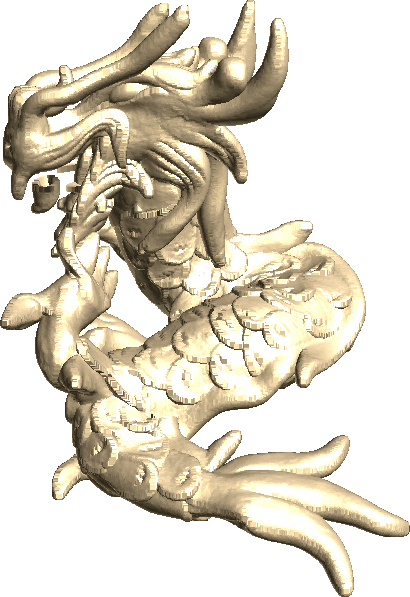}
            \includegraphics[width=0.48\textwidth]{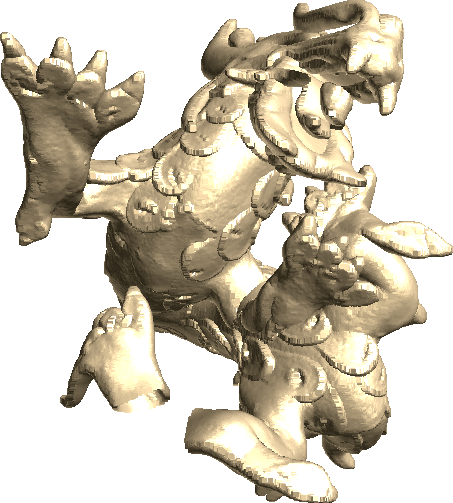}
        \caption{Skinning method of \cite{h.20201290}}
        \end{subfigure}  
    \caption{Comparison of different skinning methods on the dragon point set under two points of view.}
    \label{fig:result_dragon}
\end{figure}

Figures  \ref{fig:result_danseuse}, \ref{fig:result_aph}, \ref{fig:result_faun}, \ref{fig:result_stjean} and \ref{fig:result_dragon} show our baseline skinning results compared with common skinning algorithms.
Compared with Linear blend skinning or Dual-quaternion skinning, our baseline skinning method improves the collapse or bulge effect at joint and gives a more reasonable result for the point set. This improvement can be observed especially at joints which move a lot, such as the bending left arm of Dancer with Crotales (Figure \ref{fig:result_danseuse}), the left leg and the left arm of Aphrodite (Figure \ref{fig:result_aph}), the arms of the Dancing Faun (Figure \ref{fig:result_faun}) and the right arm and the right leg of the Saint John the Baptist (Figure \ref{fig:result_stjean}). Compared with the method in \cite{h.20201290}, which can also improve these artefacts, our method preserve the information of the original point set better. For example, we can better observe the sculpture texture on the right arm of Saint John (Fig. \ref{fig:result_stjean}) with our skinning result than the result by the method of \cite{h.20201290}. Furthermore, the method in \cite{h.20201290} does not show points that are hidden inside the sphere-mesh model after deformation and creates a discontinuity for the point set even if the skinning result is visually pleasant. 
Figure \ref{fig:result_dragon} shows a multi-layer point set example. Our baseline skinning method gives the best result among four methods, especially at the head and the tail where the details are complicate. We do not set bones for the claws so we consider the claws as a multi-layer points over the dragon's body. We can observe the cracked horns in the first point of view and a cracked back claw in the second point of view for the LBS and dual quaternion skinning methods. The cracked effect can be resolved by adjusting the influence weight, but the result is not pertinent (see the third one in Figure \ref{fig:dragon_dq}). It is difficult for LBS and dual quaternion method to deal with the multi-layer data. The method of \cite{h.20201290} performs better for multi-layer data than LBS and dual quaternion skinning method. But it also has a cracked claw because they use orthogonal projection when they compute their height field. The orthogonal projection induces a discontinuity at concave parts of the bones. 
As far as computation times are concerned, Table \ref{tab:time} shows times of the four methods. Our method is faster than the method of Fu et al. \cite{h.20201290} and is similar to Dual-quaternion skinning and Linear Blend Skinning. 

We compare our method with the skinning method presented in the paper of original sphere-mesh model~\cite{Thiery13}. Although they use the sphere-mesh model as the skeleton of the skinning, the artefacts of the classical skinning methods remains. Our method does not lose the volume and preserves well the original connectivity (See the stripes in Figure \ref{fig:comp_baseline}).).

 \begin{figure}[ht]
    \centering
    \begin{subfigure}[ht]{0.12\textwidth}
        \centering
        \includegraphics[width=\textwidth]{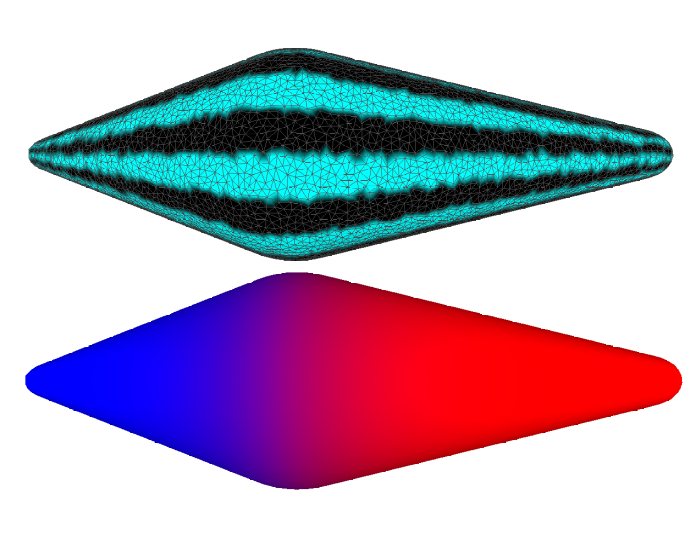}
        \caption{Skinning weights computation using \cite{Thiery13}}
        \label{fig:comp_sphermesh_init}
    \end{subfigure}
     \begin{subfigure}[ht]{0.11\textwidth}
        \centering
        \includegraphics[width=\textwidth]{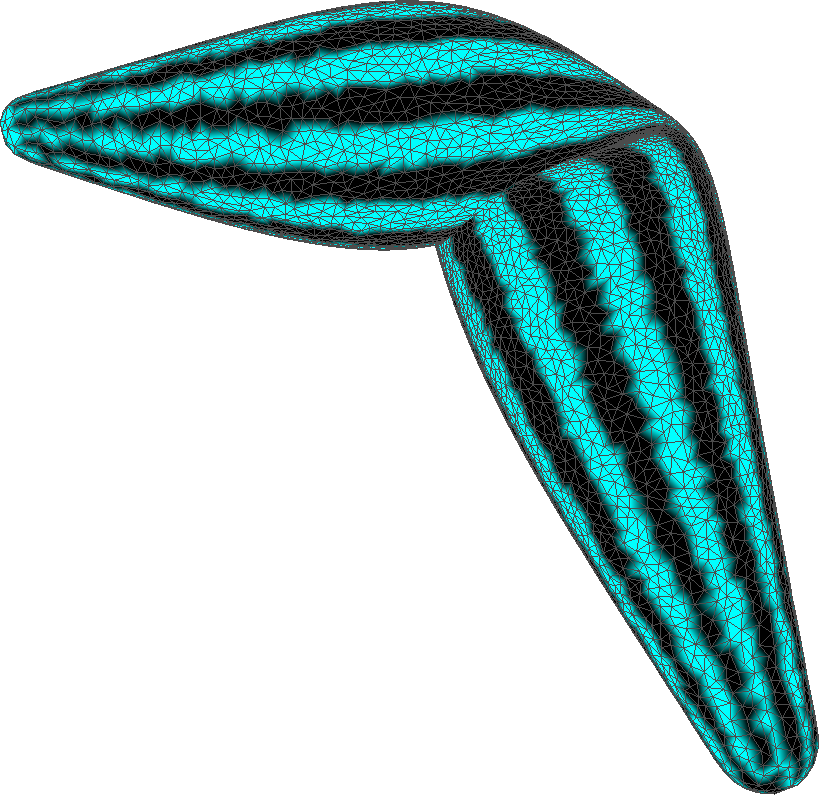}
        \caption{Sphere-mesh skinning \cite{Thiery13} by LBS}
        \label{fig:comp_spheremesh_result}
    \end{subfigure}
     \begin{subfigure}[ht]{0.11\textwidth}
        \centering
        \includegraphics[width=\textwidth]{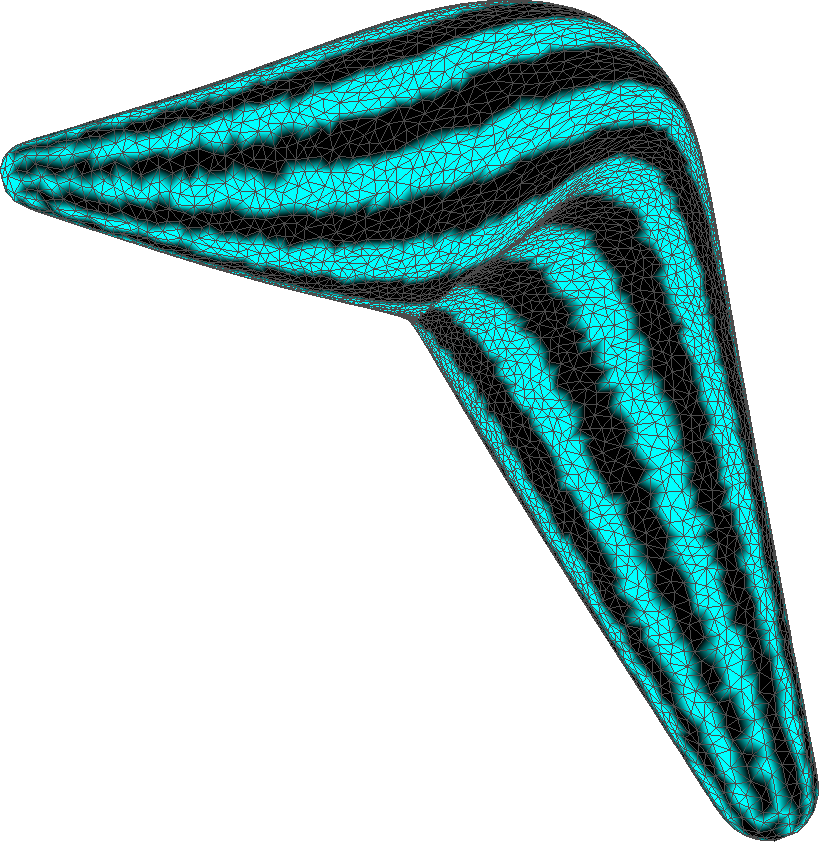}
        \caption{sphere-mesh skinning \cite{Thiery13} by DQS}
        \label{fig:comp_baseline}
    \end{subfigure}
     \begin{subfigure}[ht]{0.11\textwidth}
        \centering
        \includegraphics[width=\textwidth]{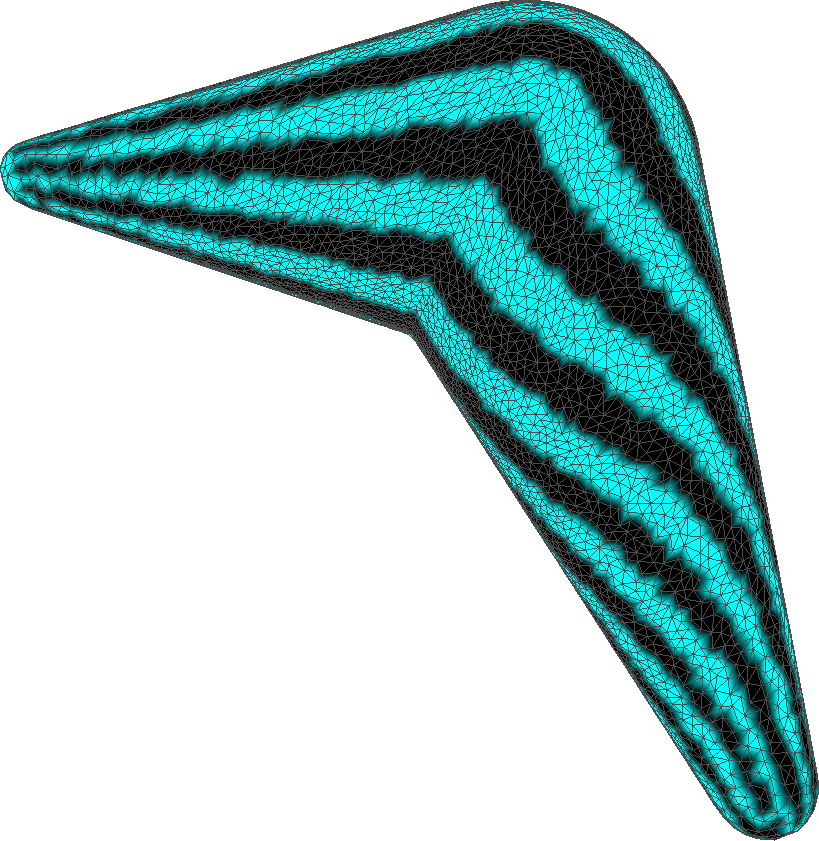}
        \caption{Baseline skinning result}
        \label{fig:comp_baseline}
    \end{subfigure}
    \caption{Comparison with the weight-based skinning approach using sphere-mesh skeletons \cite{Thiery13}. The movement of the right bone is a combination of bend and twist.}
    \label{fig:comp_thiery}
\end{figure}

\paragraph{Limitations}
Our baseline skinning method has several limitations. First, an unfolding motion raises a missing data problem since we do not have a mesh of the complete body, such as for the right arm of the Dancer (fourth column in Figure \ref{fig:danseuse_b}) or the left arm of Saint John (second column in Figure \ref{fig:stjean_b}), to alleviate this limitation, one could use a patch-based 3D inpainting process or a 1D inpainting process restricted to baselines. Another limitation is that our skinning method can only deal with points influenced by a single chain of bones and cannot be easily adapted to deal with points influenced by several different bone chains. This limitation can have an impact on examples such as the Dancer's dress which should be influenced by the motions of both legs. However, this case is clearly one where the very hypothesis of a sphere-mesh model based skinning reaches its limits, it would require cloth modeling and animation techniques.

%\raphq{PENSER A AJOUTER DES CLOSE-UPS DE LA TEXTURE GEOMETRIQUE QUAND ON PLIE DEPLIE UNE ARTICULATION OU QU'ON TOURNE UN OS SUR SON AXE.}

%\raphq{Pour l’article, il faut absolument que tu rajoutes des légendes à l’ensemble des illustrations avec des explications claires sur ce que l’on voit sur chaque image : Etat initial, Après avoir plié l’articulation, Après un dépliement, Après un twist, ... , Evolution des segments de baseline par interpolation linéraire ou quadratique, évolution de la direction du détail par évolution lineaire ou quadratique [d’ailleurs au final qu’est-ce qui est le mieux?].}

\section{Conclusion and perspectives}
We introduced a novel baseline skinning method for point sets of articulated bodies. Our method does not require a weight computation for each point and gives a realistic skinning result when changing the pose and morphology of the articulated body, even when there is no information on the underlying muscles. As a future work, we plan to design displacement fields above baselines to mimic muscles or to take into account additional folds of the skin. We also want to investigate how we can associate different bone chains in order to deal with points which are influenced by three or more bones. It would also be necessary to improve the treatment of the deformations of the junctions because for the moment the proposed solution will involve tightening/stretching effect between neighbouring baselines. We provide a solution to the case where an inpainting problem is raised in concave areas at unfolding/unbending of a joint. However, when the problem is too severe, it can be interesting to develop an other approach, such as patch-based approaches.

\section*{Acknowledgments}
This work was funded by the e-Roma project from the French Agence Nationale de la Recherche (ANR-16-CE38-0009).
The \emph{Dancer with crotales} model is a point set of the Farman Dataset \cite{ipol.2011.dalmm_ps}. The other statues data are sampled on meshes from the Sketchfab website: the \emph{Dragon} model is courtesy of Sketchfab user "3D graphics 101
", the \emph{Dancing Faun} model is courtesy of Moshe Caine and the 3 other statues models (\emph{Aphrodite}, \emph{Saint John the Baptist} and \emph{Old Fisherman}) are courtesy of Geoffrey Marchal.

\section*{References}

\bibliography{refs}

\end{document}